\documentclass[useAMS,usenatbib]{mn2e}
\usepackage{epsfig}

\title[Rotation periods in Praesepe]{Rotation periods for very low mass stars in Praesepe}

\author[Scholz et al.]{Alexander Scholz$^{1}$\thanks{E-mail:
aleks@cp.dias.ie}, Jonathan Irwin$^{2}$, Jerome Bouvier$^{3}$, Brigitta M. Sip\H{o}cz$^{4}$,
\newauthor{Simon Hodgkin$^{5}$, Jochen Eisl{\"o}ffel$^{6}$}\\
$^{1}$ School of Cosmic Physics, Dublin Institute for Advanced Studies, 31 Fitzwilliam Place, Dublin 2, Ireland\\
$^{2}$ Harvard-Smithsonian Center for Astrophysics, 60 Garden St., Cambridge, MA 02138, USA\\ 
$^{3}$ Laboratoire d'Astrophysique, Observatoire de Grenoble, BP 53, F-38041 Grenoble Cedex 9, France\\
$^{4}$ Centre for Astrophysics Research, University of Hertfordshire, Hatfield, AL10 9AB, UK\\
$^{5}$ Institute of Astronomy, University of Cambridge, Madingley Road, Cambridge CB3 0HA, UK\\
$^{6}$ Th{\"u}ringer Landessternwarte Tautenburg, Sternwarte 5, D-07778 Tautenburg, Germany}

\begin{document}

\date{Accepted. Received.}

\pagerange{\pageref{firstpage}--\pageref{lastpage}} \pubyear{2002}

\maketitle

\label{firstpage}

\begin{abstract}
We investigate the rotation periods of fully convective very low mass stars (VLM, $M<0.3\,M_{\odot}$), 
with the aim to derive empirical constraints for the spindown due to magnetically driven stellar winds. Our 
analysis is based on a new sample of rotation periods in the main-sequence cluster Praesepe (age 
600\,Myr). From photometric lightcurves obtained with the Isaac Newton Telescope, we measure rotation
periods for 49 objects, among them 26 in the VLM domain. This enlarges the period sample in this mass
and age regime by a factor of 6. Almost all VLM objects in our sample are fast rotators with periods 
$<2.5$\,d, in contrast to the stars with $M>0.6\,M_{\odot}$ in this cluster which have periods of 7-14\,d.
Thus, we confirm that the period-mass distribution in Praesepe exhibits a radical break at 
$M\sim 0.3-0.6\,M_{\odot}$. Our data indicate a positive period-mass trend in the VLM regime, 
similar to younger clusters. In addition, the scatter of the periods increases with mass. 
For the $M>0.3\,M_{\odot}$ objects in our sample the period distribution is probably
affected by binarity. By comparing the Praesepe periods with literature samples in the cluster 
NGC2516 (age $\sim 150$\,Myr) we constrain the spindown in the VLM regime. An exponential rotational 
braking law $P \propto \exp{(t/\tau)}$ with a mass-dependent $\tau$ is required to reproduce the data. 
The spindown timescale $\tau$ increases steeply towards lower masses; we derive $\tau \sim 0.5$\,Gyr for 
0.3$\,M_{\odot}$ and $>1$\,Gyr for 0.1$\,M_{\odot}$. These constraints are consistent with the current 
paradigm of the spindown due to wind braking. We discuss possible physical origins of this behaviour 
and prospects for future work.

\end{abstract}


\section{Introduction}
\label{intro}

The spin of stars is a strong function of stellar mass and age. The age-dependence for main-sequence
F-K-type stars has been empirically established in the seminal paper by \citet{1972ApJ...171..565S} 
as $\omega \propto t^{1/2}$. Originally found from rotational velocities in the Pleiades and 
Hyades, this relation still holds asymptotically on the main-sequence when evaluated with the large sets 
of rotation periods in open clusters that is currently available. Most recent tests of the Skumanich law 
tend to give slightly higher power law exponents of 0.56 (Collier Cameron et al. 2009) or 0.52 
\citep{2007ApJ...669.1167B}.

From the theory side, the Skumanich law has been reproduced in the prescription provided by 
\citet{1988ApJ...333..236K}, which is based on the analytical wind model by \citet{1984LNP...193...49M}.
Under plausible assumptions (linear dynamo, magnetic field a mixture between dipolar and radial),
the Kawaler expression simplifies to $dJ/dt \propto \omega^3$, which gives the desired 
$\omega \propto t^{-1/2}$ behaviour. Recent numerical work, however, indicates that the Kawaler-type
wind parameterisation may not be an adequate explanation for the empirically found Skumanich law
\citep{2008ApJ...678.1109M}.

F-K-type stars exhibit a well-studied rotation-mass relation on the main-sequence.
For example, at the age of the Hyades the rotation periods increase steadily towards later
spectral types, from 5\,d for late F-stars to 12\,d for late K-stars \citep{1987ApJ...321..459R}.
This relation is remarkably tight and it seems possible to explain the few outliers as tidally
locked binaries or as objects with specific spot configurations resulting in a wrong period
measurement. Thus, for these objects mass and age essentially fix the rotation rate, which 
allows for the possibility of 'gyrochronology', i.e. measuring ages from rotation periods 
\citep{2007ApJ...669.1167B}.

Observations have not been able yet to establish similarly robust age/mass-rotation 
dependencies for the very low mass stars in the M-type regime. It is clear that the F-K-type 
period-mass relation breaks down in the early-M regime, corresponding to a mass threshold 
of 0.3-0.5$\,M_{\odot}$. This is most readily seen from the M-dwarfs periods in Praesepe 
\citep{2007MNRAS.381.1638S}, which are 1-3\,d, much shorter than in the K-type
regime, and from the rotational velocity data, which indicates a significant increase in the
rotation rate between early to mid M-types \citep[e.g.,][]{1998A&A...331..581D,2009ApJ...704..975J}.

Similarly, the Skumanich-type rotational braking does not hold anymore for VLM objects with 
$M<0.3\,M_{\odot}$. While angular momentum losses occur in this mass regime as well, the stars 
tend to maintain high rotation rates over Gyrs, which is not consistent 
with the $\omega \propto t^{-1/2}$ spindown. Most commonly the VLM spindown is empirically described
with an exponential braking law $\omega \propto \exp{(-t/\tau)}$.

This exponential behaviour is primarily motivated by the theoretical framework by 
\citet{1988ApJ...333..236K}, see above. In a modification suggested by \citet{1995ApJ...441..865C},
stars above a critical threshold $\omega >\omega_{\mathrm{crit}}$ are treated with
$dJ/dt \propto \omega_{\mathrm{crit}}^2 \omega$, which results in an exponential spindown
law. To be able to match the period data in open clusters, $\omega_{\mathrm{crit}}$ has
to be assumed to be a function of mass \citep[e.g.][]{1997ApJ...480..303K,2007MNRAS.377..741I}. 
Empirically, however, the form of the spindown law is poorly constrained. For recent reviews on 
these subjects, see \citet{2009AIPC.1094...61S} and \citet{2009IAUS..258..363I}.

In this paper we set out to investigate the period-mass and period-age relation for fully
convective very low mass stars based on a new set of rotation periods measured for members
of the open cluster Praesepe. Praesepe, at an age of $\sim 600$\,Myr, is an important cluster
to constrain the spindown law, because the effect of wind braking can be studied in isolation. 
So far, however, only a very small sample of 4 periods was available for evolved VLM stars in 
open clusters \citep{2007MNRAS.381.1638S}. Our goal here is to provide a quantitative 
measurement of the VLM spindown law based on a significantly larger sample of periods.

\section{Photometric monitoring}

\subsection{Observations}

Photometric monitoring observations of Praesepe were made with the 2.4\,m Isaac Newton Telescope 
(INT, La Palma), during the nights of 19th-27th January 2010. Data were obtained on 8 out of the 
9 nights. The INT was equipped with the Wide Field Camera (WFC), mounted at prime focus. The INT 
WFC provides a field of view of approximately 34'$\times$34' over a mosaic of four 2k$\times$4k 
pixel CCDs, with $\sim0\farcs33$ pixels. We selected 4 contiguous fields (with minimal overlap) 
around the centre of the cluster, such that the central pixels of chip 
4 fell on the positions given in Table \ref{fields}.

All observations were made with the SDSS-$i$ filter which suffers rather less from fringing than 
the other, broader, I-band filters available for the INT WFC. For every exposure, we tried to 
ensure that the same guide-star was centred on the same guide-pixel. This adds a small overhead to 
our observing cadence, but has the advantage of minimizing photometric systematics arising from 
drifts in the telescope pointing between and during exposures. Exposures were alternated between 
short and long integration times of 20\,s and 300\,s, allowing us to measure the more 
massive members of the cluster which would saturate in the long exposures. Our average observing 
cadence per-star is 29.9\,min (i.e. the time to complete a cycle of short and long exposures round 
all four fields).  On one night the full moon was very close to Praesepe, and 3$\times$100\,s
exposures were used in place of a single long exposure. Conditions on the whole were good for La 
Palma in winter, with a median seeing of 1\farcs36 arcseconds for our Praesepe observations, and a 
median ellipticity of 0.09 measured from the stellar images. Most nights were reasonably clear, 
with intermittent thin cirrus appearing throughout the run, and occasional periods of high humidity 
forcing us to close the dome. Photometric standards were measured at the start and end (and occasionally 
in the middle) of each night to enable us to transform our $i$-band photometry into the Johnson-Cousins 
system. Our observations are summarised on a night-by-night basis in Table \ref{obs}. 

\begin{table}
\caption{Field central coordinates and number of long/short exposures per field
\label{fields}}
\begin{tabular}{ccccc}
\hline
Field & RA (J2000) & DEC (J2000) & Nlong$^*$ & Nshort$^*$ \\
\hline
1 & 8:42:10.4 & 19:27:54 & 96 & 99 \\
2 & 8:39:44.8 & 19:27:54 & 97 & 95 \\
3 & 8:42:10.4 & 20:02:06 & 96 & 94 \\
4 & 8:39:44.8 & 20:02:06 & 93 & 94 \\
\hline
\end{tabular}

$^*$ number of epochs in the final lightcurves after rejecting a few images 
obviously affected by bad guiding or clouds
\end{table}

\begin{table}
\caption{Dates of observations with number of exposures for each integration
time and comments
\label{obs}}
\begin{tabular}{ccccl}
\hline
Date         & N300 & N20 & N100 & Comments \\
\hline
2010-01-19   & 53 & 56 &  0 & clear \\
2010-01-20   & 57 & 63 &  0 & clear \\
2010-01-21   & 57 & 57 &  0 & cirrus \\
2010-01-22   & 47 & 46 &  0 & cirrus, humid \\
2010-01-23   & 60 & 64 &  0 & clear \\
2010-01-24   & 60 & 68 &  0 & clear \\
2010-01-25   & 12 & 12 &  0 & clear \\
2010-01-26   &  0 &  0 &  0 & clear, humid, bad seeing \\
2010-01-27   &  0 & 20 & 49 & cirrus, variable seeing \\
\hline
\end{tabular}
\end{table}

\subsection{From images to lightcurves}

The data reduction steps are described in full elsewhere \citep{2007MNRAS.375.1449I}. In summary, for the 2-D 
processing (bias correction, flatfielding, defringing), and astrometric and photometric 
calibration of each WFC exposure, we used a pipeline developed by the Cambridge Astronomical Survey Unit (CASU). 
Next, two \emph{master images} (one for the short exposures and one for the long exposures) were made for each 
of the four Praesepe pointings by stacking 10 of our best images. From these we generated {\em master catalogues}
which contain all the sources and their coordinates for which we measure list-driven aperture photometry from each 
of the short and long exposures. Each source has an associated morphological classification, and a flag indicating 
the degree to which it is blended (based on an analysis of overlapping isophotes).

Lightcurves were constructed for $\sim$ 47000 objects, of which $\sim 13500$ have stellar shape parameters. 
To calibrate each epoch, we fit a 2-D quadratic polynomial to the residuals in each frame (measured for 
each object as the difference between its magnitude on the frame in question and the median calculated 
across all frames) as a function of position, for each of the 4 WFC CCDs separately. Subsequent removal of 
this function accounted for effects such as varying differential atmospheric extinction across each frame. 
Over a single WFC CCD, the spatially-varying part of the correction remains small, typically 0.02 mag 
peak-to-peak \citep{2007MNRAS.375.1449I}.

\begin{figure*}
\includegraphics[width=8.5cm,angle=0]{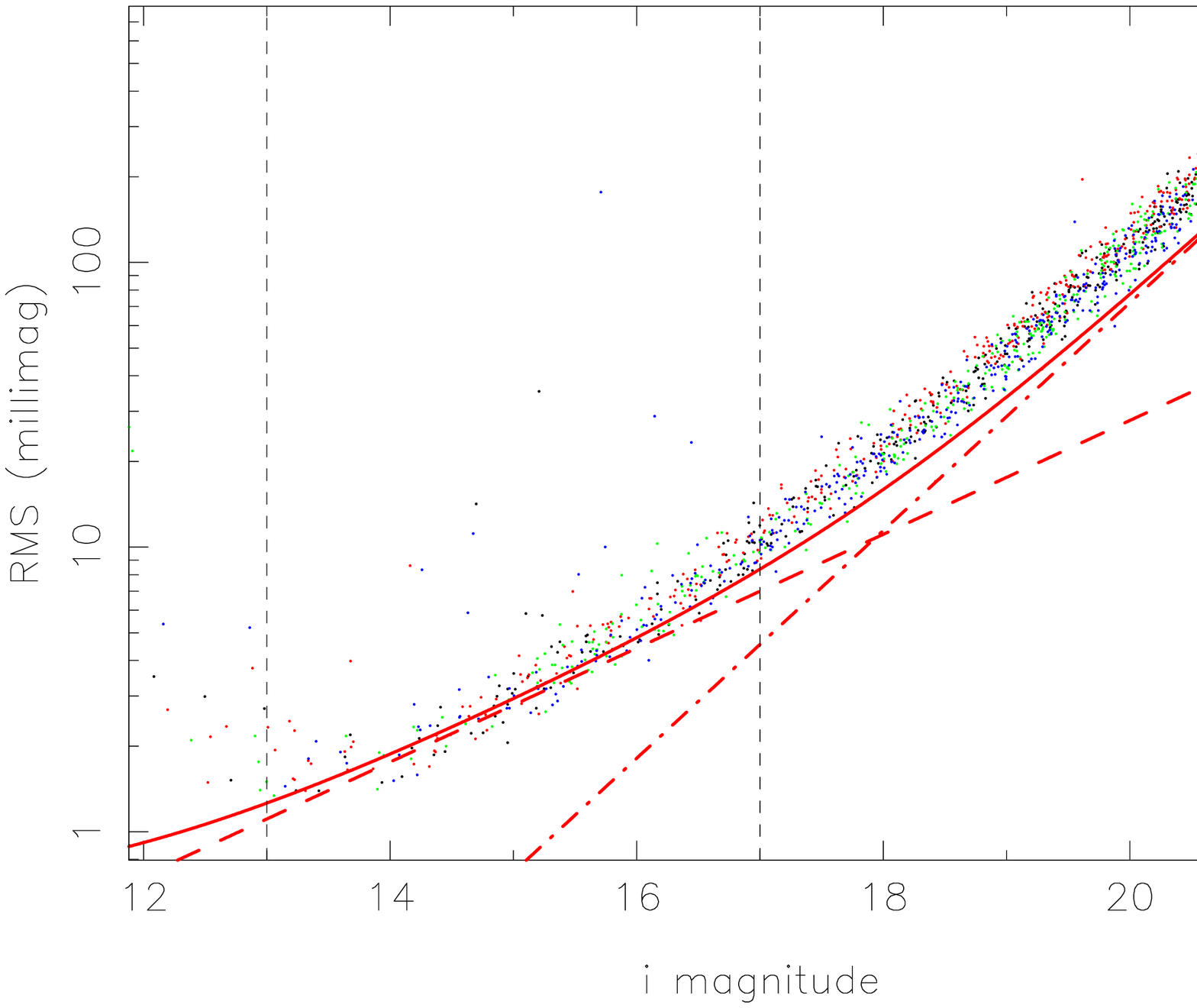} \hfill
\includegraphics[width=8.5cm,angle=0]{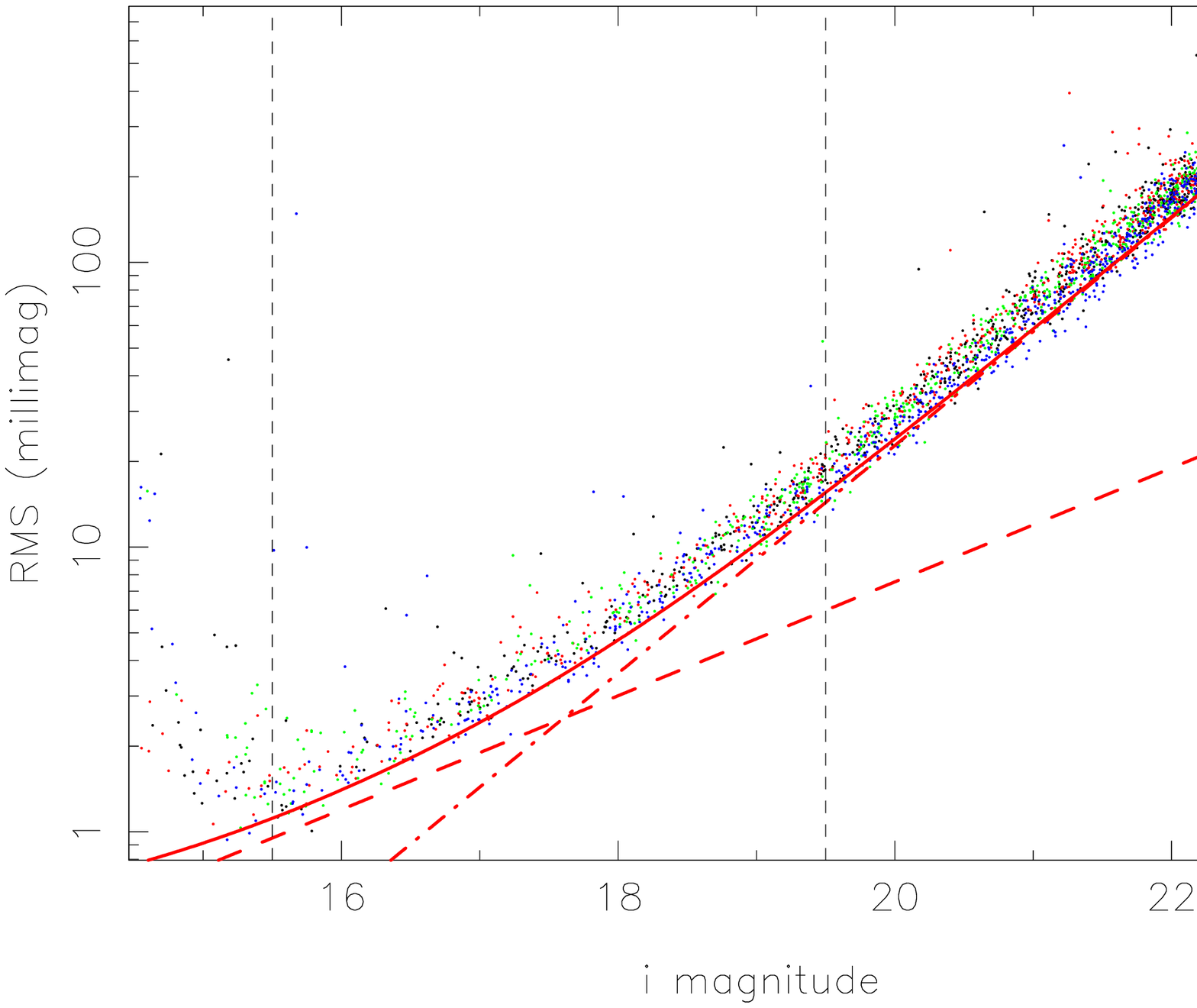} 
\caption{{\it RMS} versus magnitude for $i$-band photometric monitoring in one of the Praesepe fields 
for the short (20s, left) and long (300s, right) exposures. Only unblended objects with stellar morphology 
are shown as dots, colour-coded by detector. The solid line shows the overall predicted {\it RMS}, combining 
contributions from sky noise (dot-dashed line) and object photon counting noise (dashed line). The error
in the sky background fitting is not included here which explains why the predicted {\it RMS} underestimates
the measurements at the faint end. The vertical lines indicate the magnitude range from which we identify 
non-variable stars for performing the frame-to-frame photometric calibration. 
 \label{f1}}
\end{figure*}

We show the {\it RMS} diagrams for the stellar sources in the long and short exposures in Fig. \ref{f1}. 
For the long exposures we achieve a photometric precision of 1-2\,mmag for the brightest objects at $i=15.5$ 
(stars brighter than this are saturated), with {\it RMS} scatter $<1$\% for $i \leq 19$. For the short 
exposures we achieve {\it RMS}=1-2\,mmag at $i=13.0$ and about 4\,mmag at $i=15.5$.

We cross-checked our lightcurve database with the member lists published in \citet{1995A&AS..109...29H}
and \citet[][KH in the following]{2007AJ....134.2340K}. 381 of our lightcurves belong to one of their member 
candidates. This total sample contains 170 duplications, which are detected in the long and the short 
exposures, i.e. the number of cluster members covered by our survey is 211.

\section{Period search}
\label{perser}

The main goal of this study is to derive rotation periods from photometric monitoring. Therefore we
searched for the presence of periodicities in the lightcurves of the 211 objects which are known
members based on colours and proper motion. The duplicates which are measured in 
short and long exposures provide a useful cross-check for the periods. 

We use four independently developed period search routines. Three of the algorithms are based on 
Fourier-like periodograms, but they differ in the implementation of the periodograms and the 
way the best periods are selected. The fourth one is based on the string length method. In addition, 
each of the four period searches includes a visual check of the phased lightcurves. In our study, this 
procedure is done by three researchers separately, which should minimize the subjectiveness of 
the results. In the following we will briefly explain the different routines.

\subsection{Monitor}

Here we follow the method described in \citet{2006MNRAS.370..954I}. We adopt as the null hypothesis a 
model of a constant magnitude, and compare this to the alternate hypothesis that the light curve 
contains a sinusoidal modulation, of the form:
\begin{equation}
m_1(t) = m_{\rm dc} + \alpha \sin(\omega t + \phi)
\end{equation}
where $\alpha$ and $\phi$ represent the semi-amplitude and phase of the sinusoid, and 
$\omega = 2 \pi / P$ is the angular frequency corresponding to rotation period $P$. In order to 
determine the period, we fit this model using standard linear least-squares, at discrete values 
of $\omega$ sampled on a uniform grid in frequency from $0.01$ to $100$\,d. To judge the 
significance of each period, we calculate the $\chi^2$ improvement of the sinusoidal model relative to
the null hypothesis ($\Delta \chi^2$).

Due to the small sample size, we omit the cut in $\Delta \chi^2$ used by \citet{2006MNRAS.370..954I}, 
and simply subjected all of the light curves to eyeball examination, to define the final sample of
periodic variables. $65$ light curves fulfill these
criteria. Several objects are identified which appeared to be varying, but the period was 
poorly-determined due to less than a cycle being seen; these are excluded, as are the remaining 
cluster members, the majority of which consistent with no variation.
The sample of 65 periods contains 18 duplicates; this leaves 47 good periods, all shorter than 6\,d. 
For the duplicates we adopt the period with maximum $\Delta \chi^2$. 


\subsection{SE2004}
\label{se2004}

This set of routines is based on the period search as first published in 
\citet[][SE2004]{2004A&A...419..249S} and further discussed in 
\citet{2004A&A...421..259S,2005A&A...429.1007S}. In a first step, the routine eliminates 
3$\sigma$ outliers from the lightcurves. Next, we identify the highest peak in the Scargle 
periodogram for unevenly sampled lightcurves \citep{1982ApJ...263..835S}. For this peak we 
calculate a false alarm probability following the relation by \citet{1986ApJ...302..757H}. For the
best period from the Scargle periodgram, a phaseplot is calculated (lightcurve as a function of
phase). We subtract the period modeled as a sinecurve and compare the variance in the original
lightcurve with the variance in the residuals using the F-test. Finally the CLEANed periodogram 
is computed using the routines by \citet{1987AJ.....93..968R}. We accept a period if 
the Scargle FAP is $<1$\%, the F-test FAP is $<5$\%, and the highest peak in the Scargle 
periodogram is among the highest in the CLEAN periodogram. In addition, the phaseplots 
for all 381 lightcurves are visually inspected.

Using the method as outlined in SE2004, we estimate the range of periods we are able to detect.
For 10 non-variable objects with Gaussian noise, we artificially add periods ranging from 0.05 to 10\,d
and try to recover them using the same combination of Scargle periodogram and F-test as outlined
above. From this test it is obvious that the analysis is sensitive to periods $P<6$\,d and becomes 
unreliable (deviation between imposed and measured period $>10$\%) for longer periods.

In the catalogue of 381 objects, 222 fulfill the FAP criteria given above. We exclude the 99 periods 
longer than our upper limit of 6\,d and 15 periods for bright sources affected by saturation 
($J<10$\,mag). From the remaining 108, 82 pass the visual inspection. 25 of the objects with periods 
are duplicates; two of them have inconsistent periods and are excluded. For the duplicates with 
consistent period, we adopt the one measured with higher Scargle power. In total, this method 
yields 55 good periods. 

\subsection{CLEAN and String-Length}

Here we discuss the period search using the CLEAN discrete Fourier transform \citep[CLN,][]{1987AJ.....93..968R} 
and string-length \citep[SL,][]{1983MNRAS.203..917D} methods. Periods are searched over the range from 
$P_{\mathrm{min}}=0.1$ to $P_{\mathrm{max}}=6$\,d by sampling the frequency range 
uniformly with increments of $5\cdot 10^{-3}$\,cycles/day. Light curves folded in 
phase with the best periods derived from the 2 methods are visually inspected to 
estimate the reliability of the reported period. We rank the results in 3 groups: 
clearly periodic light curves, possibly periodic ones, and non-periodic ones.

The CLN analysis of 381 light curves yields robust periods for 47 light curves, 
while possible periods are found for 23 additional light curves. The remaining 311 
light curves do not show evidence for periodic variations. After duplicate removal, 
we are left with 54 periods from CLN, from which 12 are measured twice consistently. 
The majority of these periods (38/54) are classified as robust after eyeballing. 

The SL method applied to our sample yields robust periods for 18 light curves, of which 
15 are also found and classified as robust from the CLN method.  The SL method additionally 
yields 13 possible periods, of which 5 are detected as robust periods in the CLN analysis, 
sometimes as aliases. After duplicate removal 24 periods remain from the SL method, from which 
7 are measured twice consistently. 



\subsection{Combining the four period samples}
\label{comb}

The application of four different period search routines and the presence of a significant sample of 
duplicates measured in short and long exposures gives us a good handle on the robustness of our results. 
In general, when two lightcurves are available for the same object, the agreement is usually excellent 
within the same period search routine. Typical deviations are $<2$\% with few outliers. Thus, all four 
period routines are internally consistent and provide accuracies in the range of 1\%.

\begin{figure*}
\includegraphics[width=3.0cm,angle=-90]{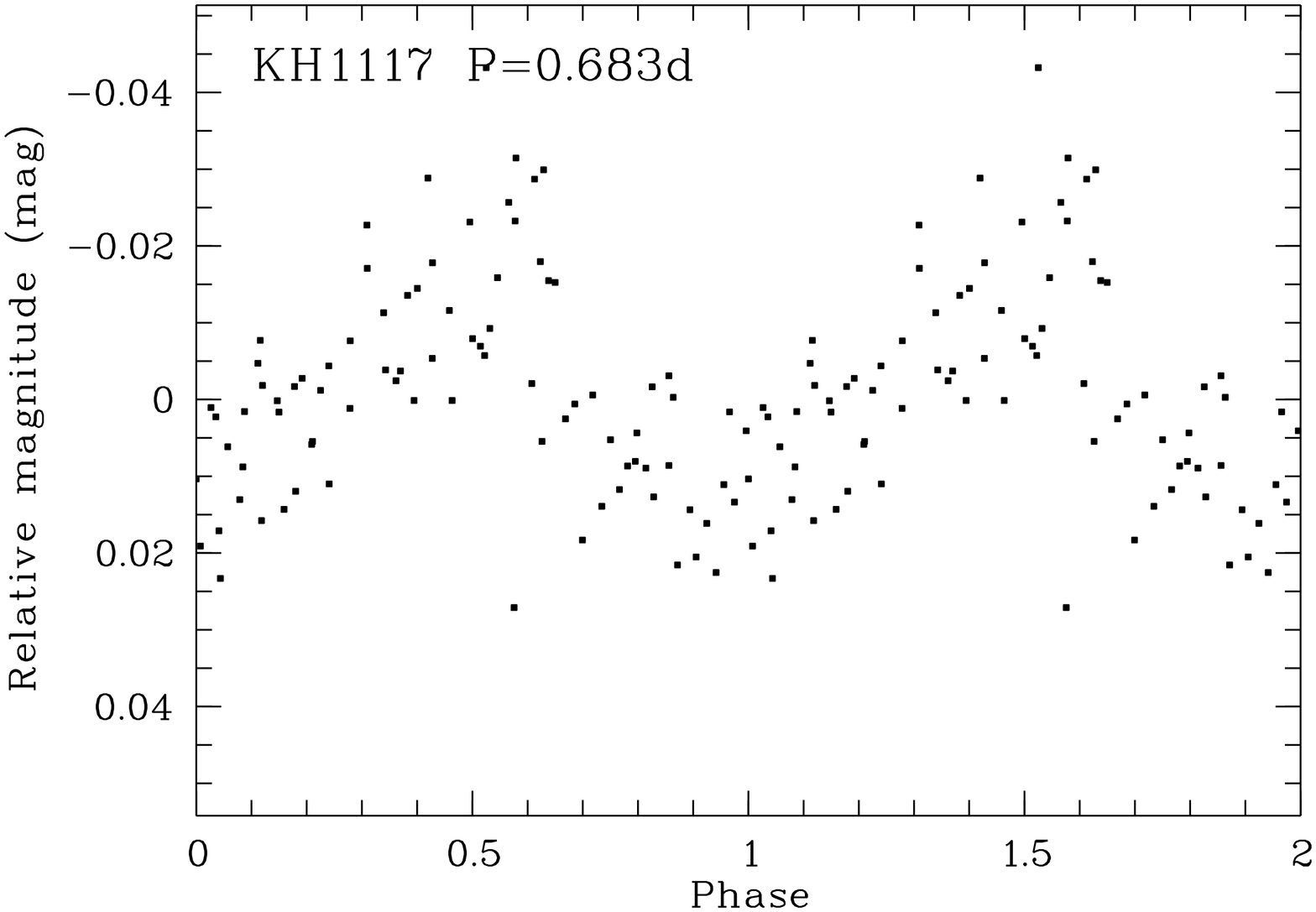} \hfill
\includegraphics[width=3.0cm,angle=-90]{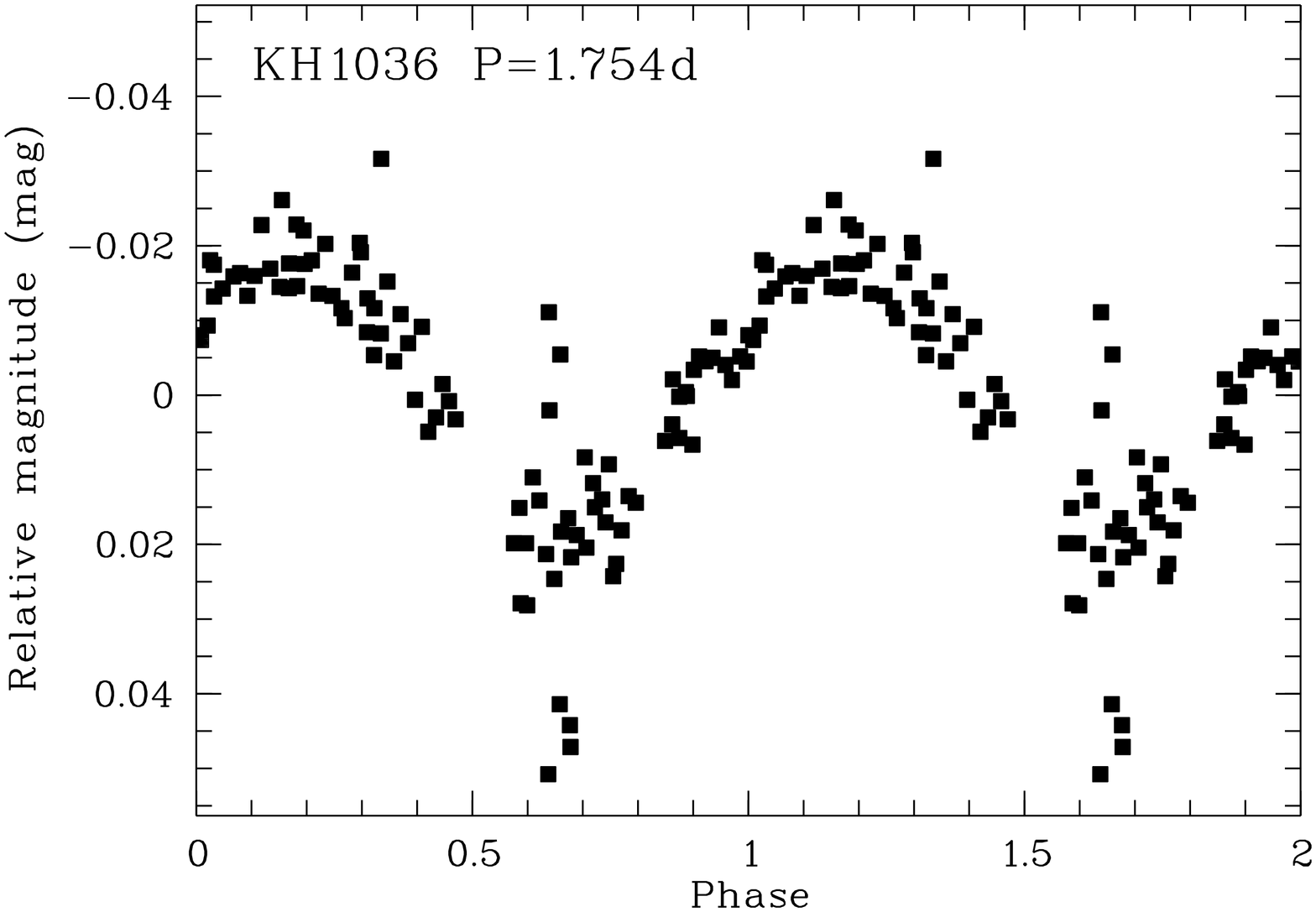} \hfill
\includegraphics[width=3.0cm,angle=-90]{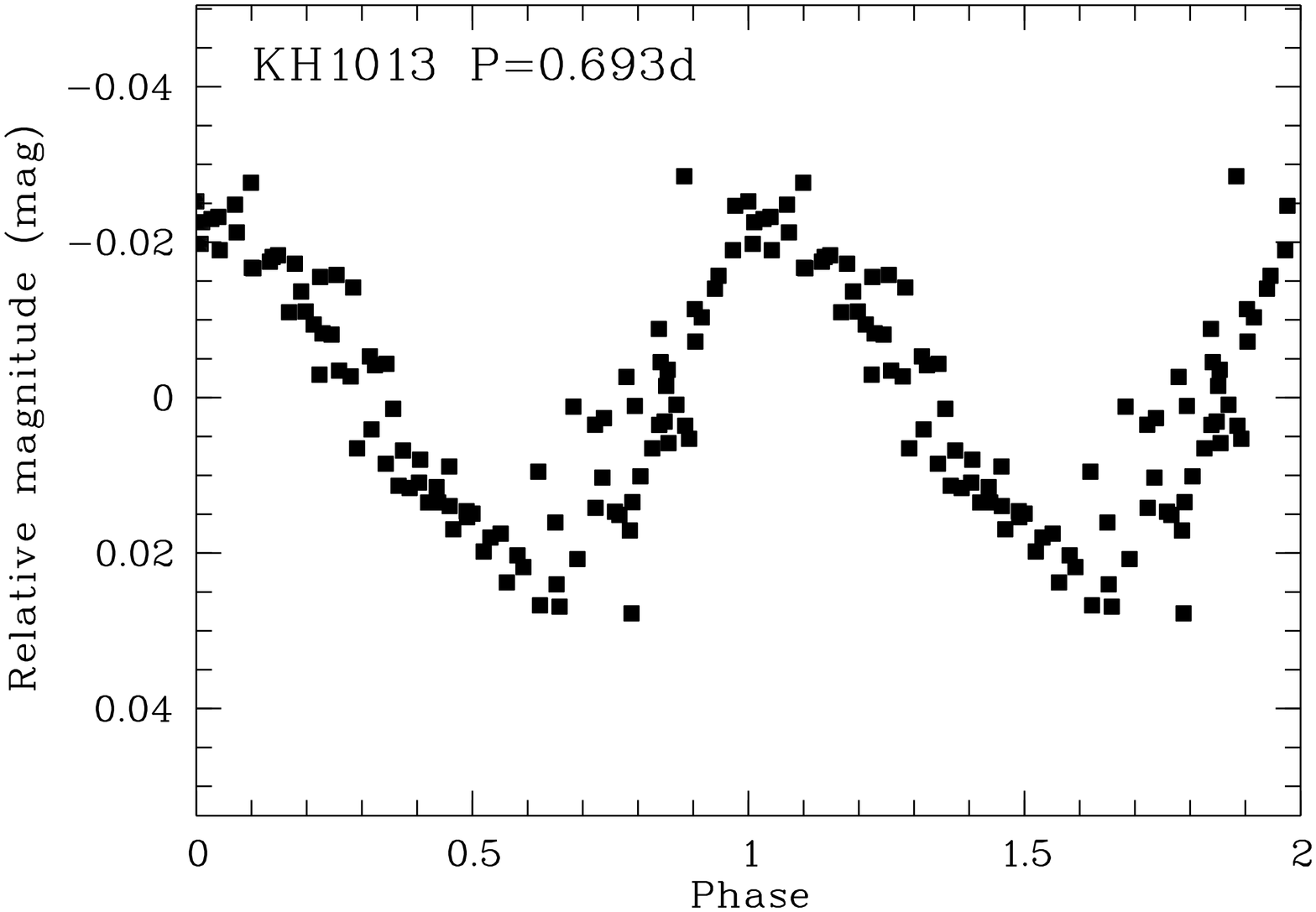} \hfill
\includegraphics[width=3.0cm,angle=-90]{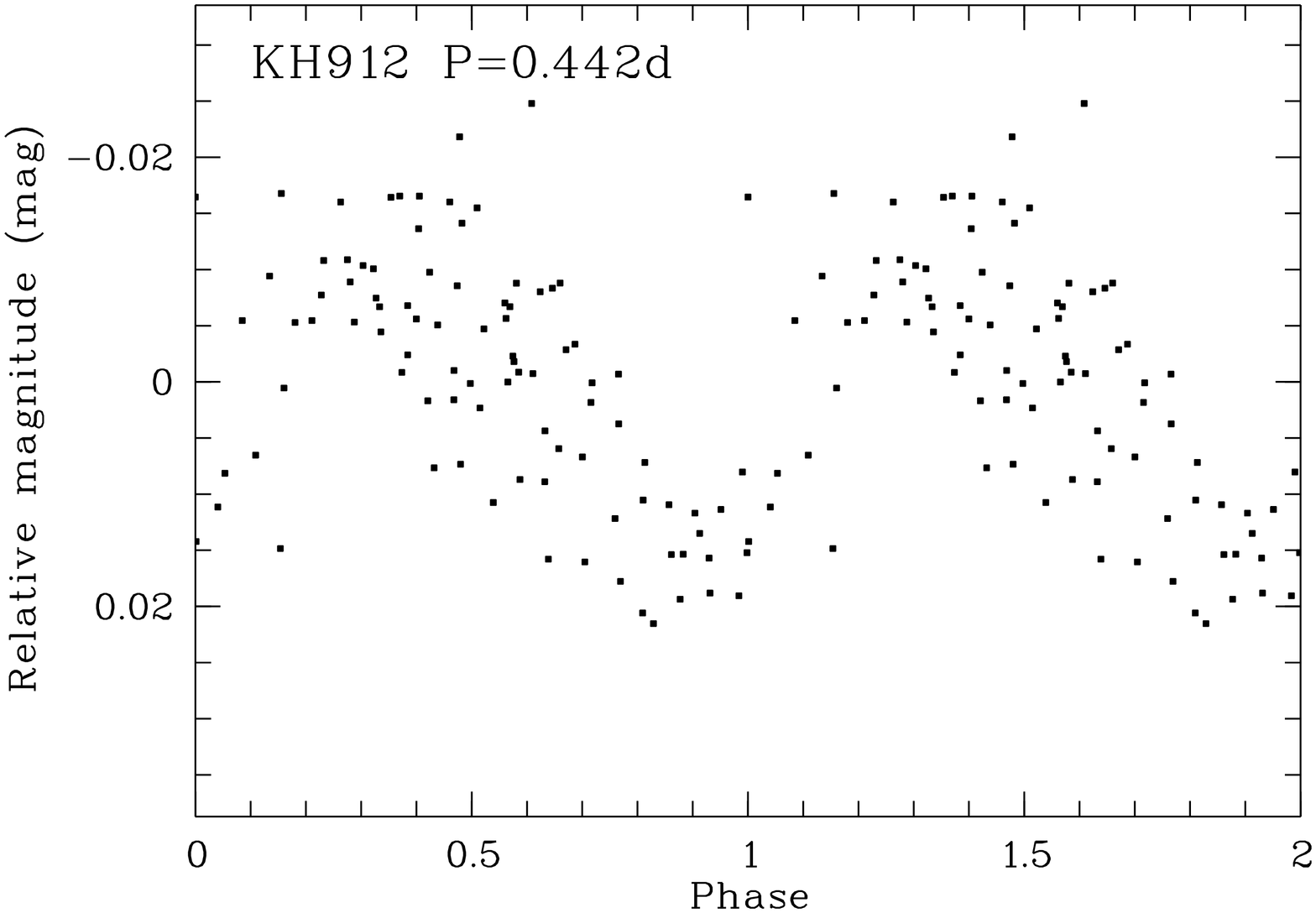} \\
\includegraphics[width=3.0cm,angle=-90]{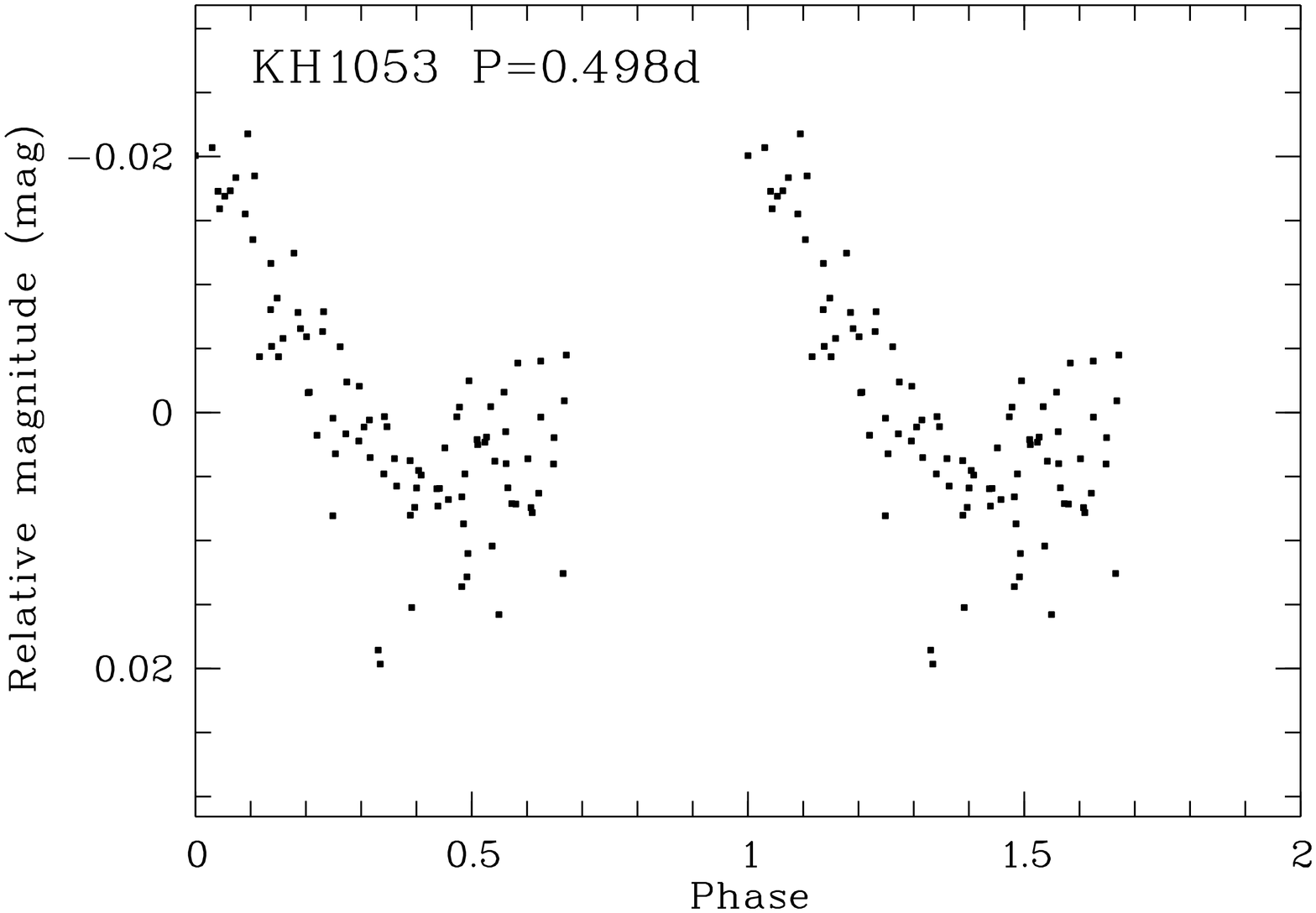} \hfill
\includegraphics[width=3.0cm,angle=-90]{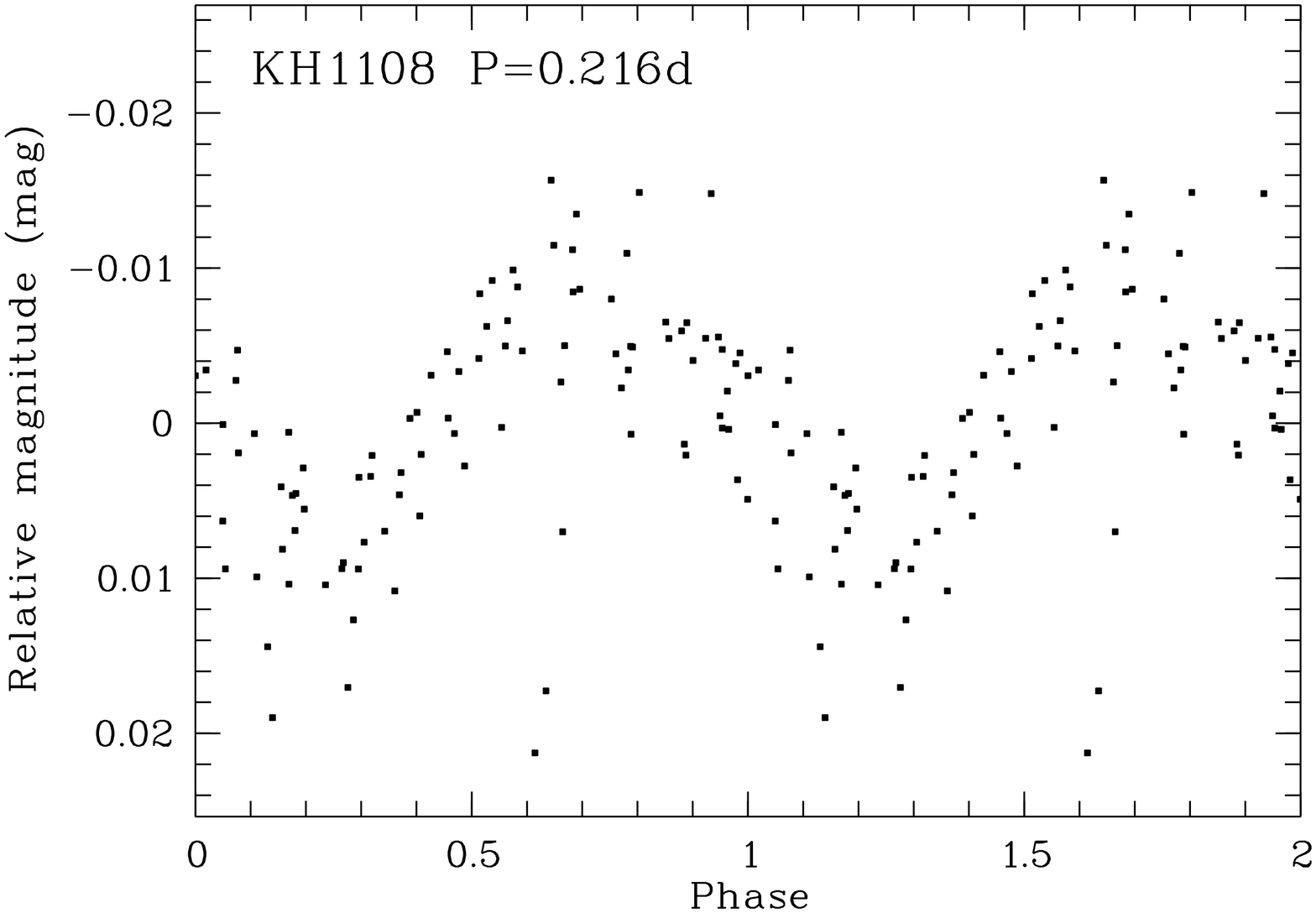} \hfill
\includegraphics[width=3.0cm,angle=-90]{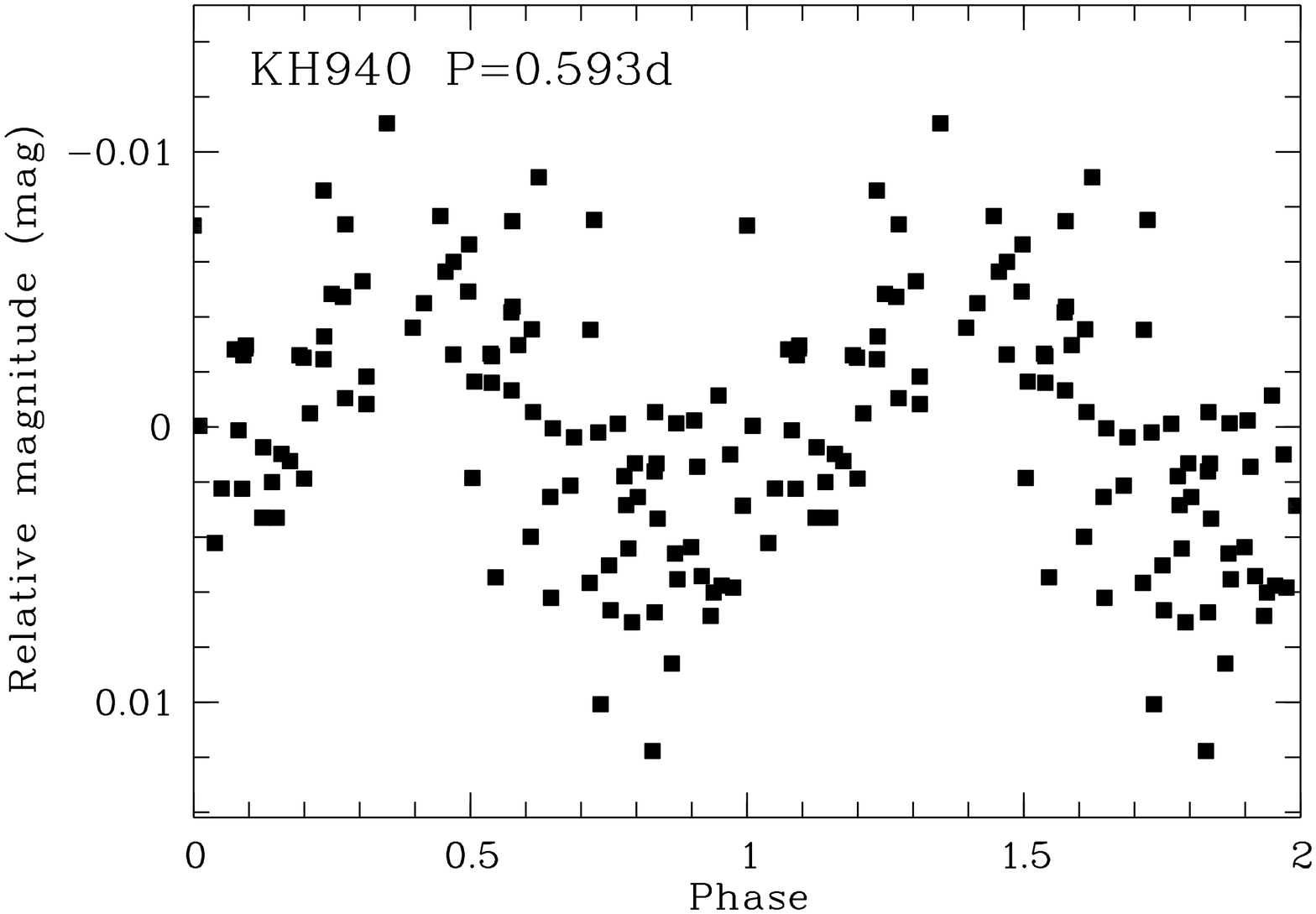} \hfill
\includegraphics[width=3.0cm,angle=-90]{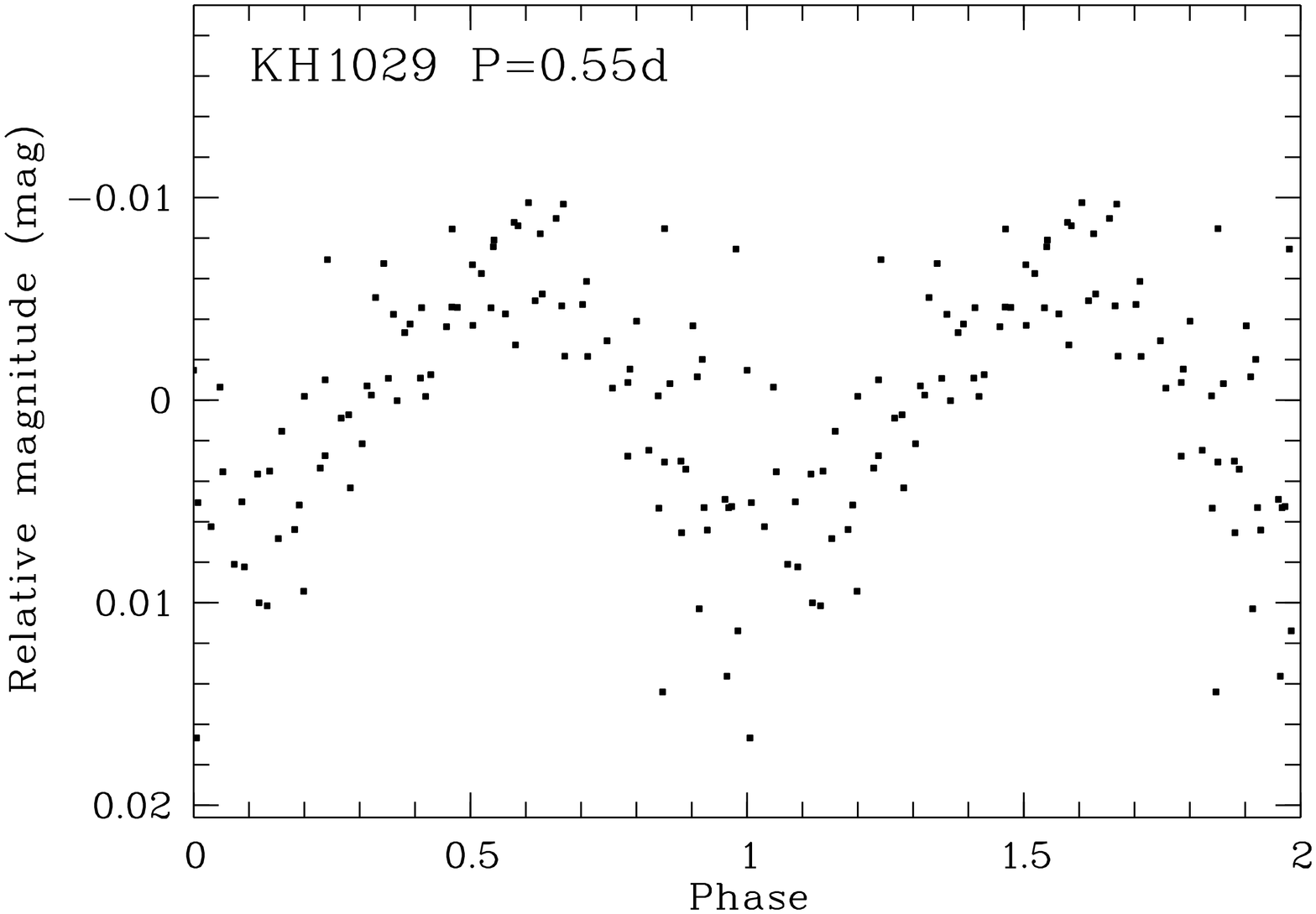} \\
\includegraphics[width=3.0cm,angle=-90]{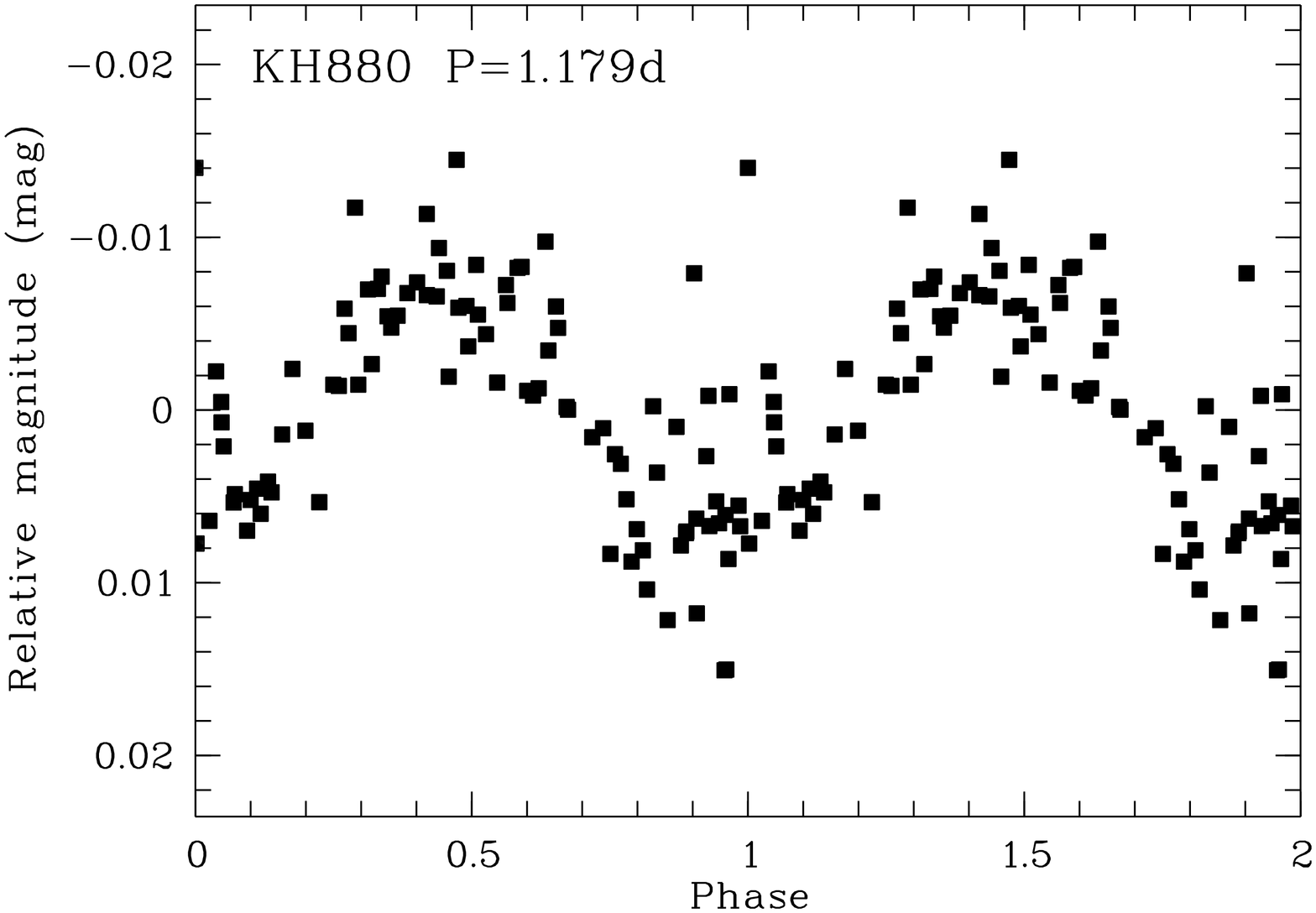} \hfill
\includegraphics[width=3.0cm,angle=-90]{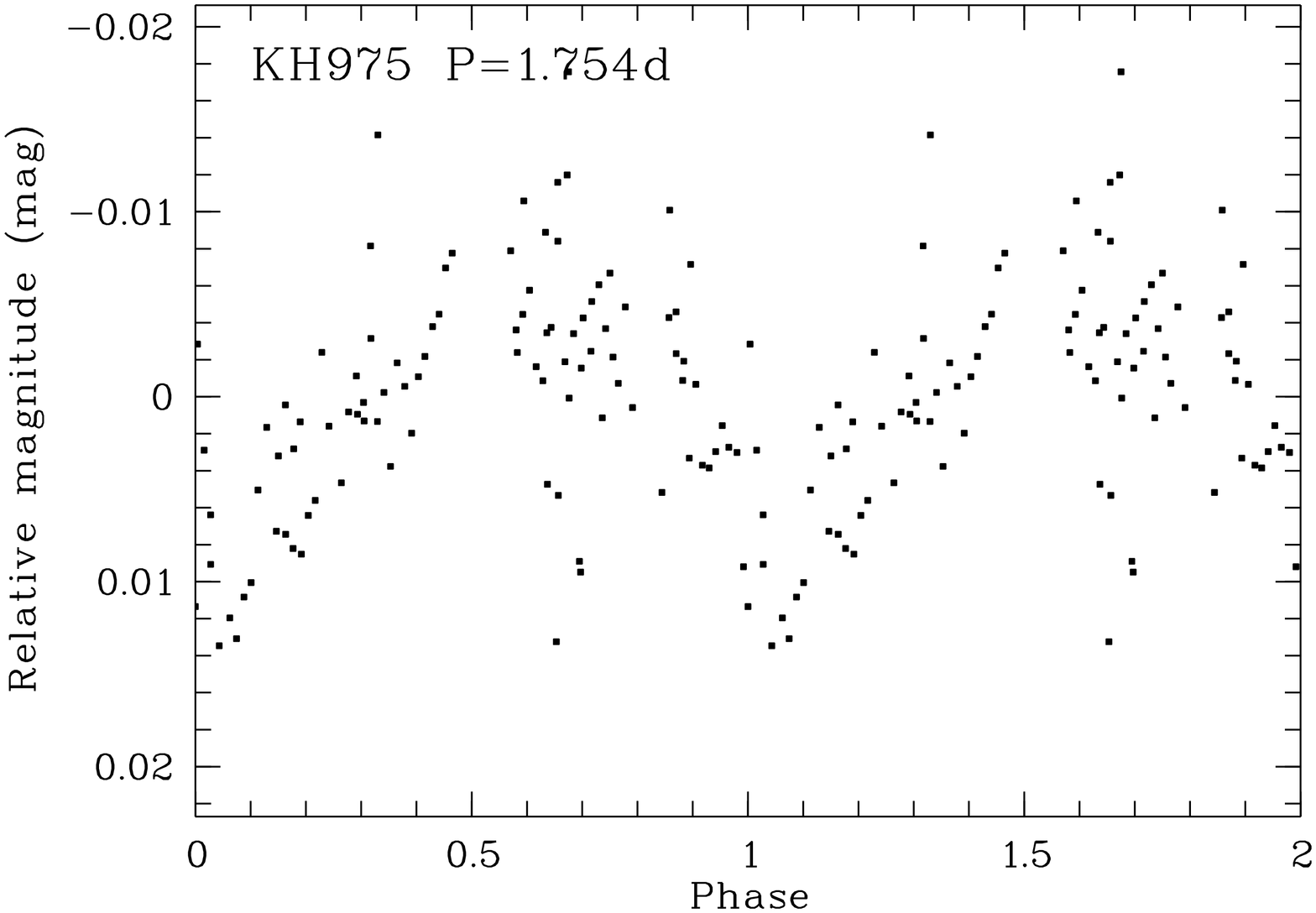} \hfill
\includegraphics[width=3.0cm,angle=-90]{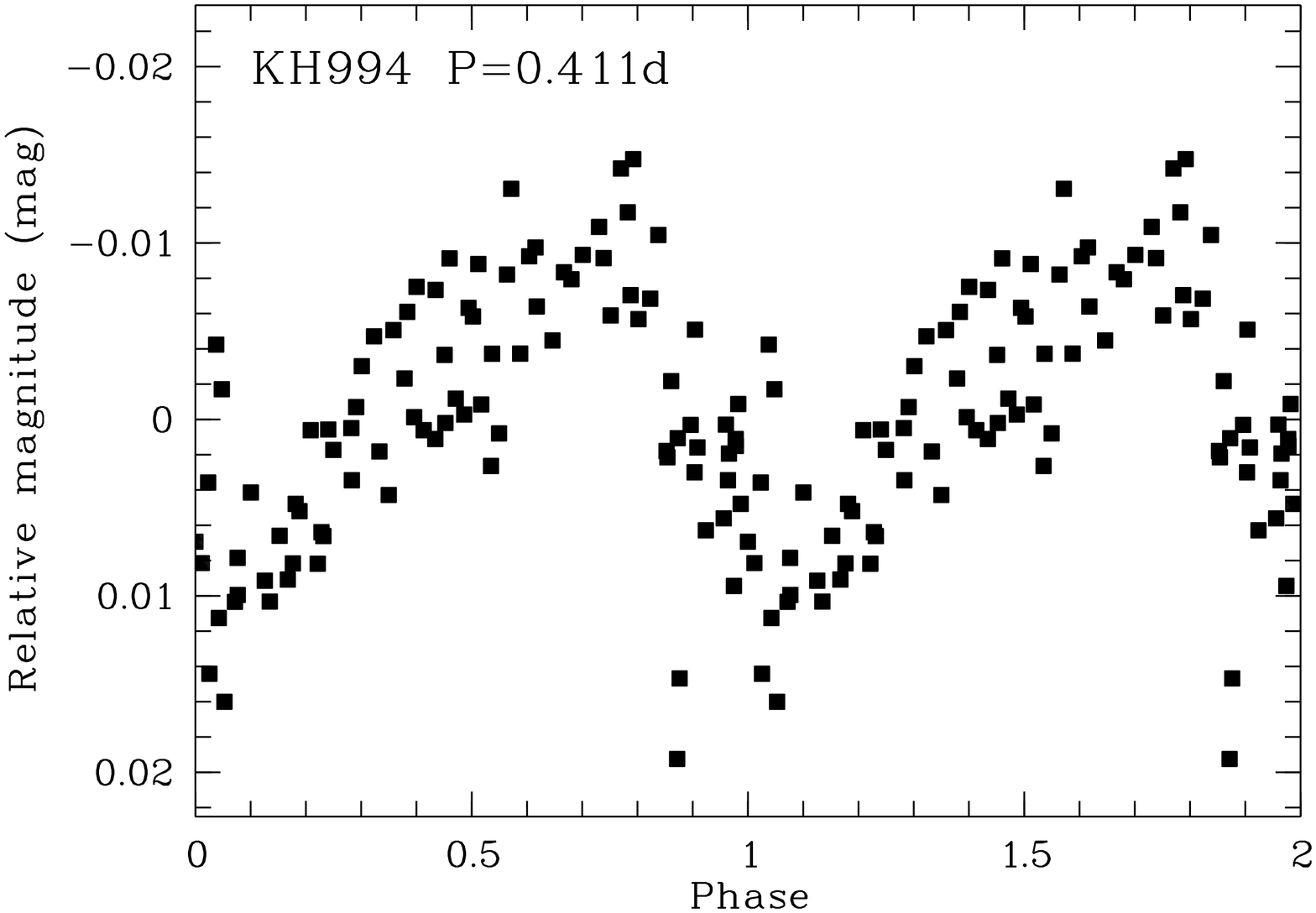} \hfill
\includegraphics[width=3.0cm,angle=-90]{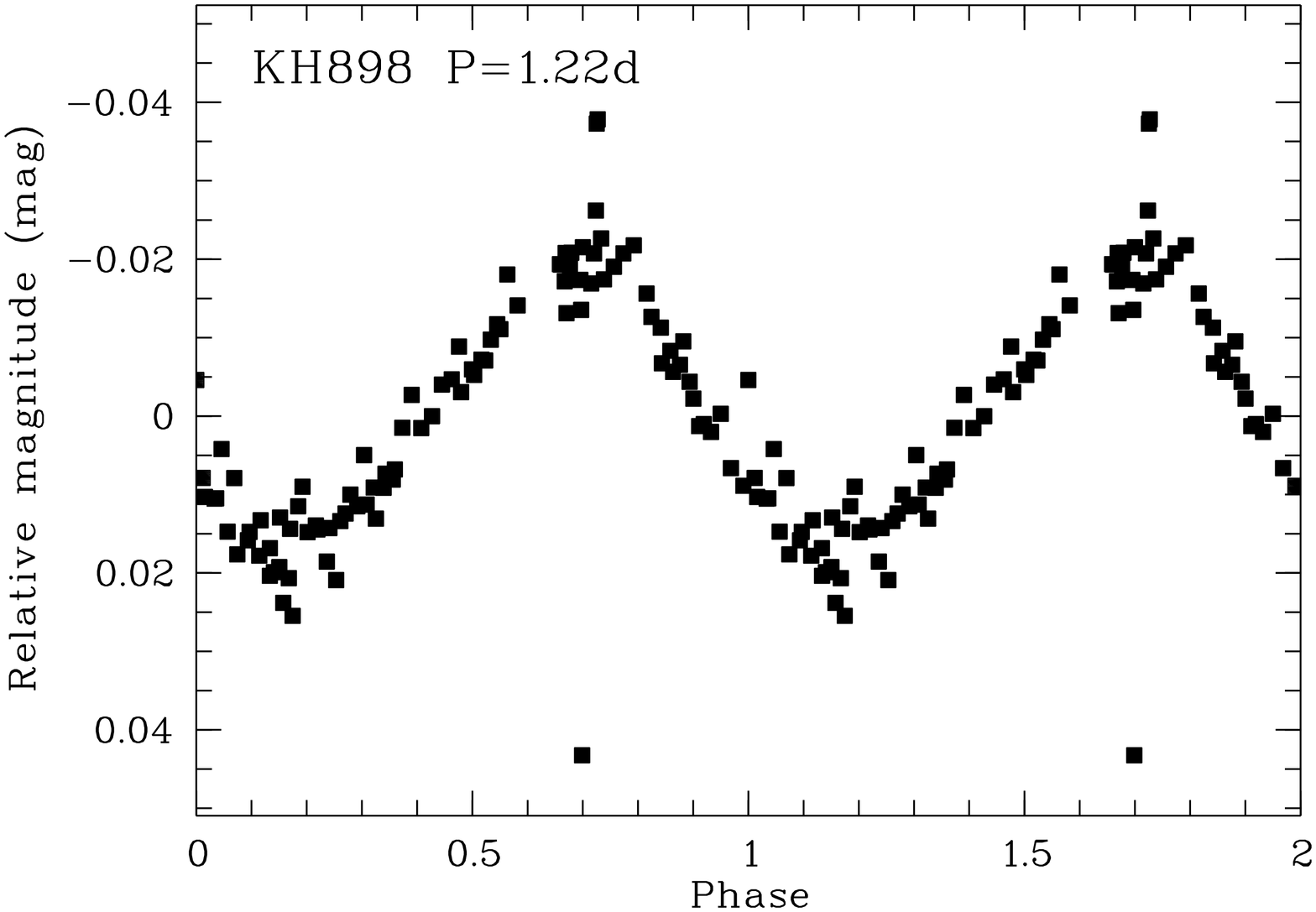} \\
\includegraphics[width=3.0cm,angle=-90]{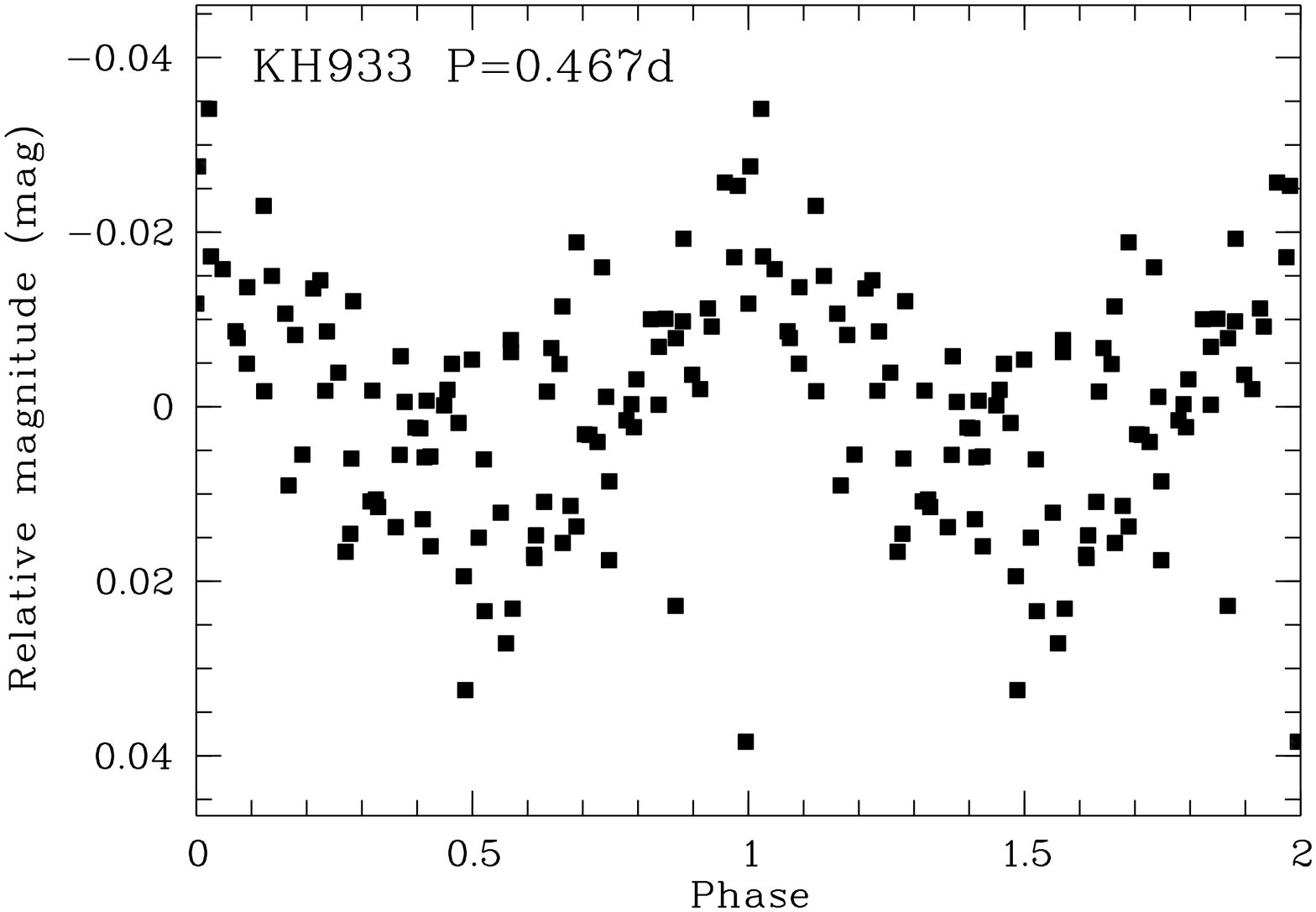} \hfill
\includegraphics[width=3.0cm,angle=-90]{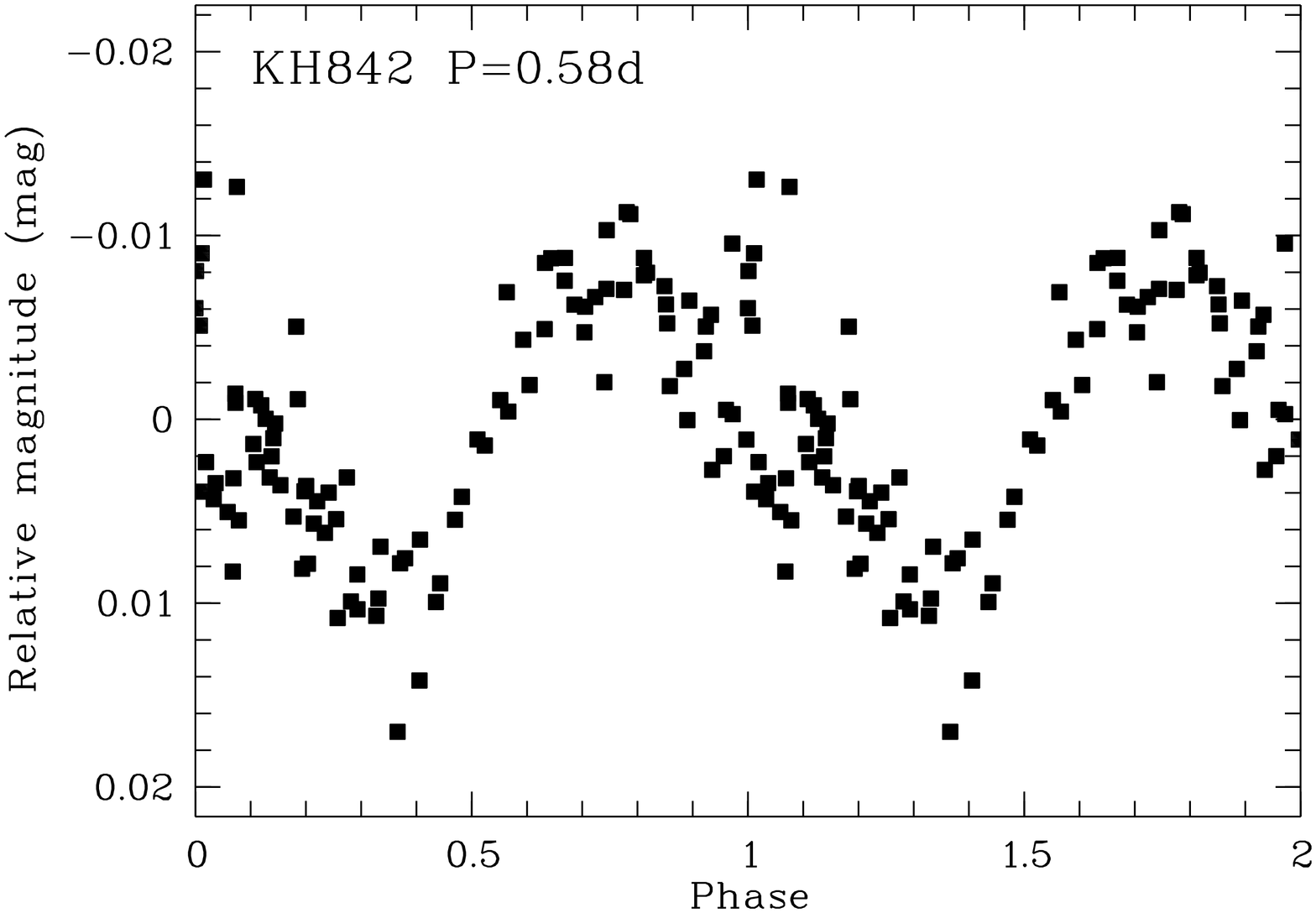} \hfill
\includegraphics[width=3.0cm,angle=-90]{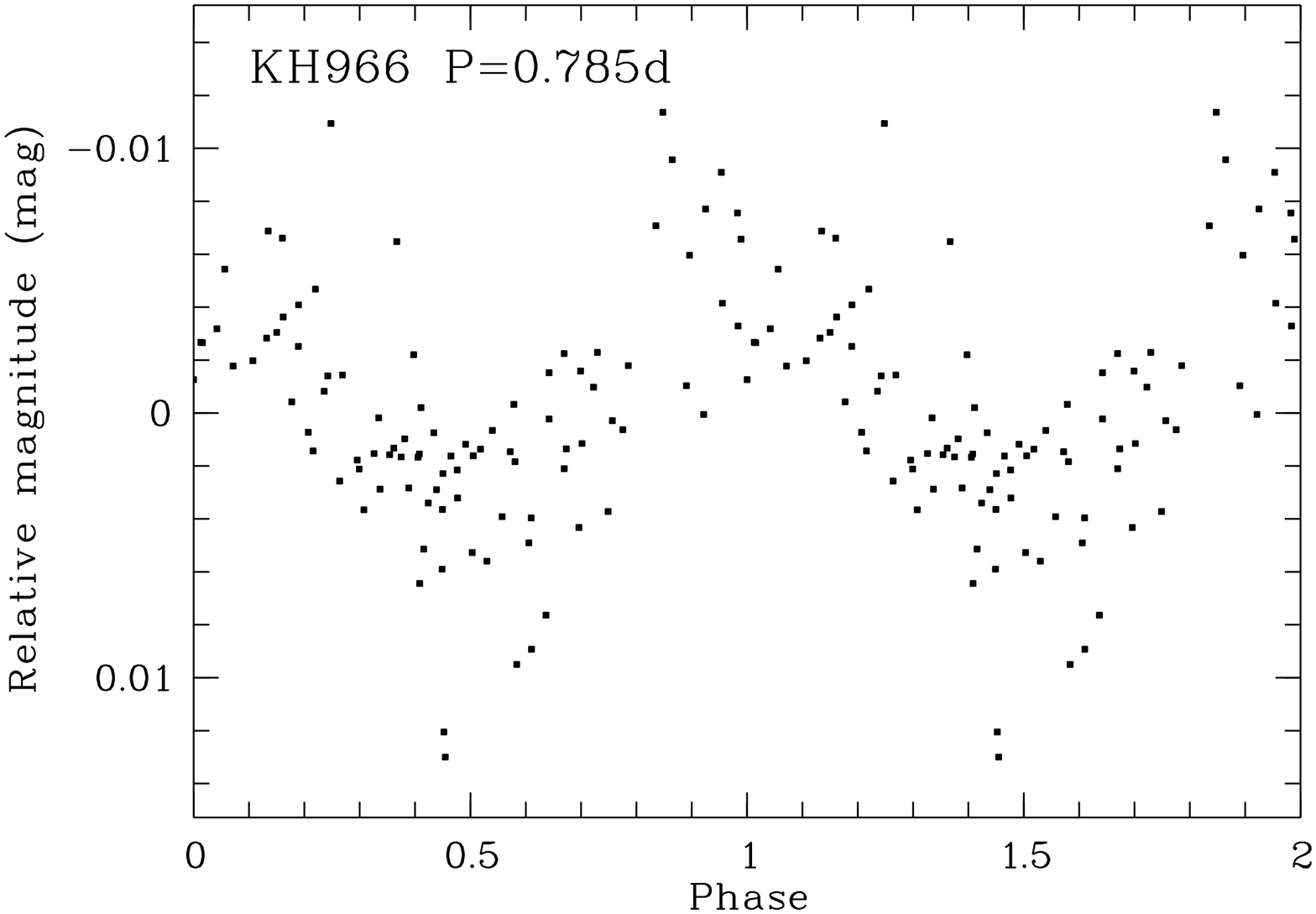} \hfill
\includegraphics[width=3.0cm,angle=-90]{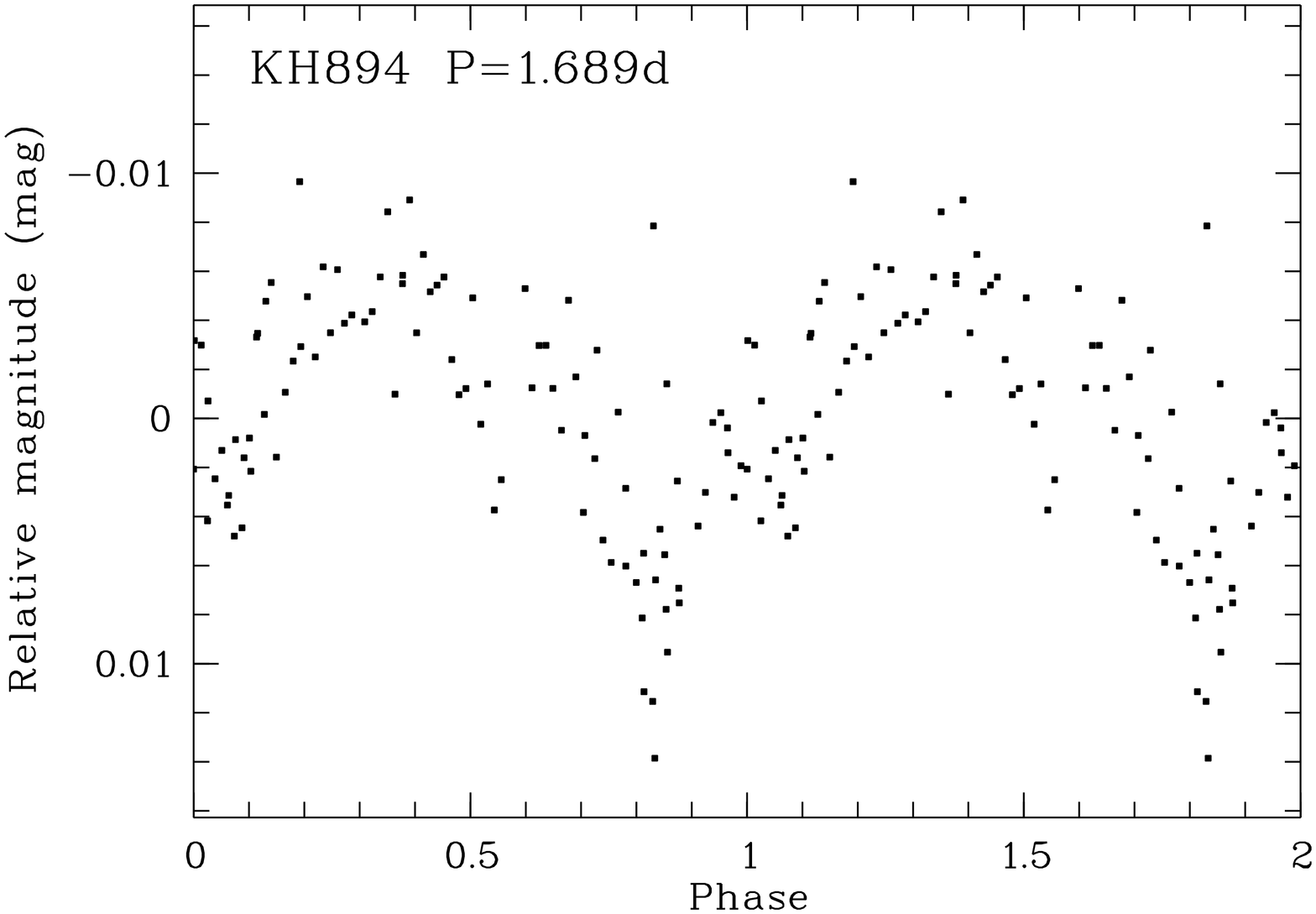} \\
\includegraphics[width=3.0cm,angle=-90]{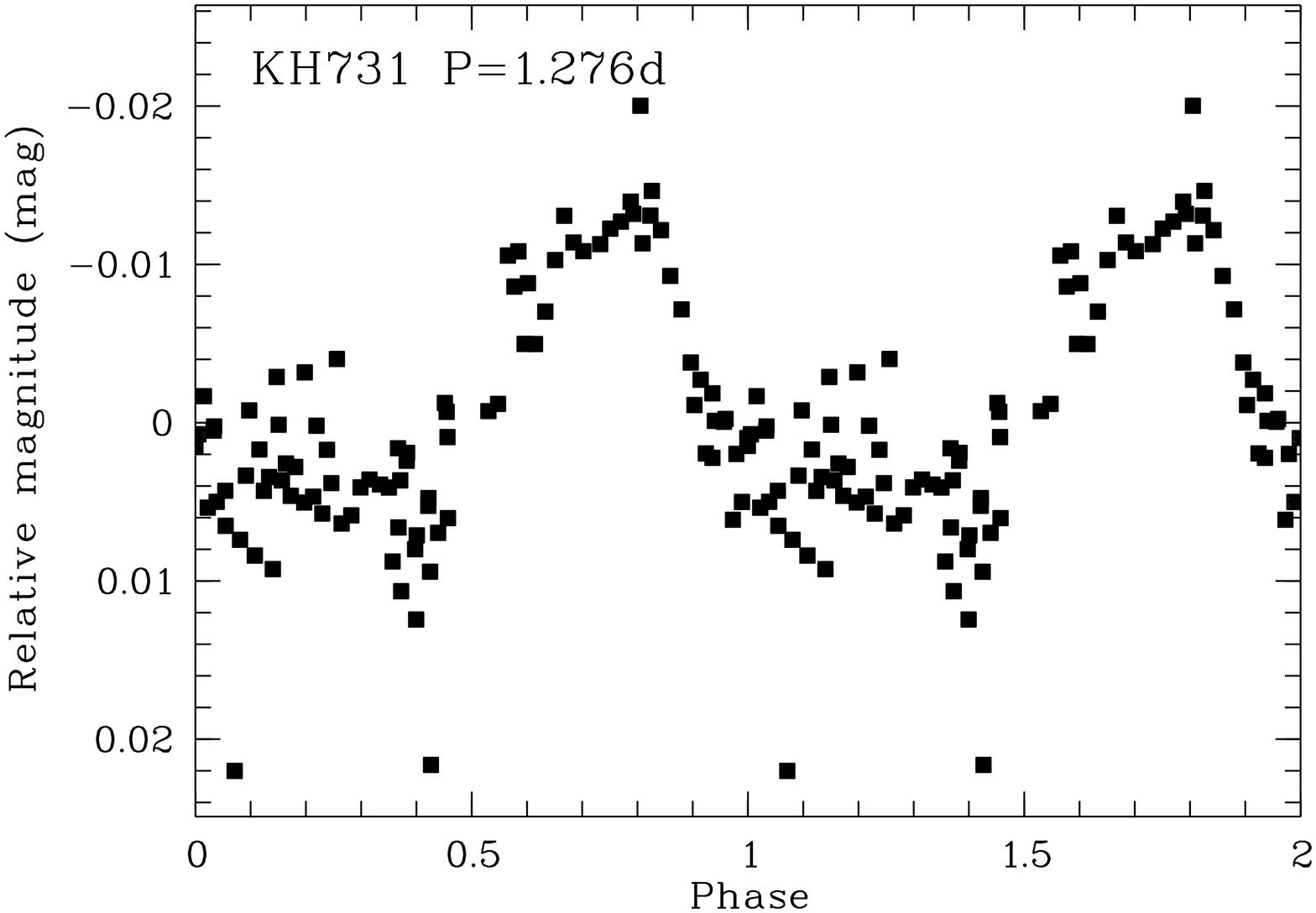} \hfill
\includegraphics[width=3.0cm,angle=-90]{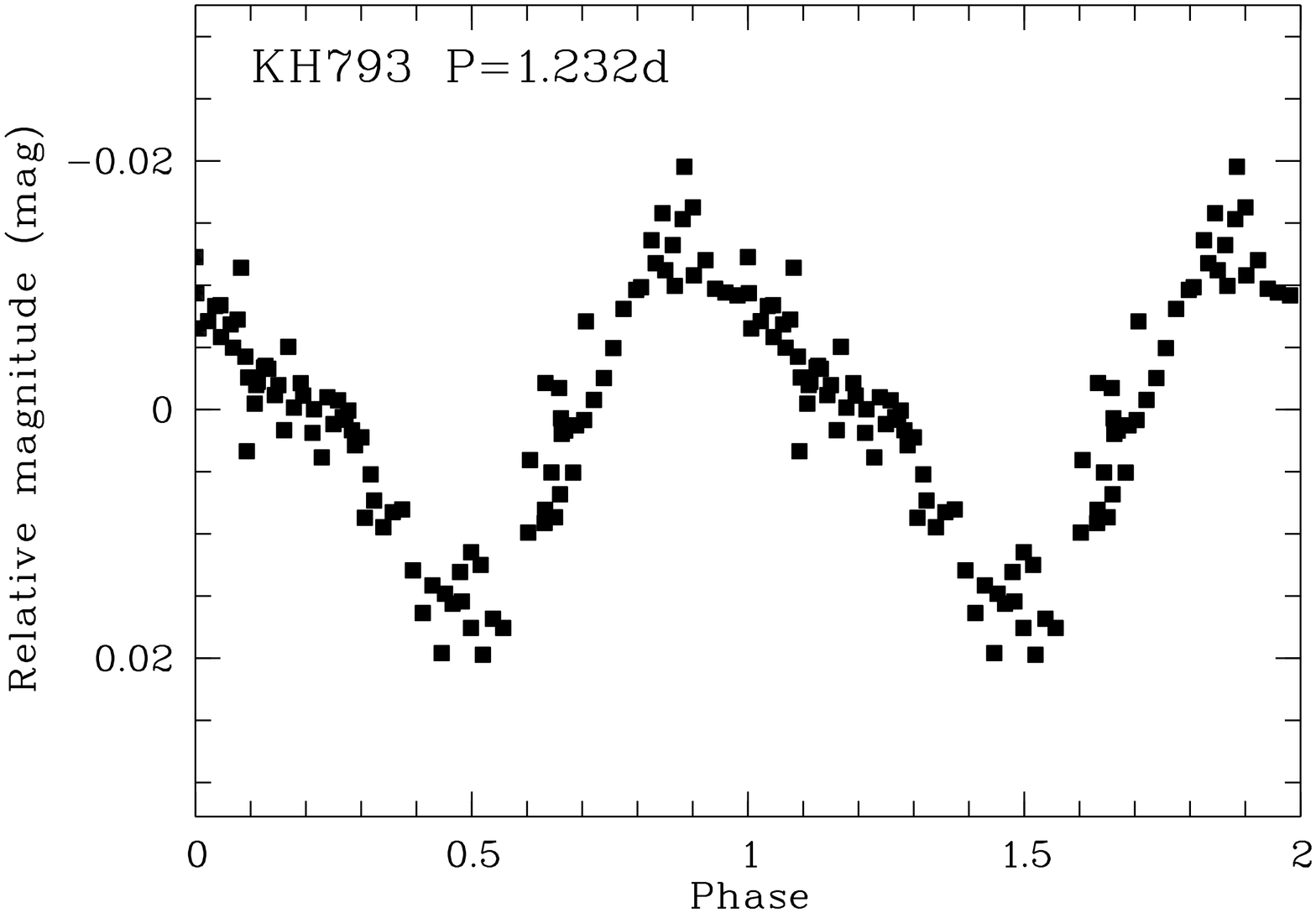} \hfill
\includegraphics[width=3.0cm,angle=-90]{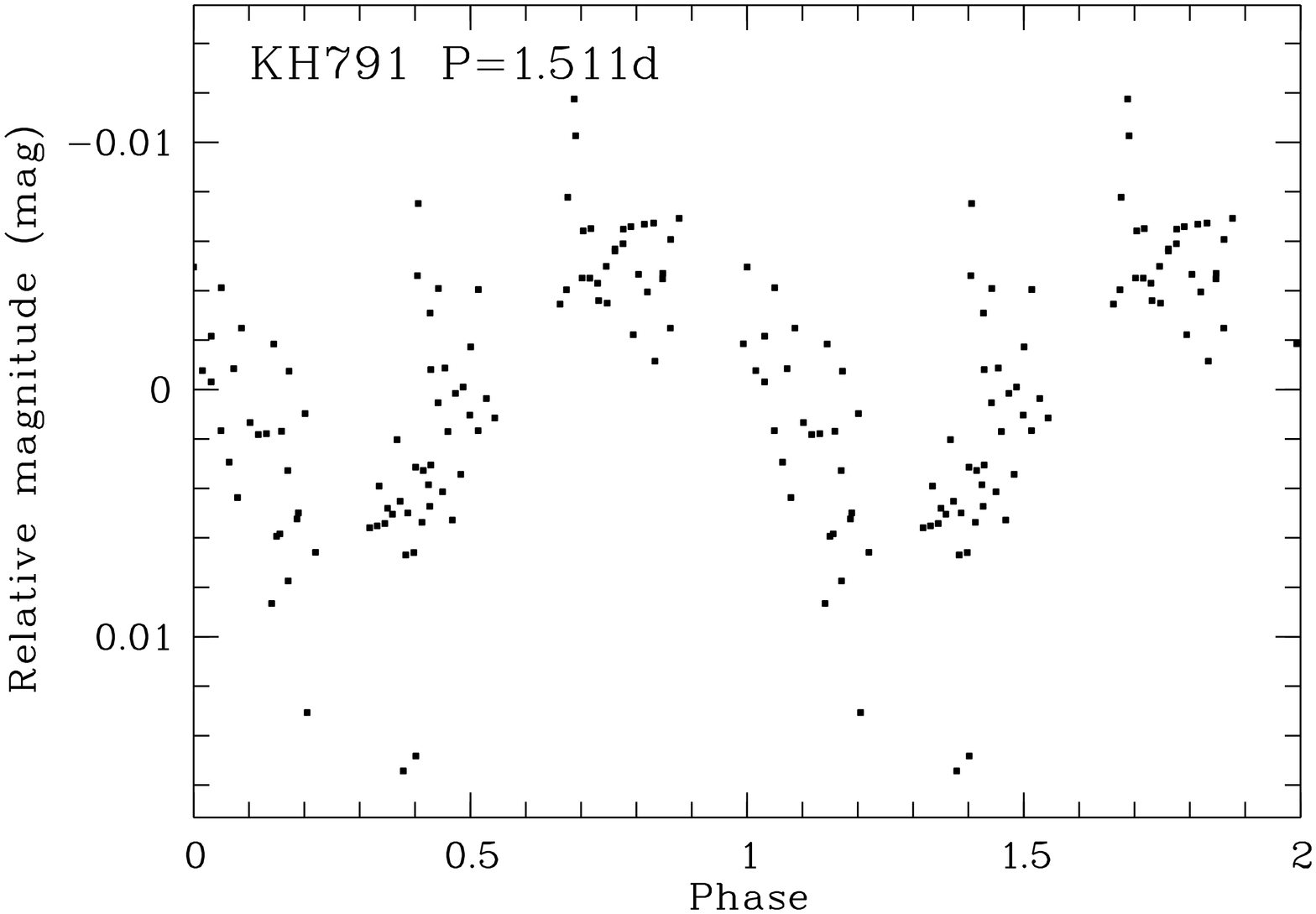} \hfill
\includegraphics[width=3.0cm,angle=-90]{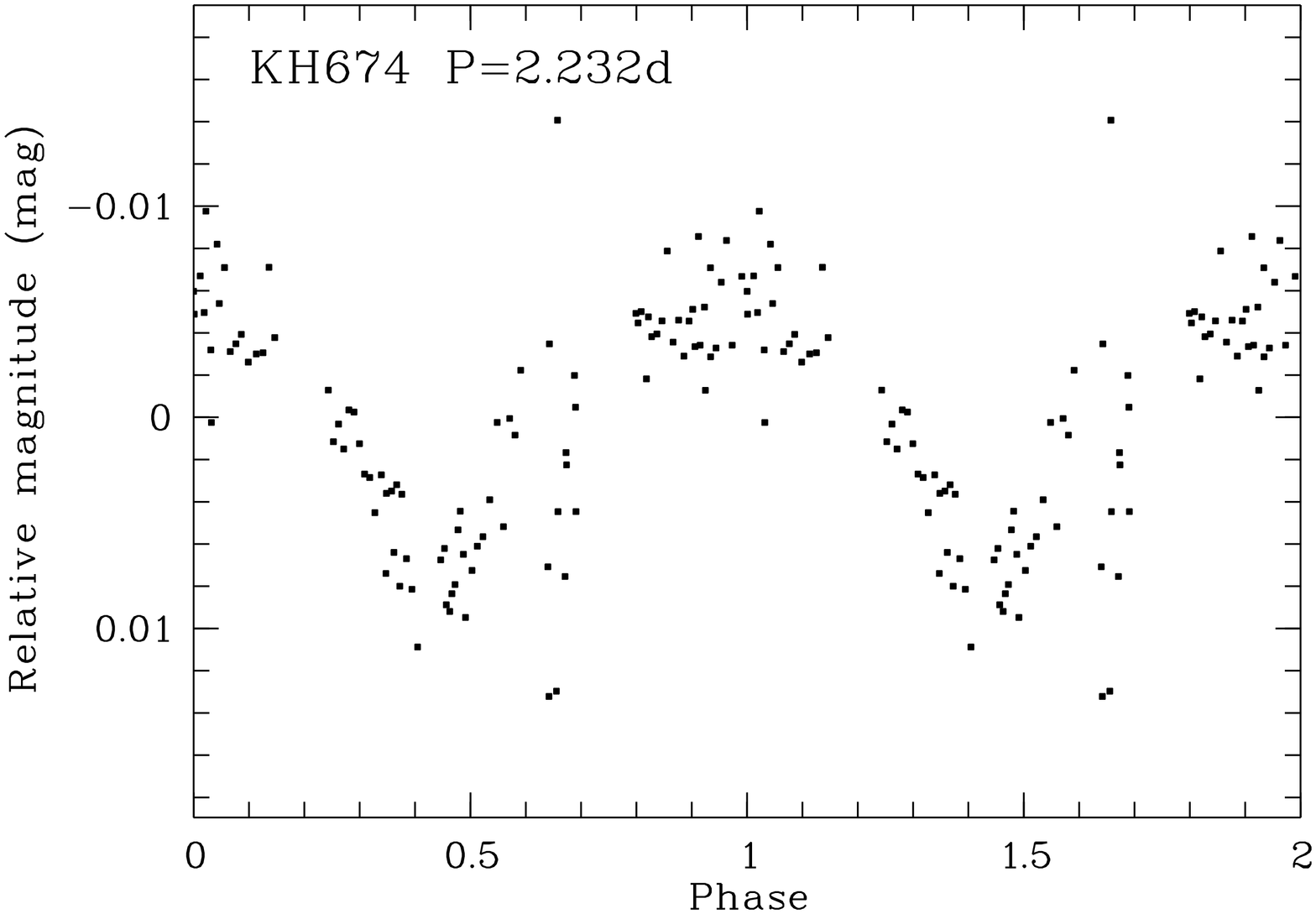} \\
\includegraphics[width=3.0cm,angle=-90]{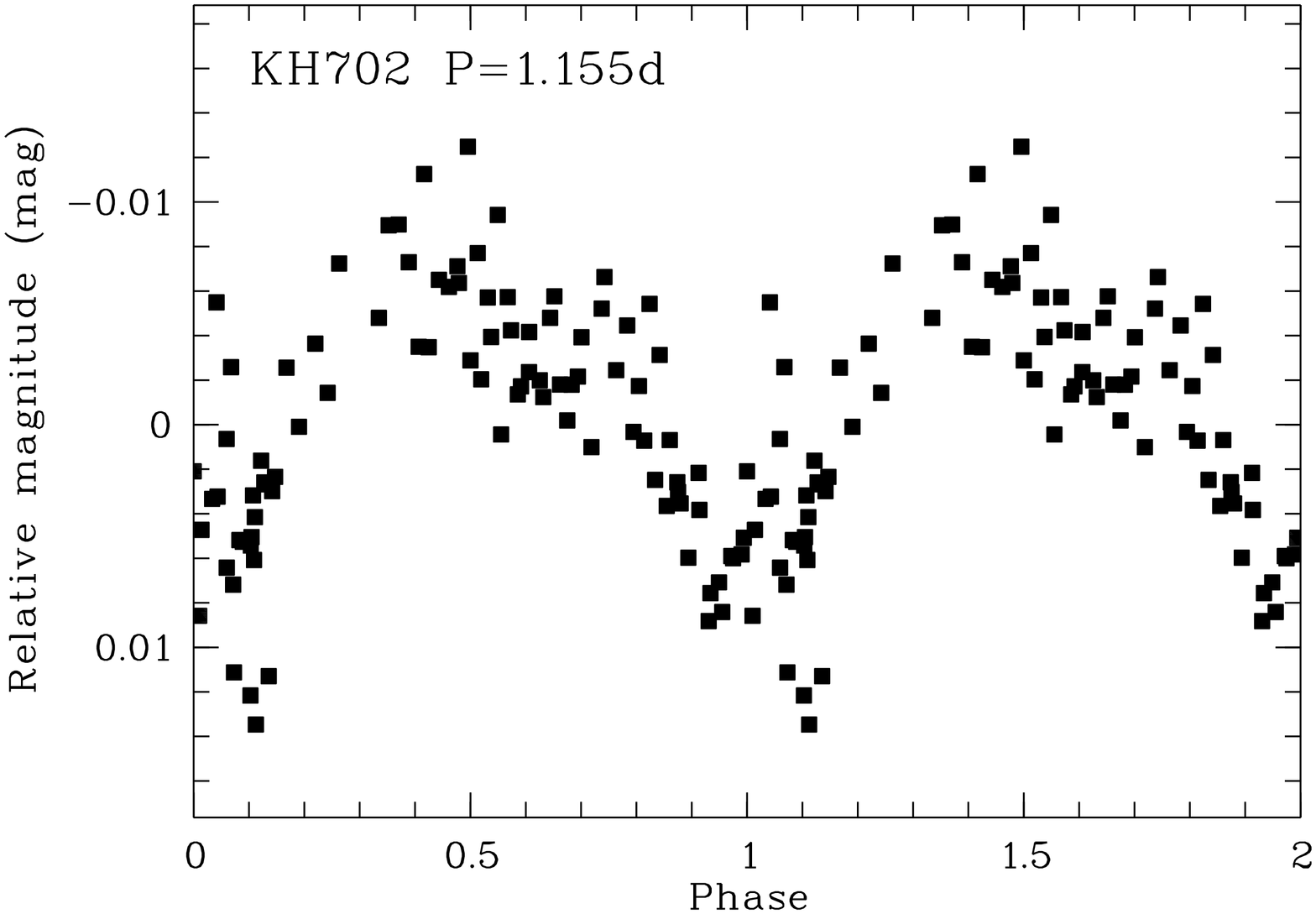} \hfill
\includegraphics[width=3.0cm,angle=-90]{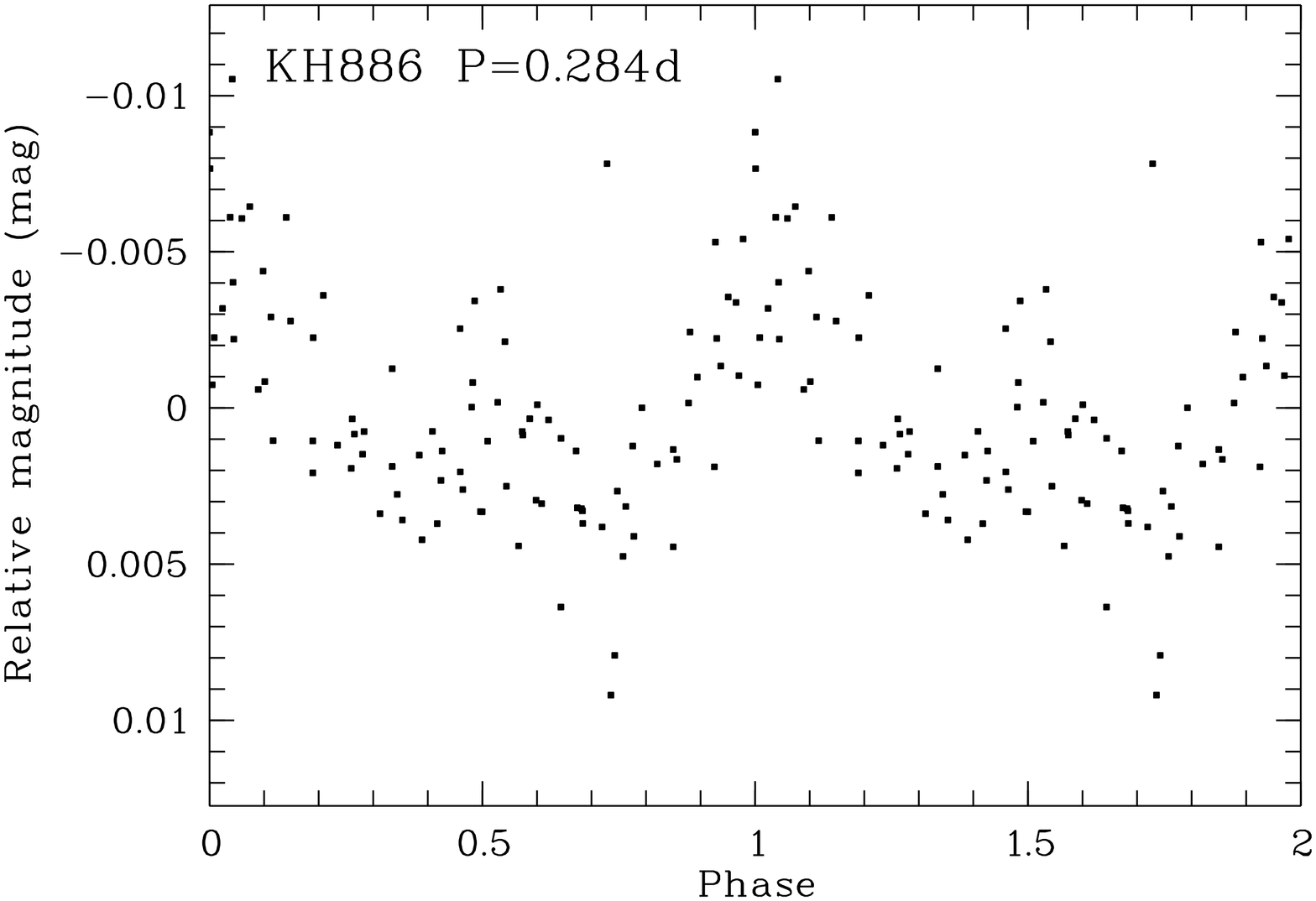} \hfill
\includegraphics[width=3.0cm,angle=-90]{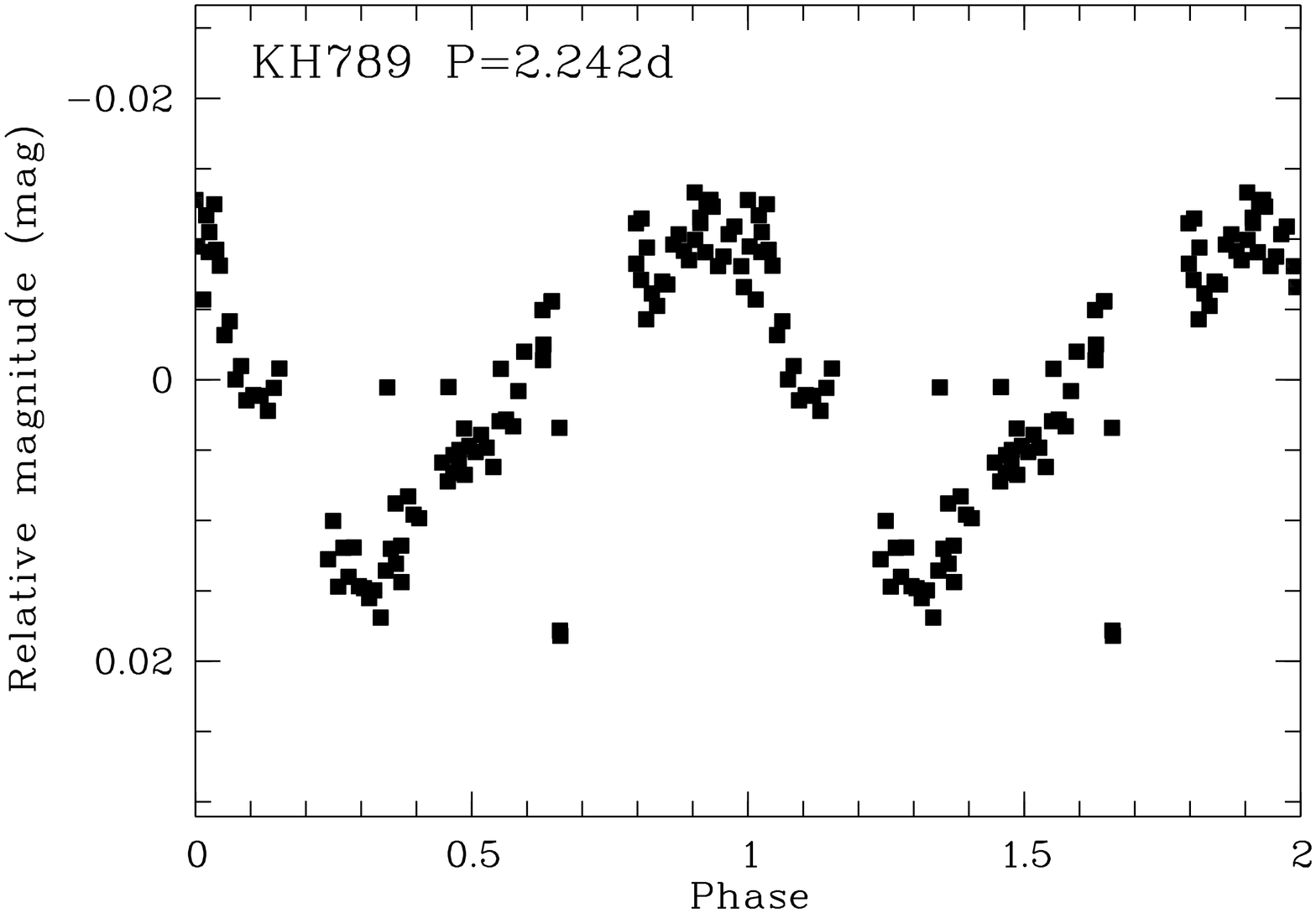} \hfill
\includegraphics[width=3.0cm,angle=-90]{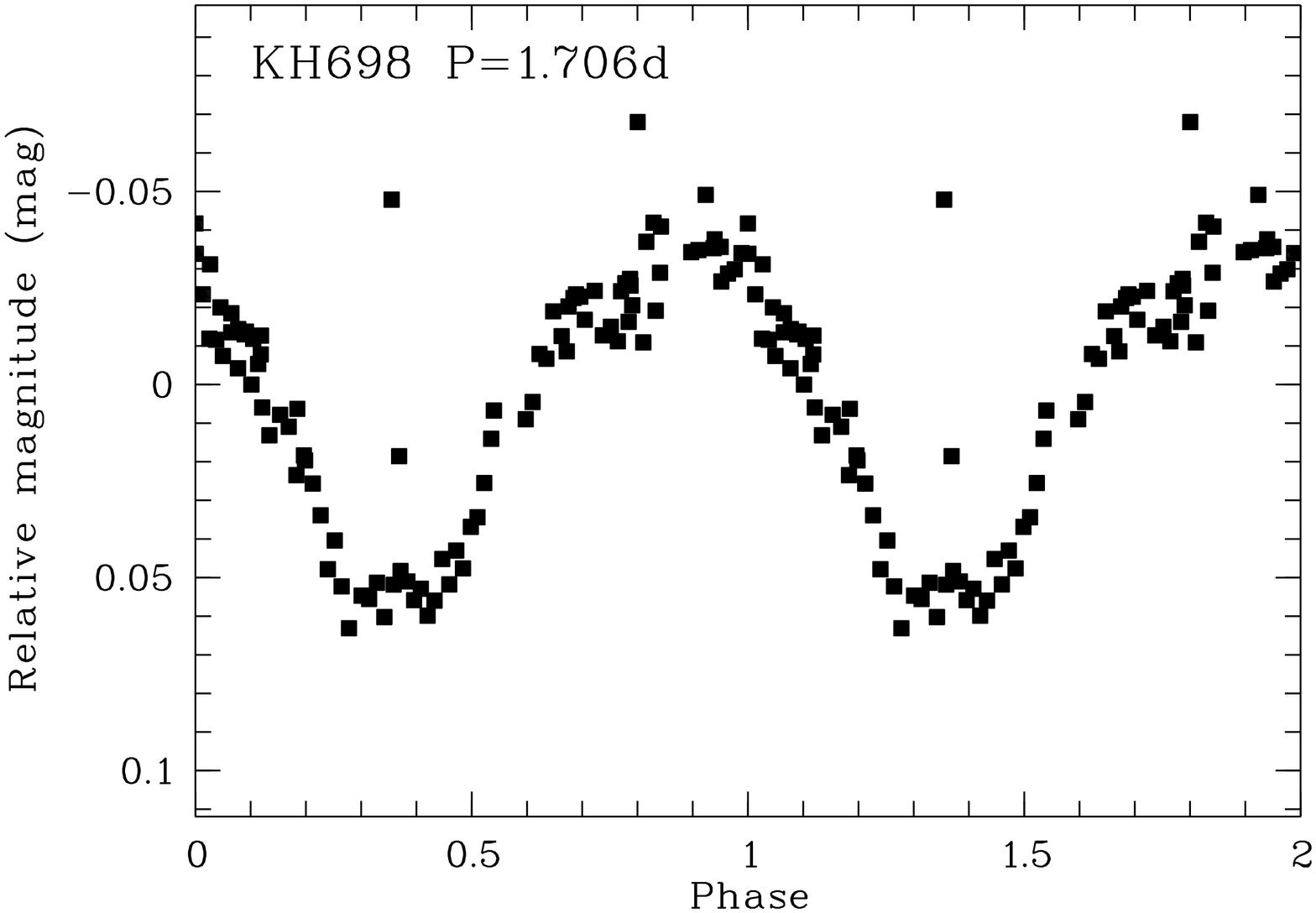} \\
\includegraphics[width=3.0cm,angle=-90]{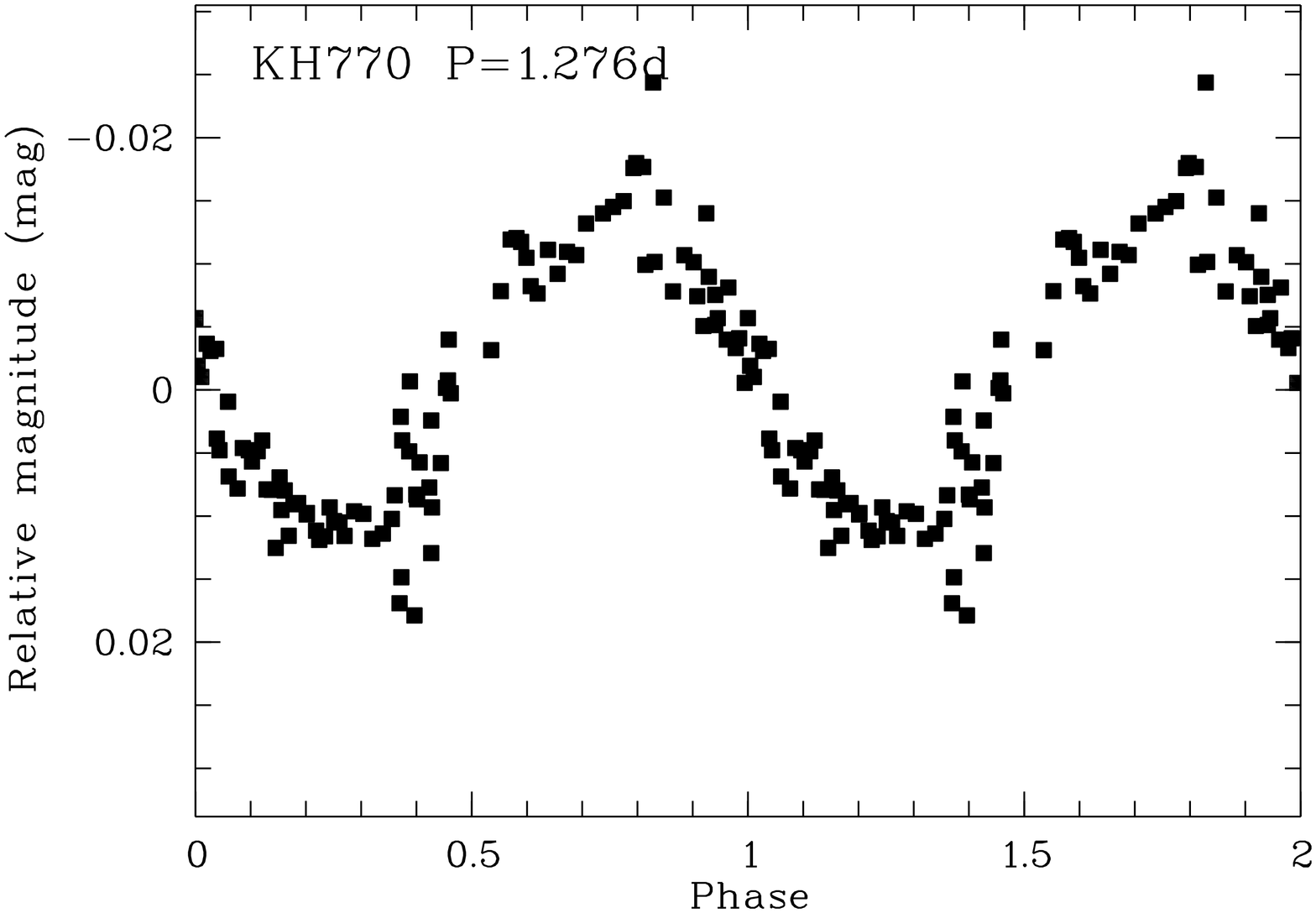} \hfill
\includegraphics[width=3.0cm,angle=-90]{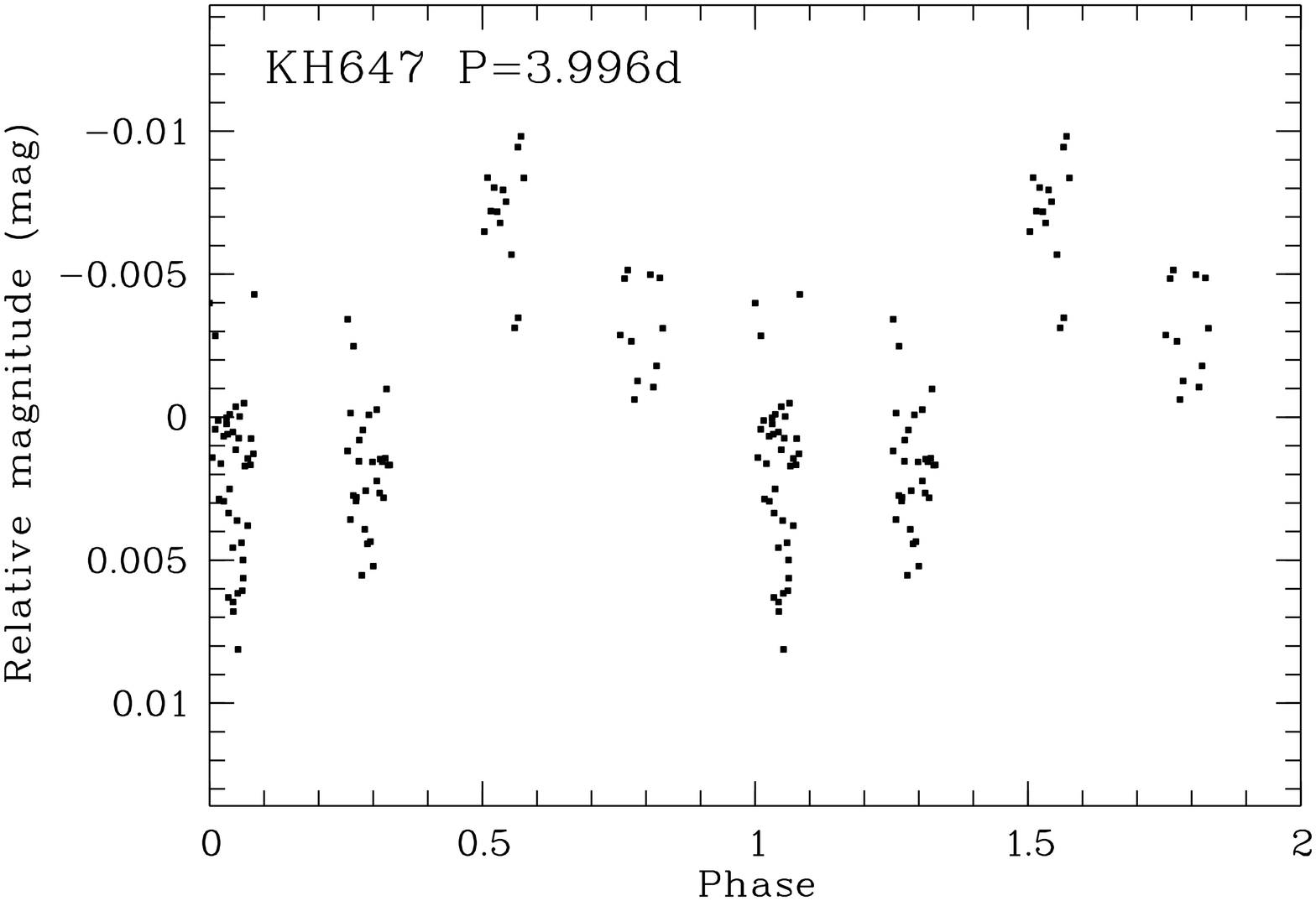} \hfill
\includegraphics[width=3.0cm,angle=-90]{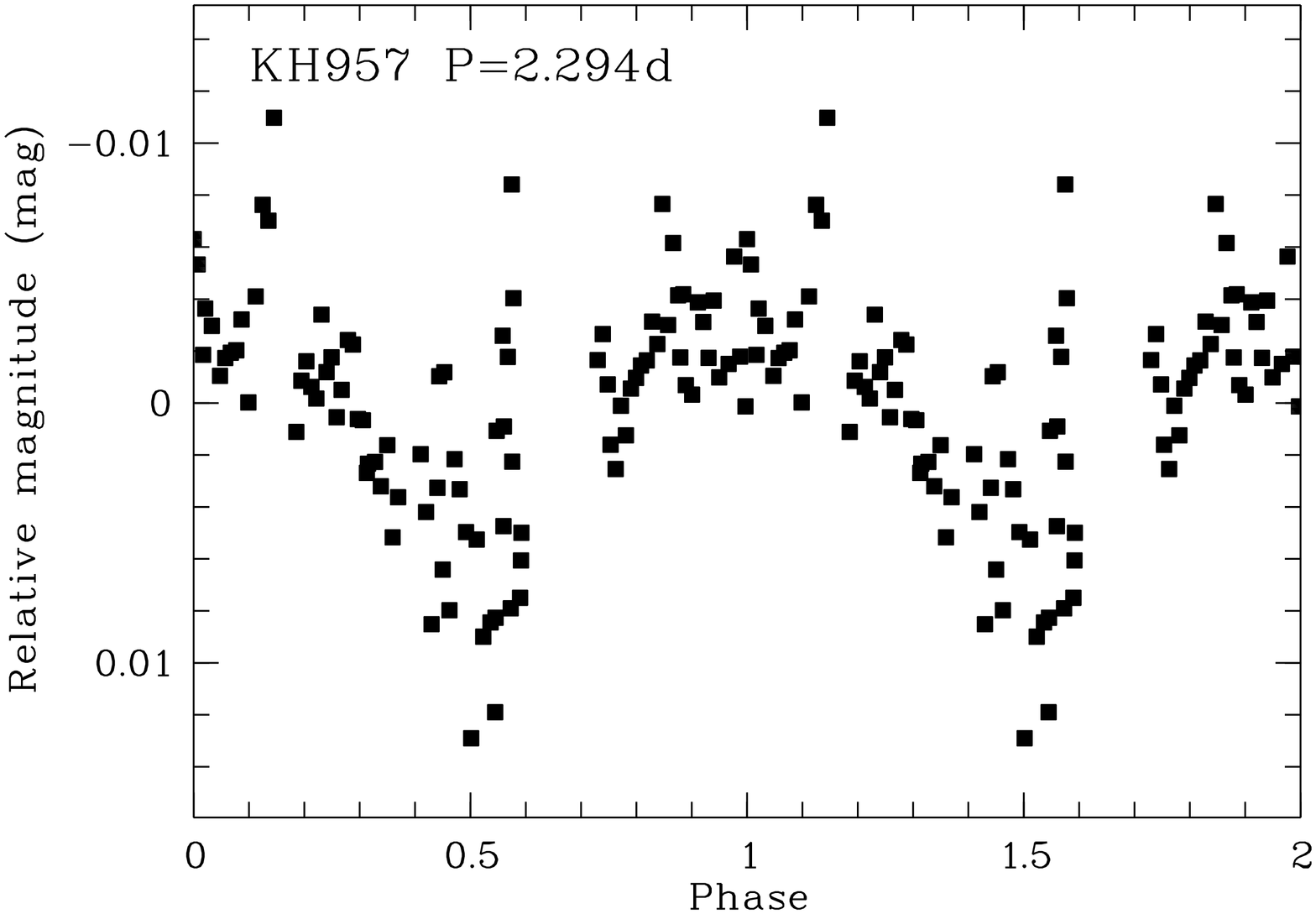} \hfill
\includegraphics[width=3.0cm,angle=-90]{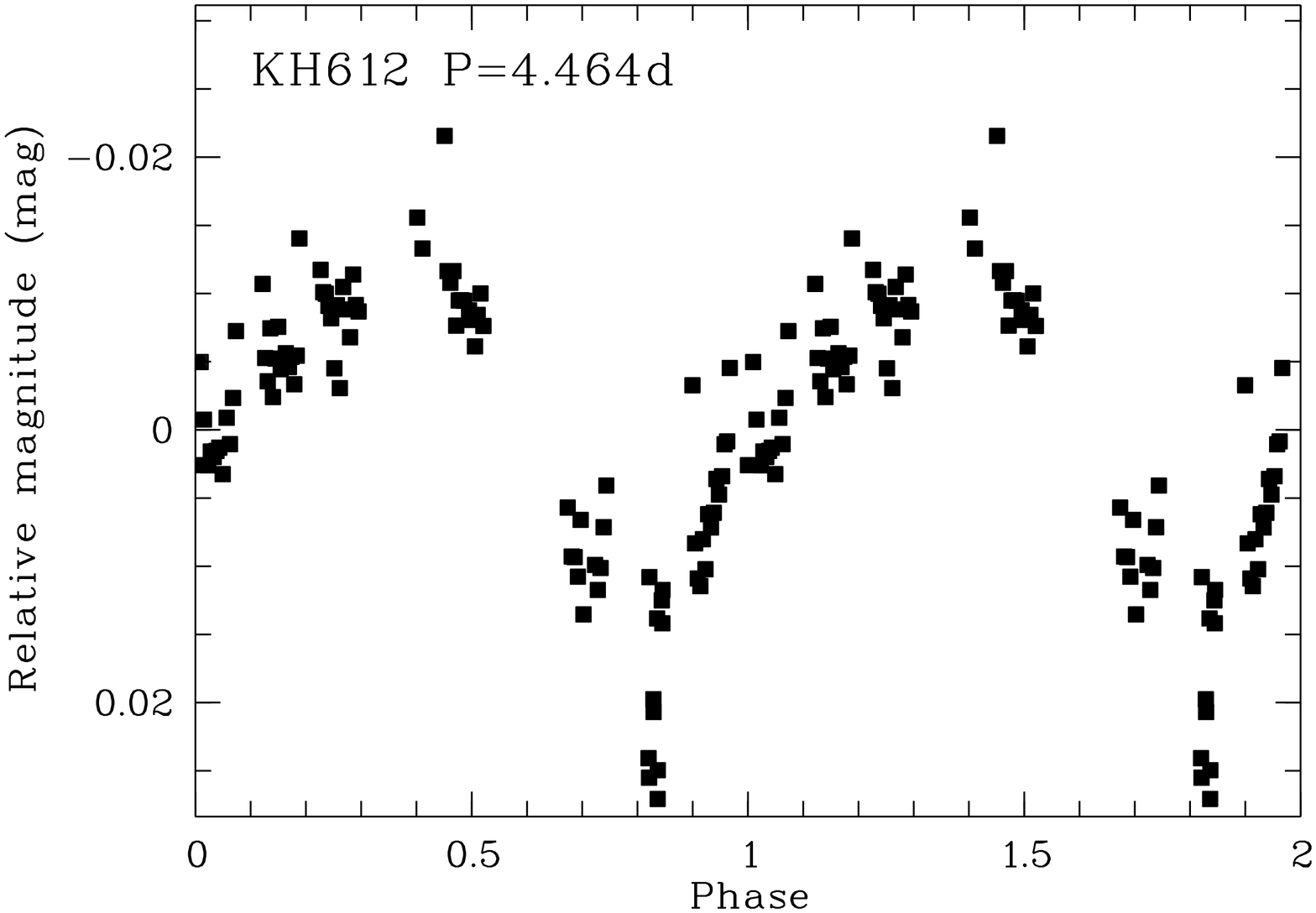} \\
\caption{Phased lightcurves for the 49 objects with periods in the order as listed in Table \ref{periods}, part 1. 
Ids from \citet{2007AJ....134.2340K} and adopted periods are indicated. The most robust periods (flag $\ge 4$)
are plotted with bold symbols. \label{f6}}
\end{figure*}

\begin{figure*}
\includegraphics[width=3.0cm,angle=-90]{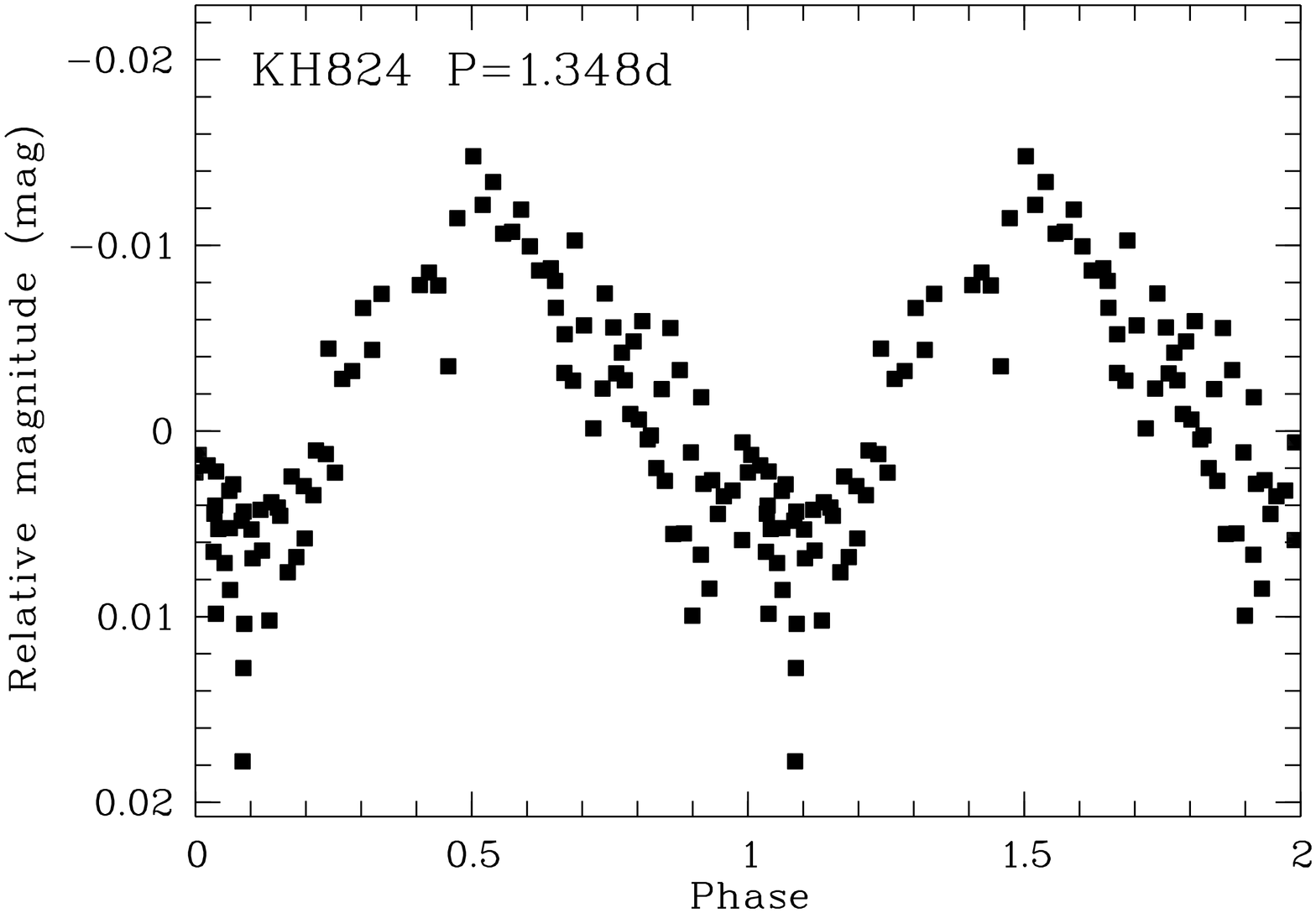} \hfill
\includegraphics[width=3.0cm,angle=-90]{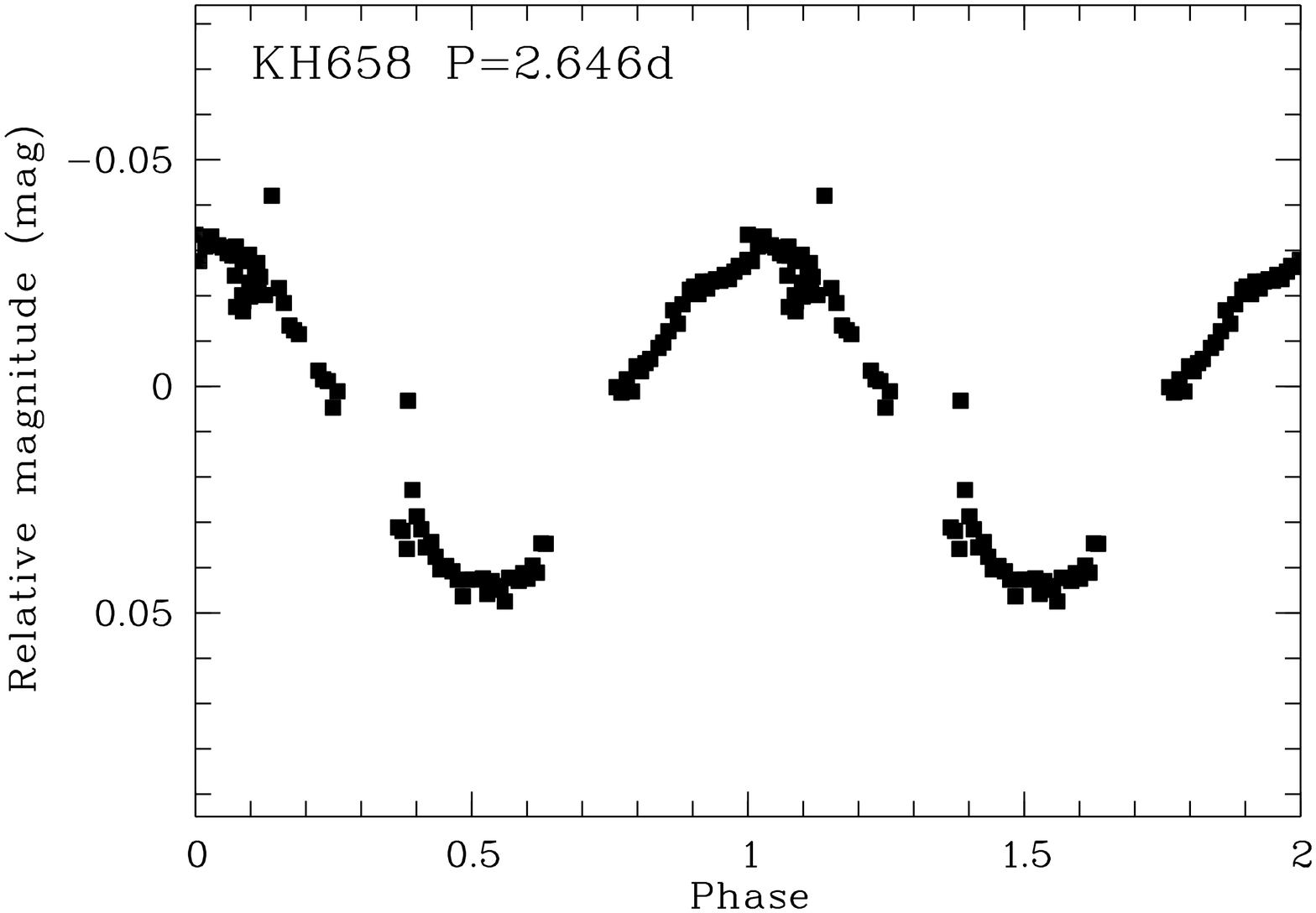} \hfill
\includegraphics[width=3.0cm,angle=-90]{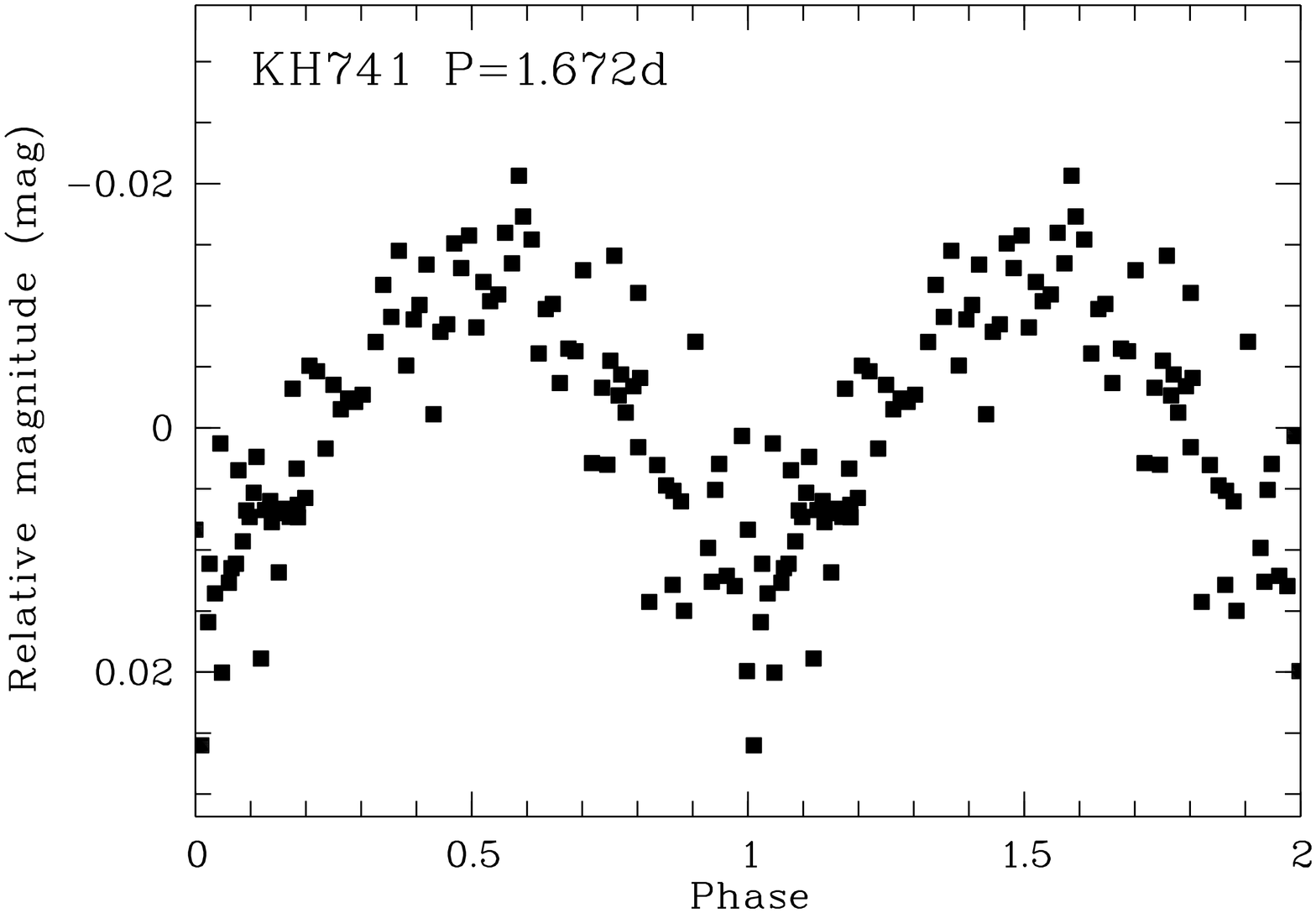} \hfill
\includegraphics[width=3.0cm,angle=-90]{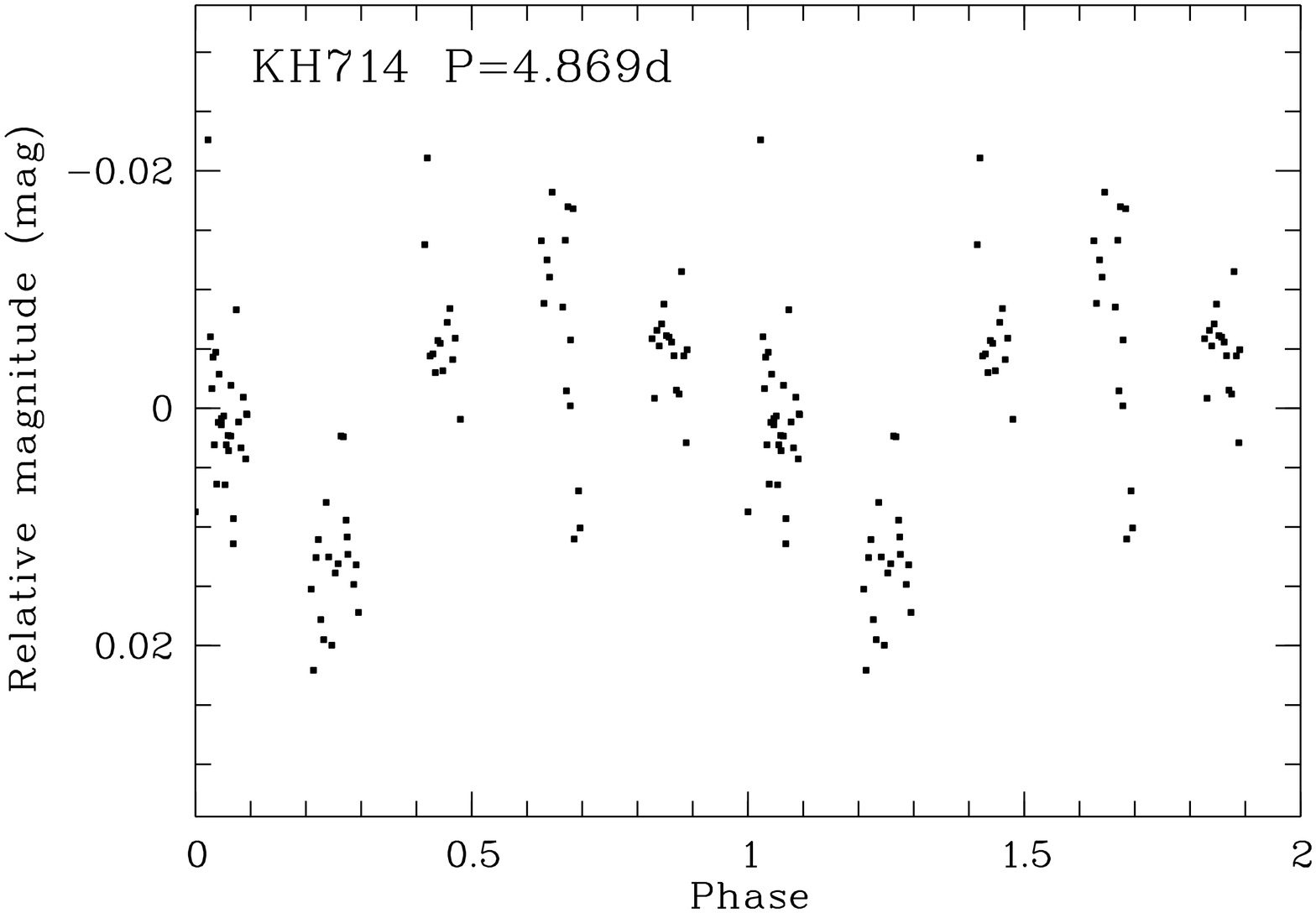} \\
\includegraphics[width=3.0cm,angle=-90]{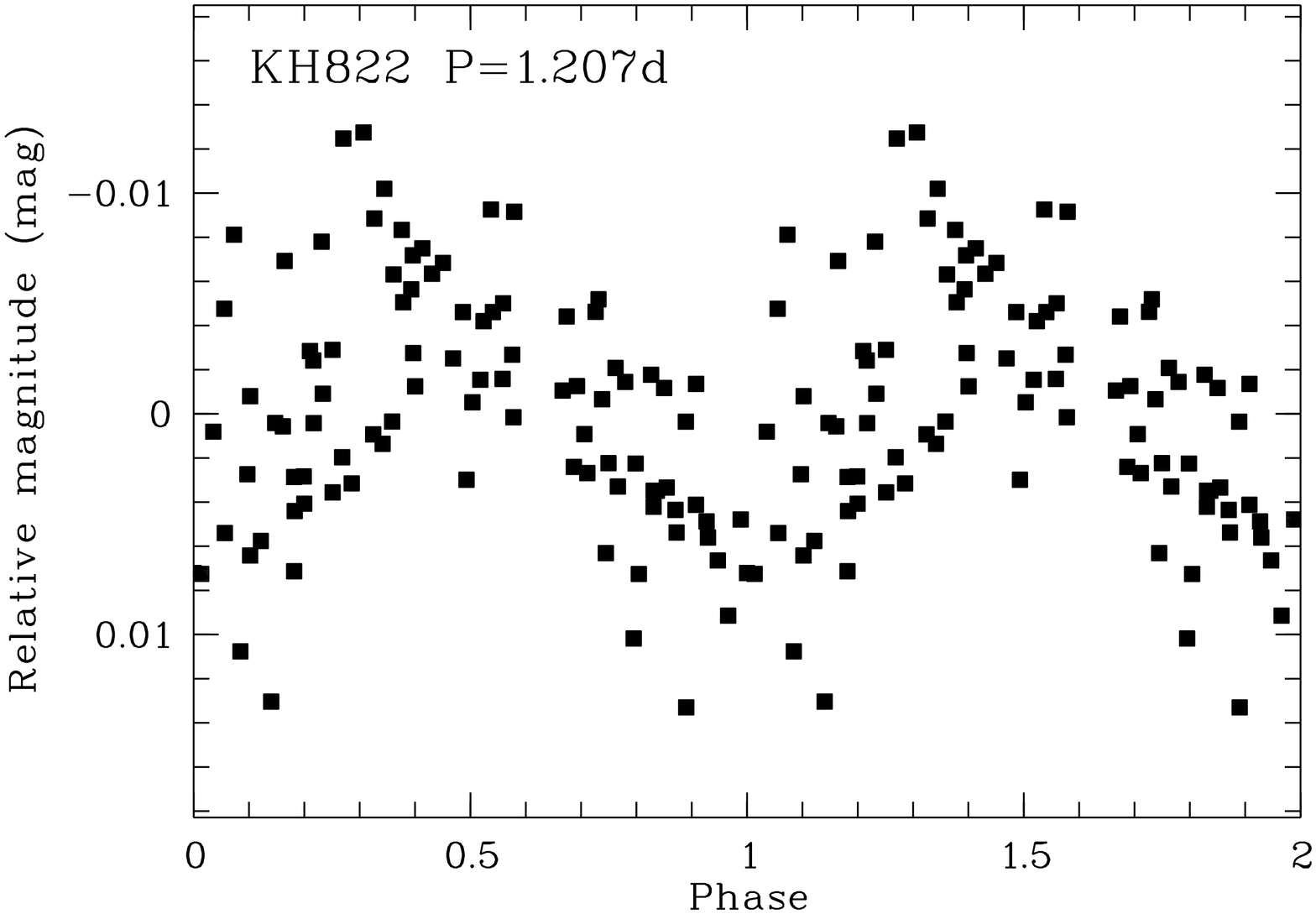} \hfill
\includegraphics[width=3.0cm,angle=-90]{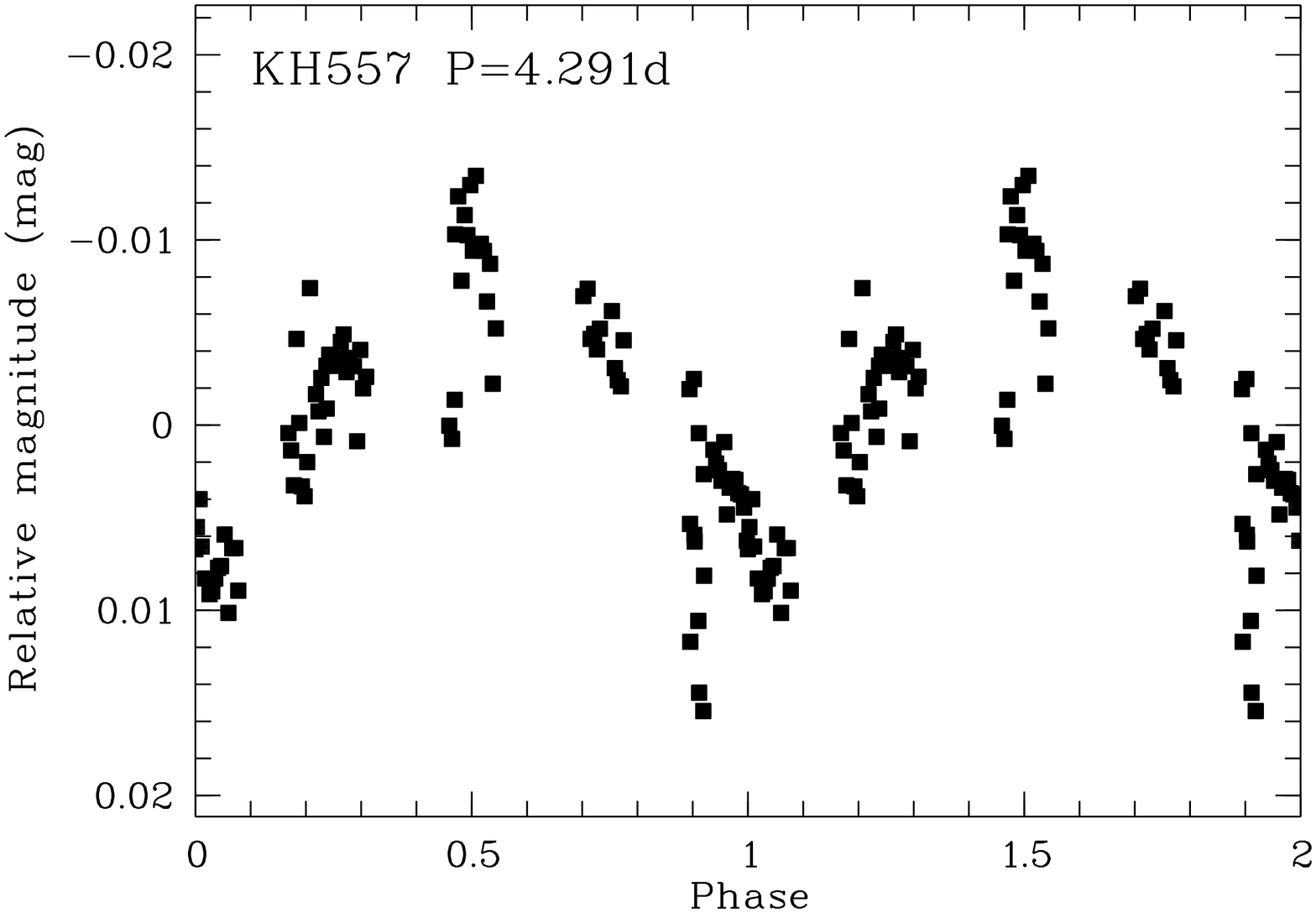} \hfill
\includegraphics[width=3.0cm,angle=-90]{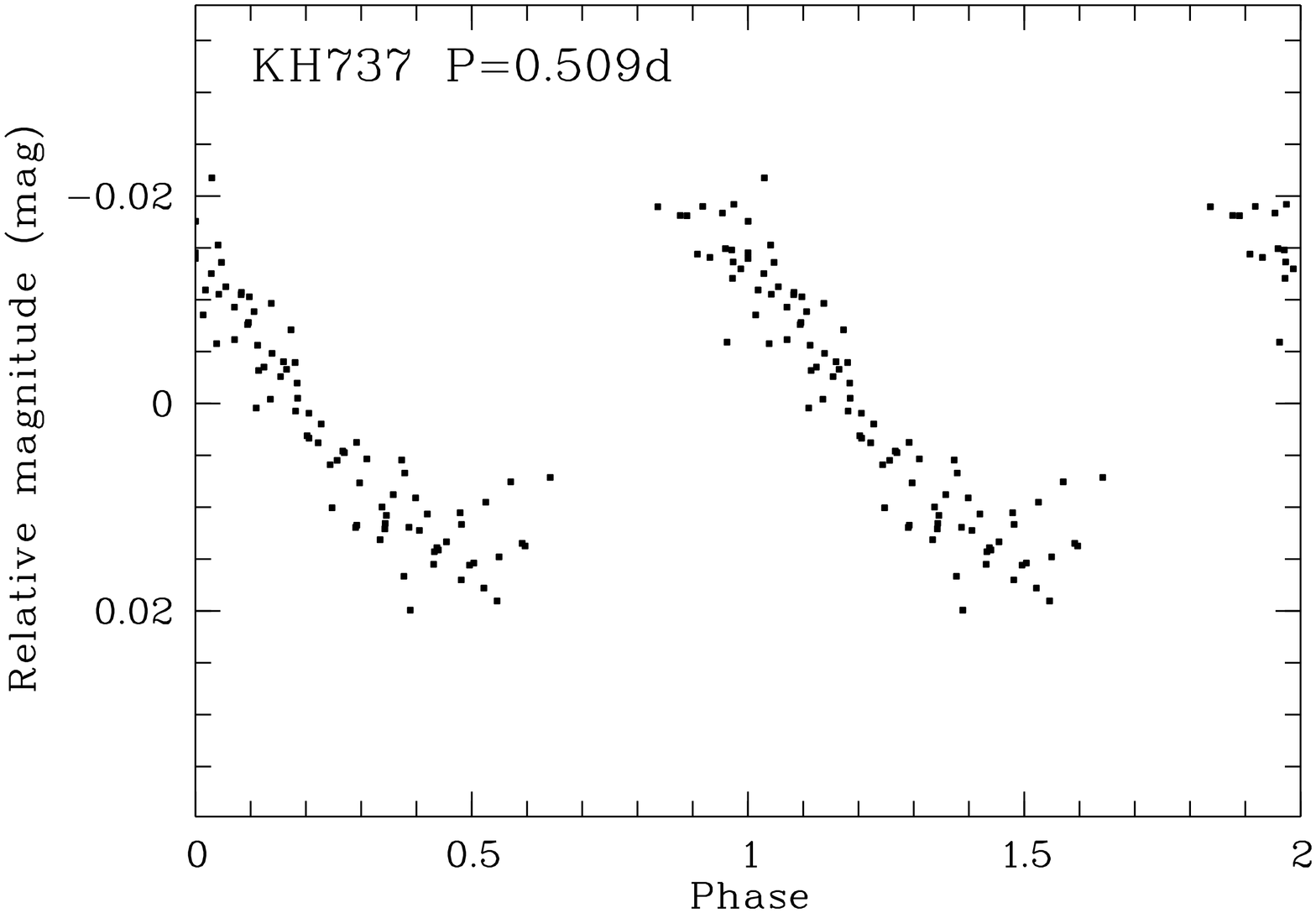} \hfill
\includegraphics[width=3.0cm,angle=-90]{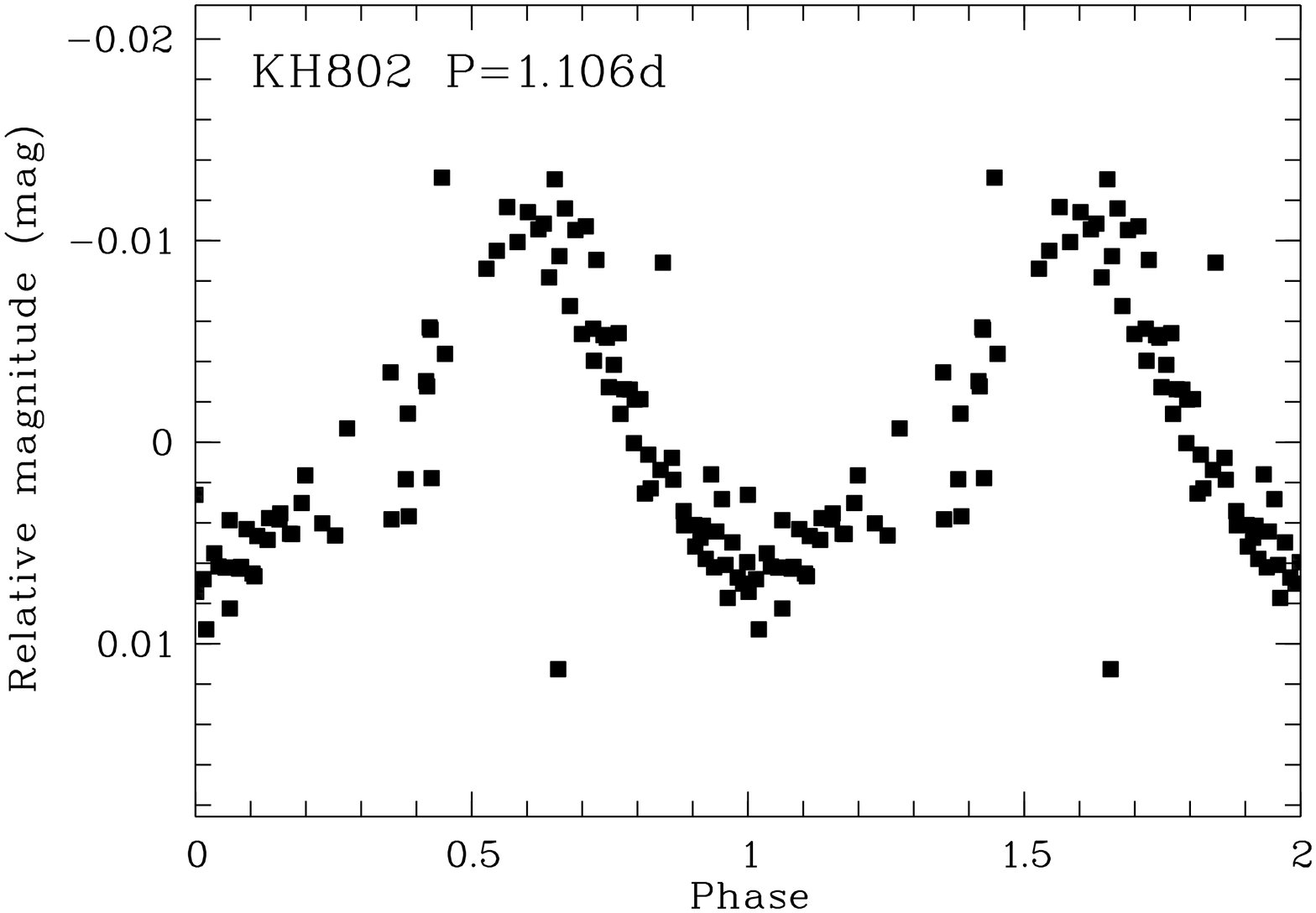} \\
\includegraphics[width=3.0cm,angle=-90]{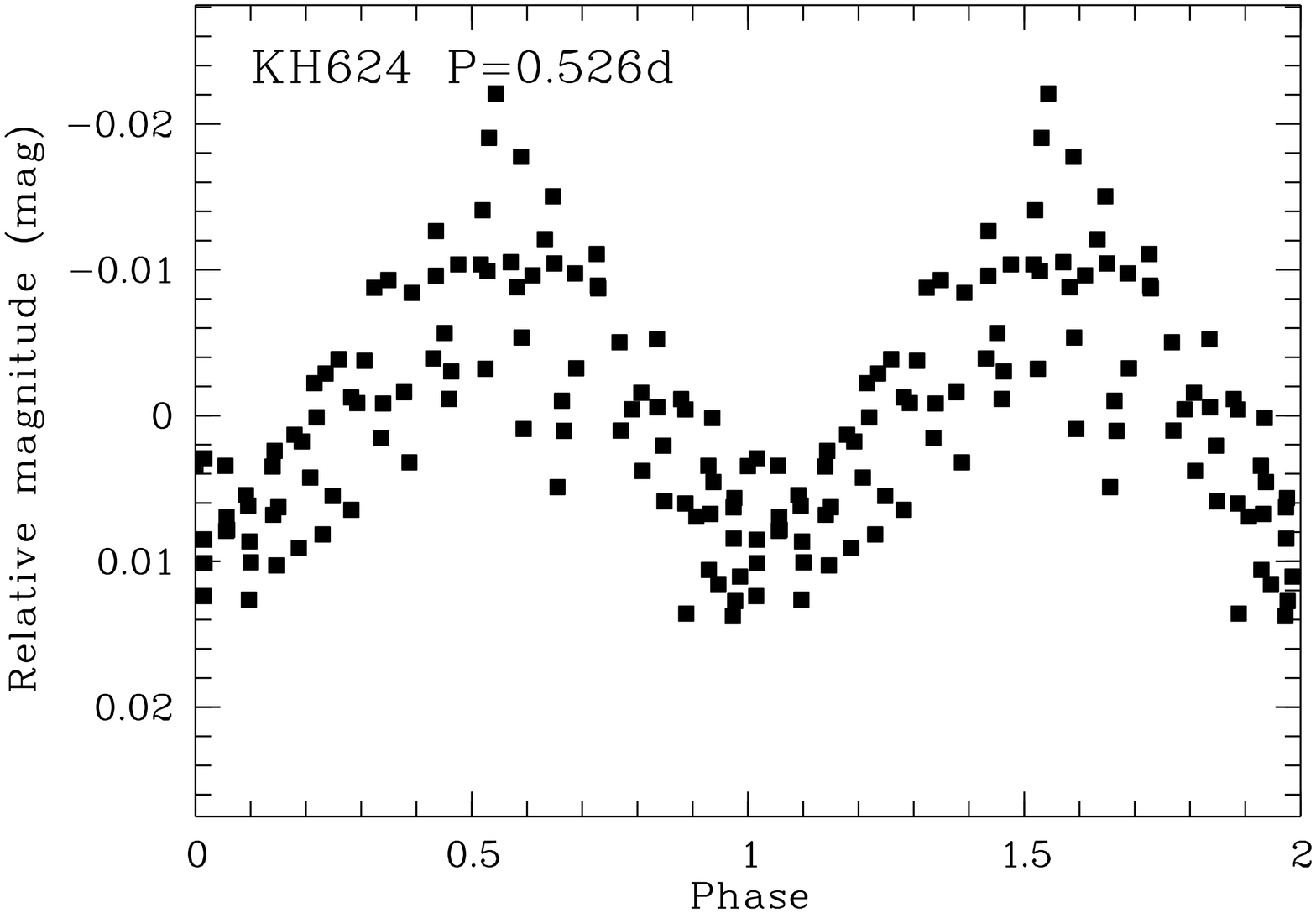} \hfill
\includegraphics[width=3.0cm,angle=-90]{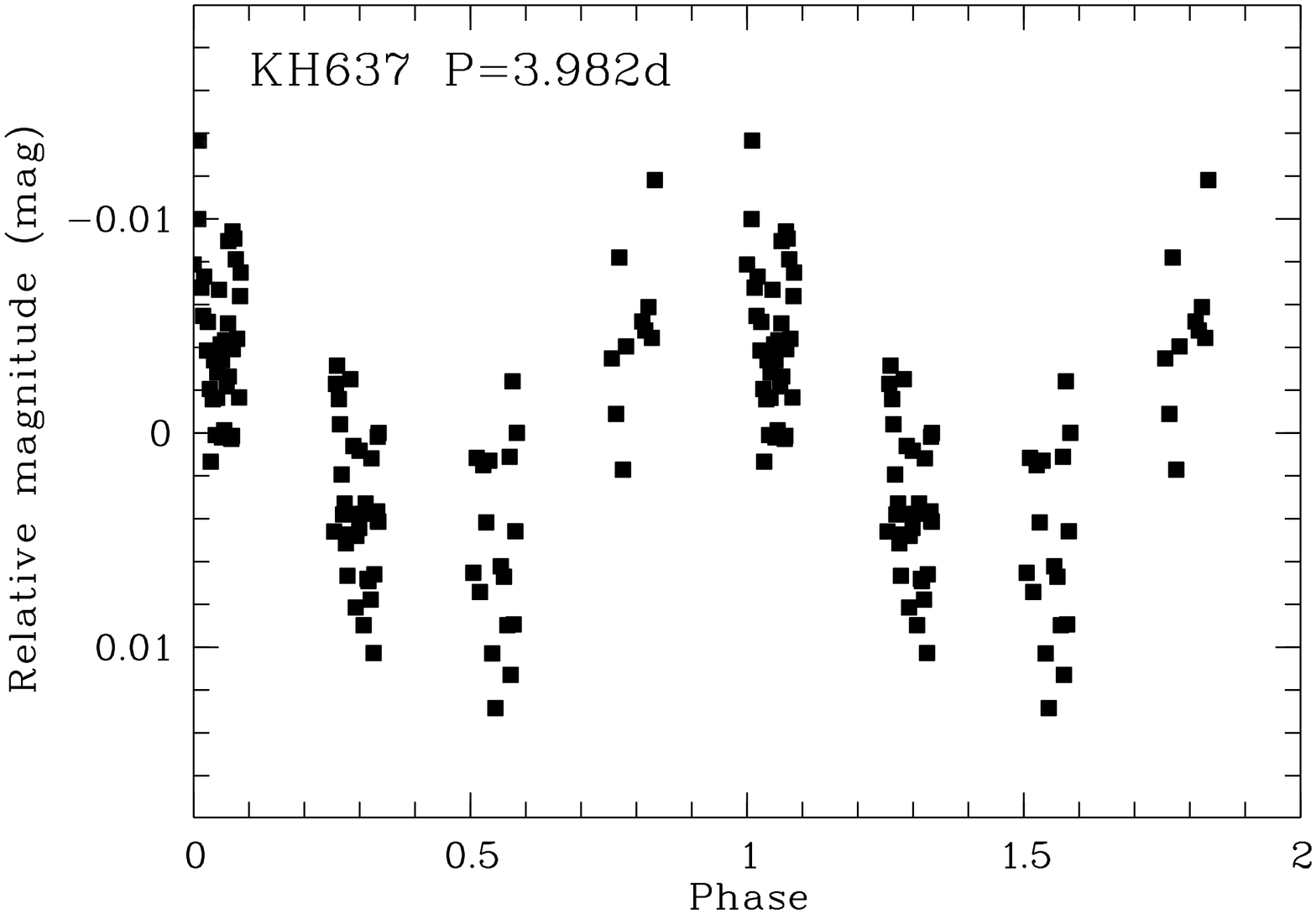} \hfill
\includegraphics[width=3.0cm,angle=-90]{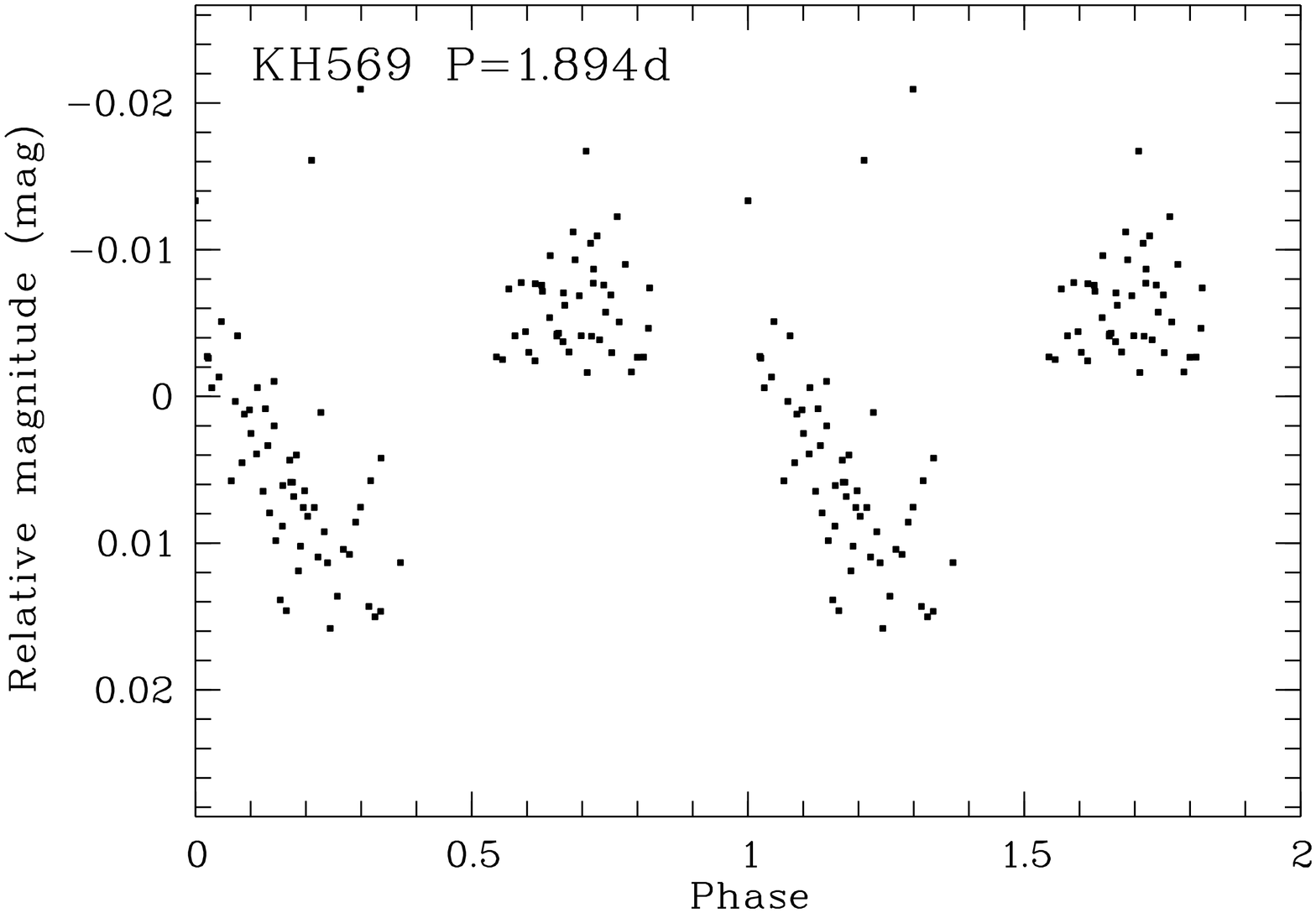} \hfill
\includegraphics[width=3.0cm,angle=-90]{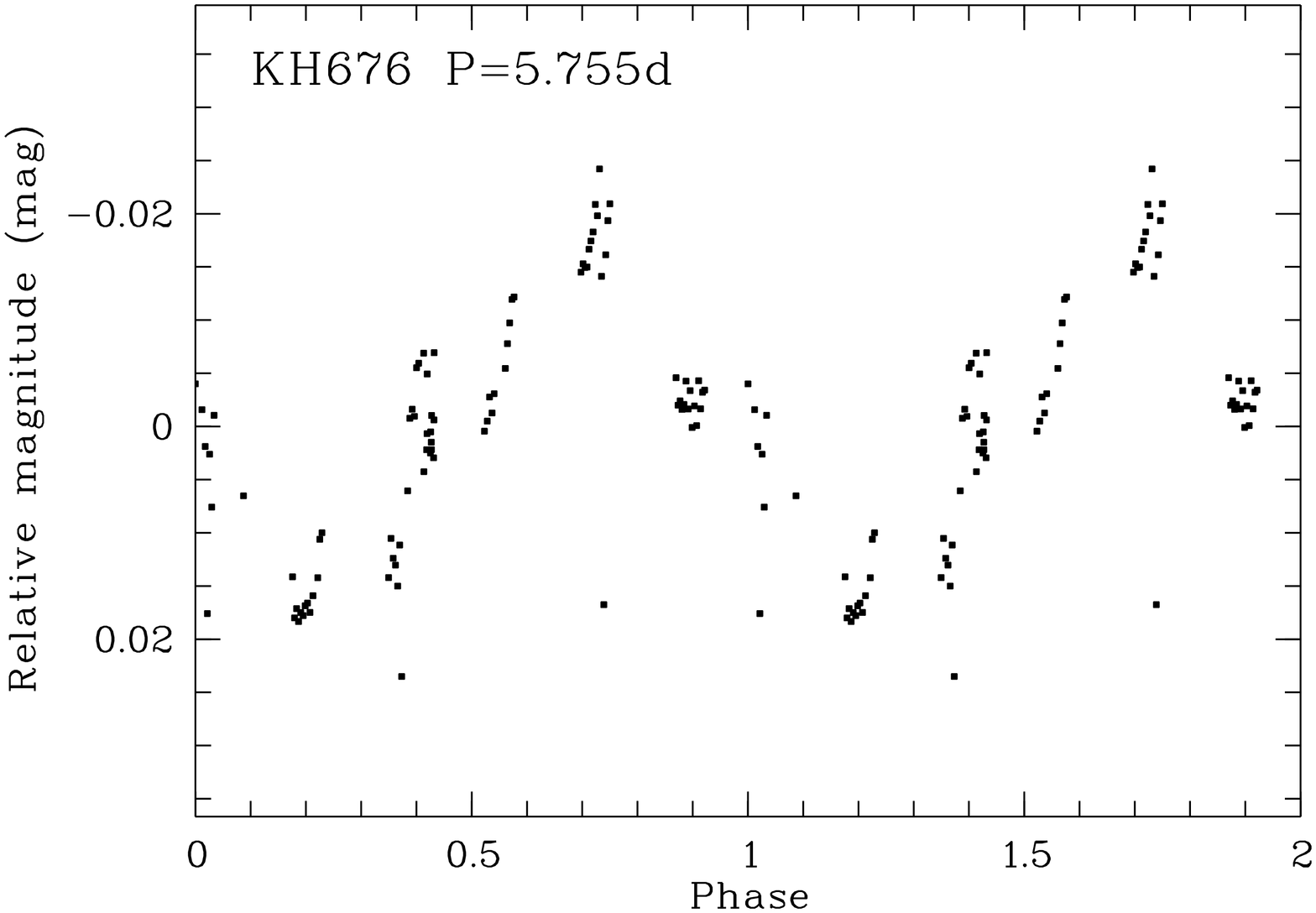} \\
\includegraphics[width=3.0cm,angle=-90]{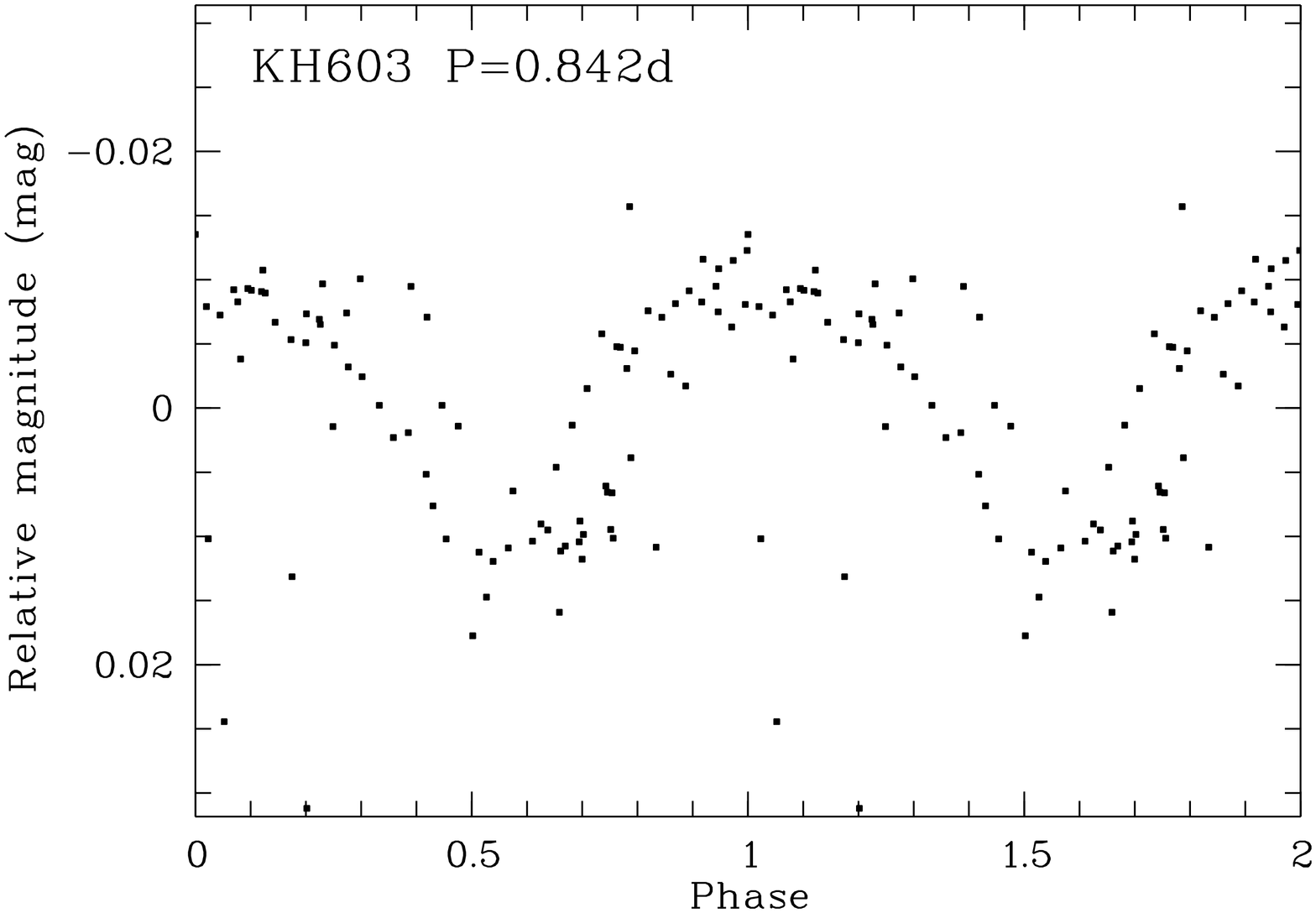} \hfill
\includegraphics[width=3.0cm,angle=-90]{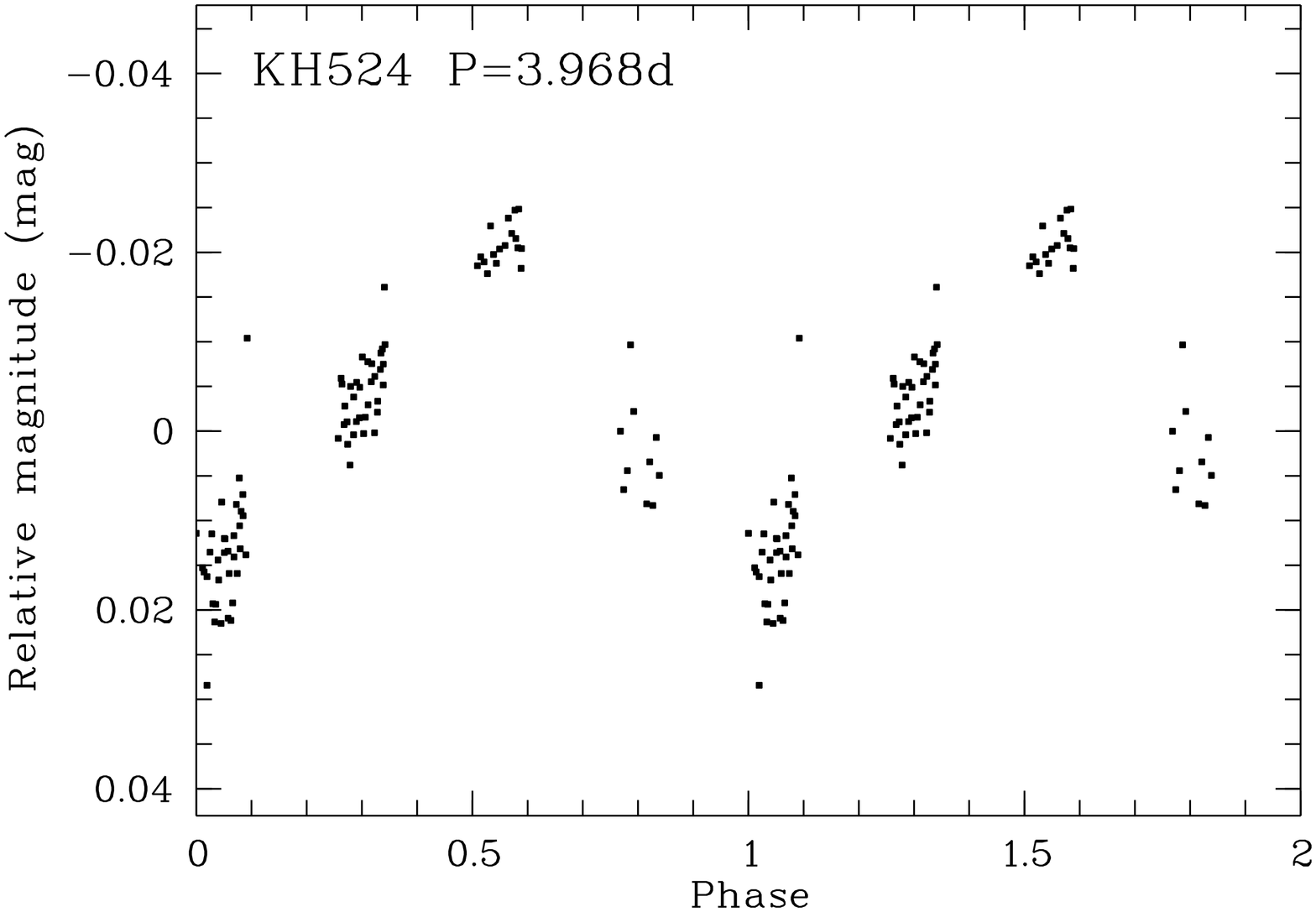} \hfill
\includegraphics[width=3.0cm,angle=-90]{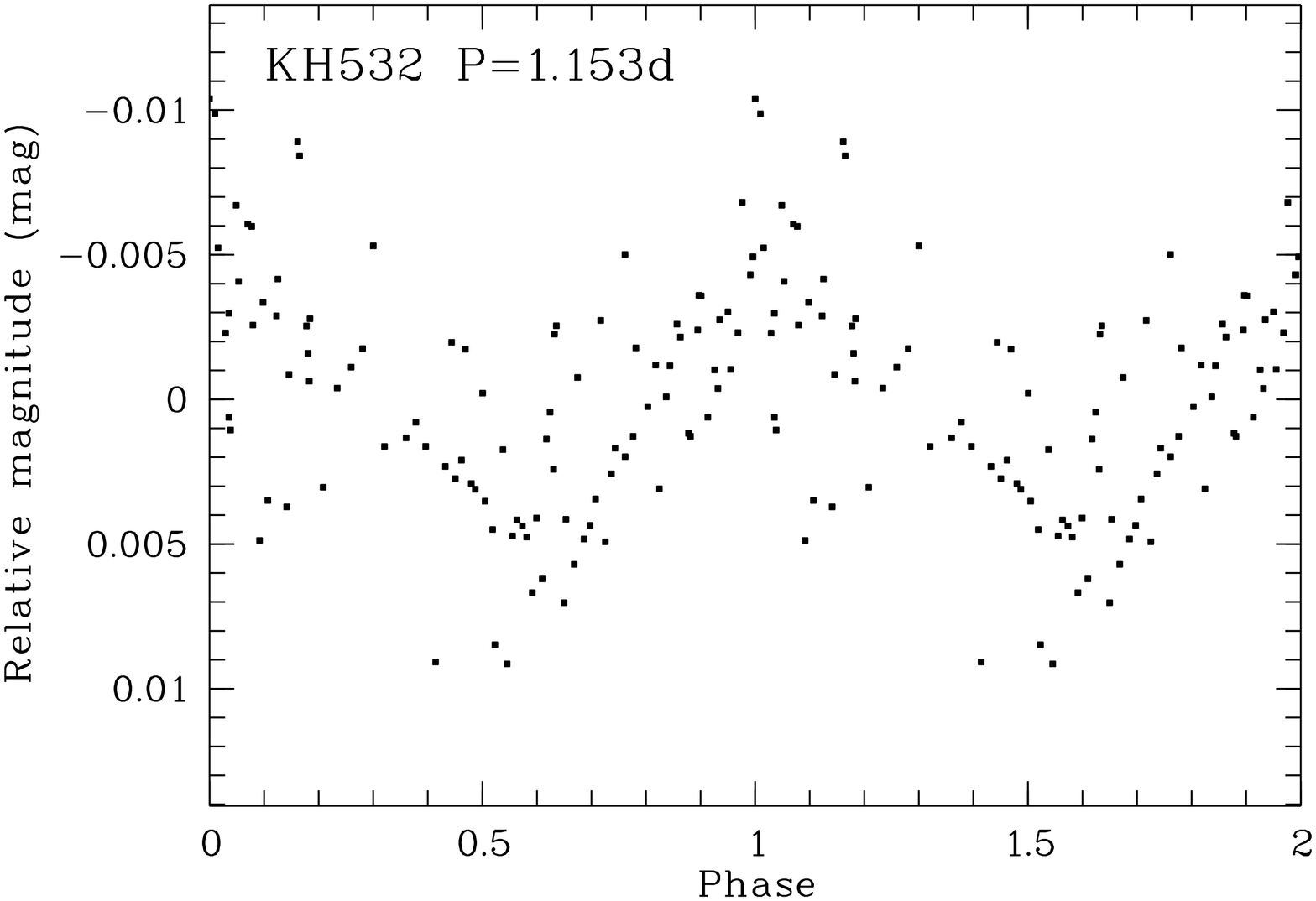} \hfill
\includegraphics[width=3.0cm,angle=-90]{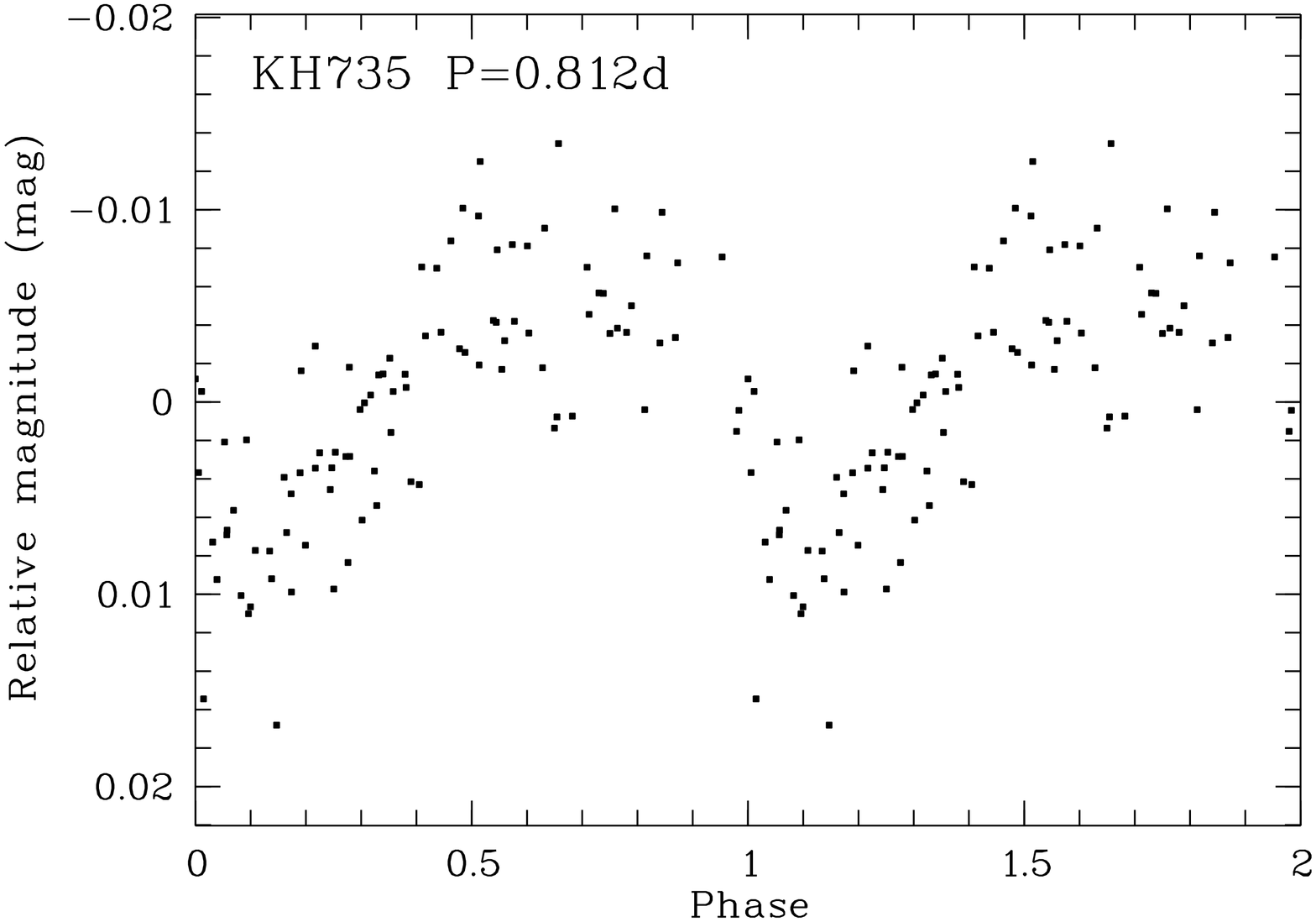} \\
\includegraphics[width=3.0cm,angle=-90]{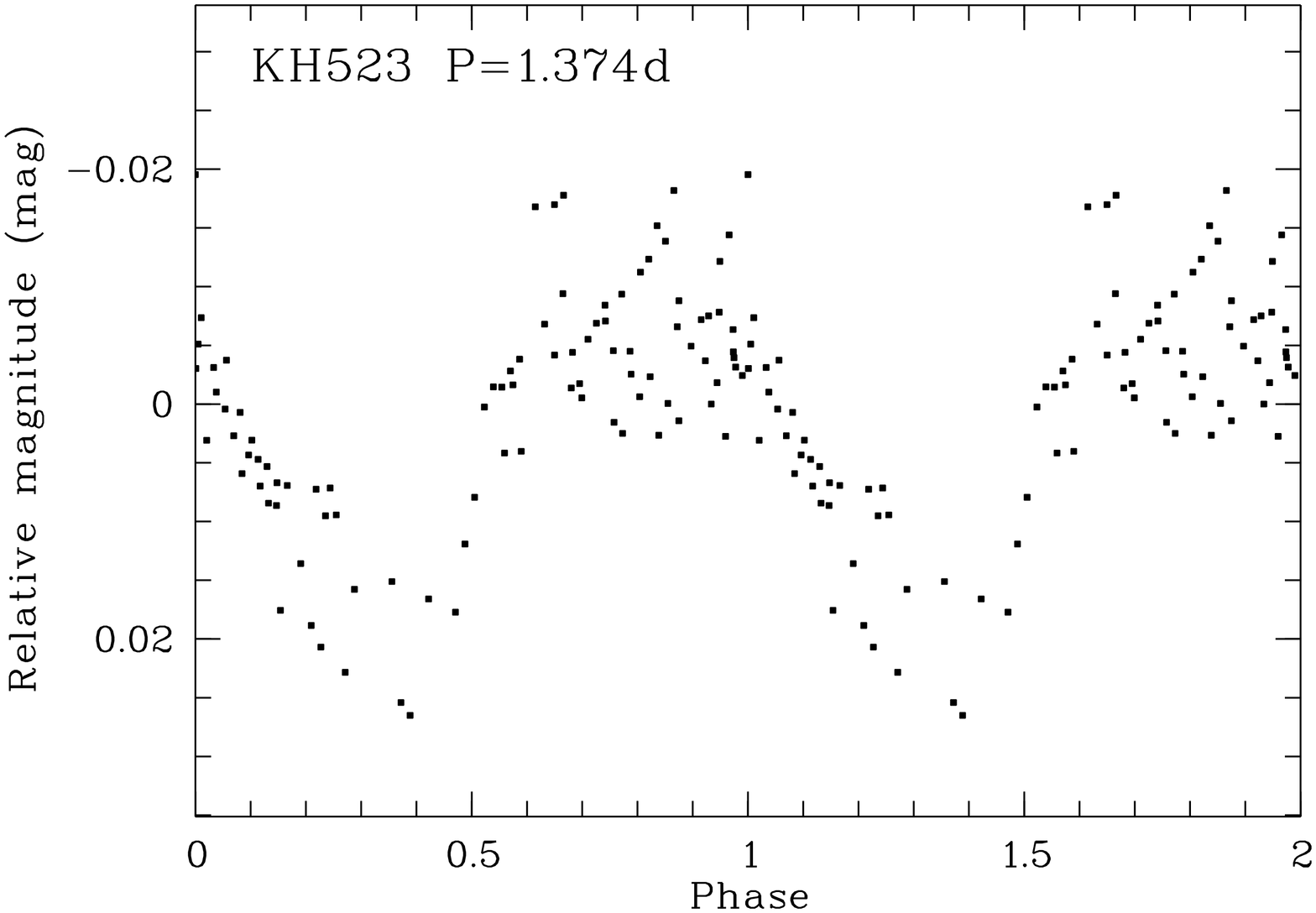} \hfill
\includegraphics[width=3.0cm,angle=-90]{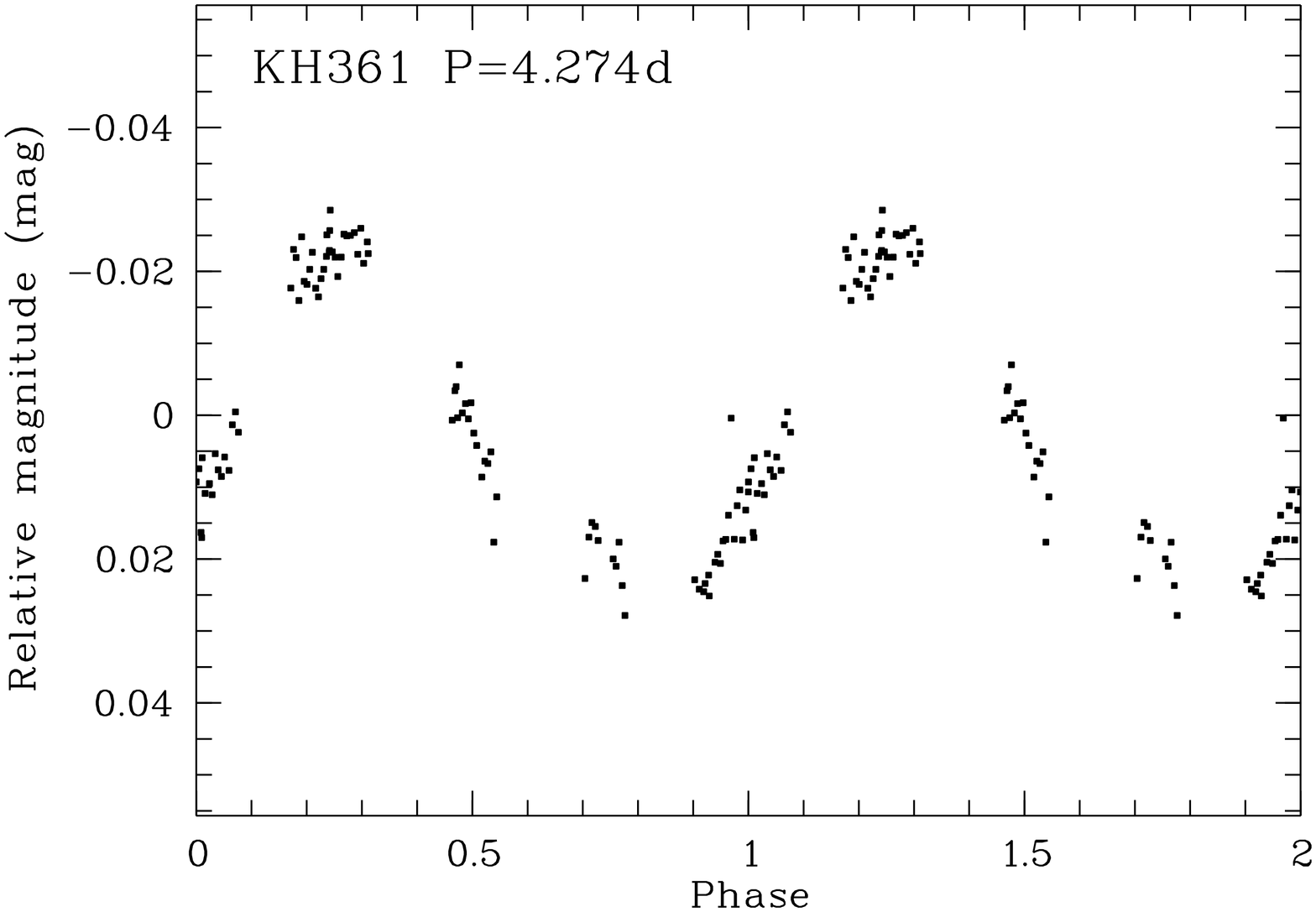} \hfill
\includegraphics[width=3.0cm,angle=-90]{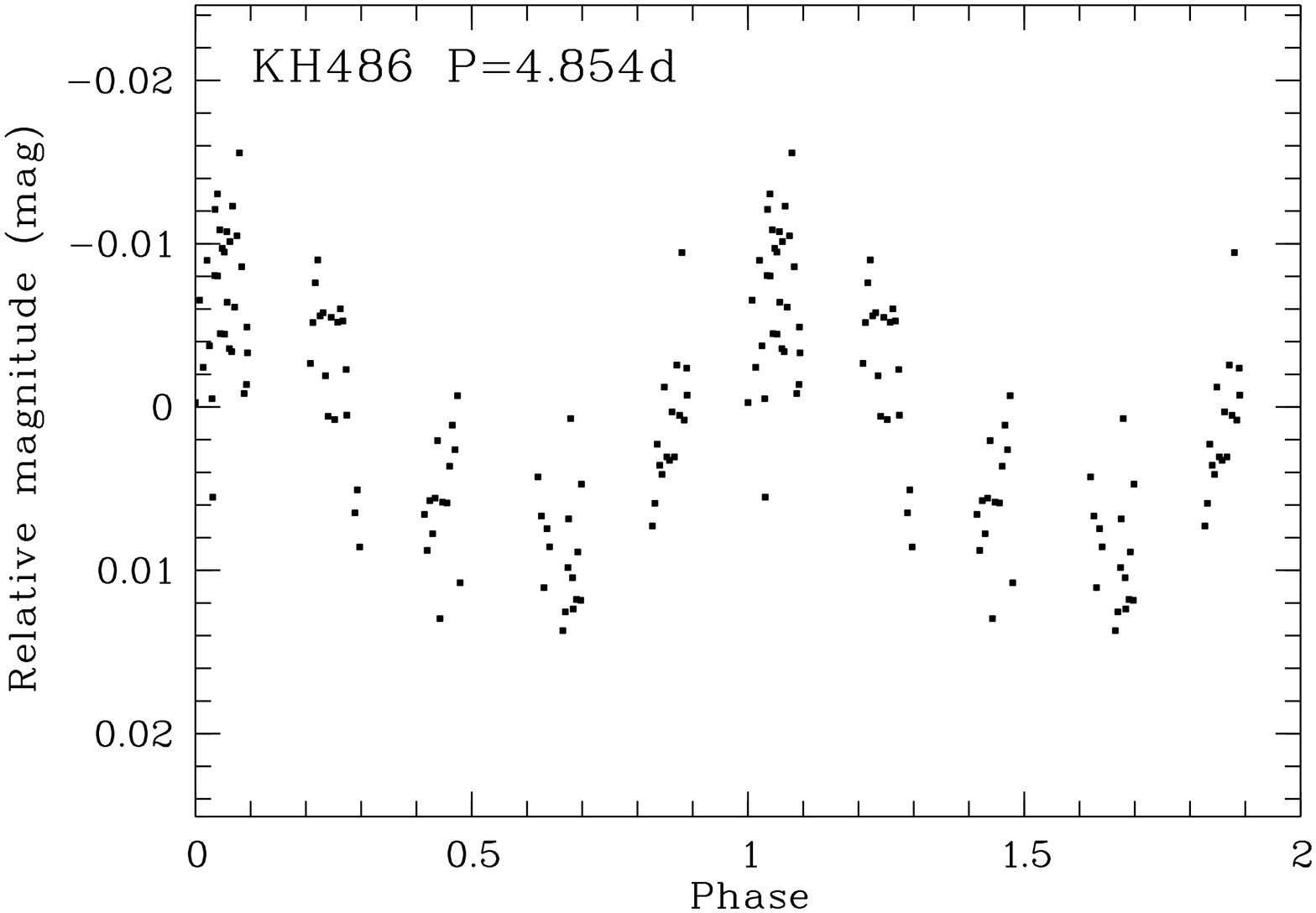} \hfill
\includegraphics[width=3.0cm,angle=-90]{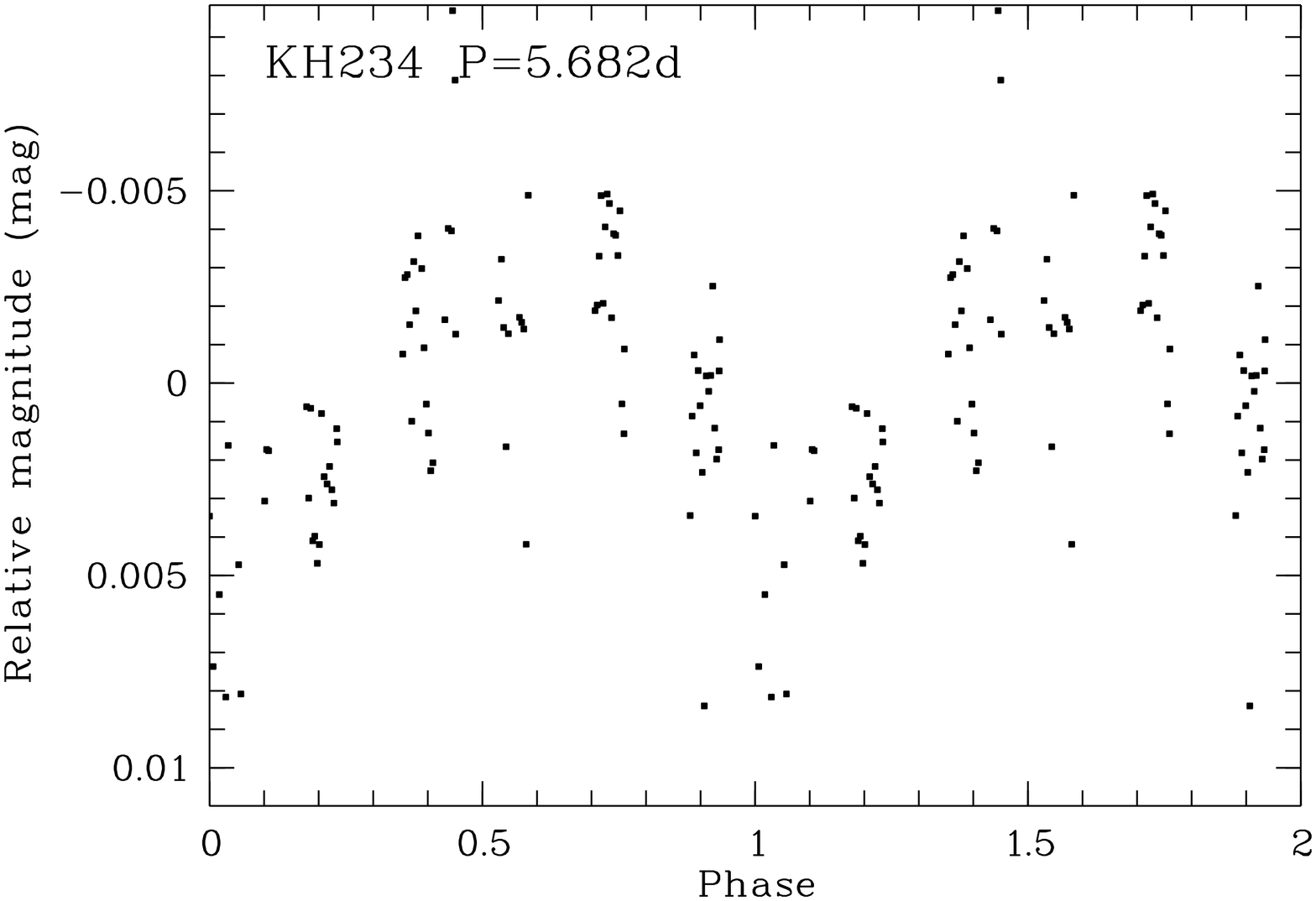} \\
\includegraphics[width=3.0cm,angle=-90]{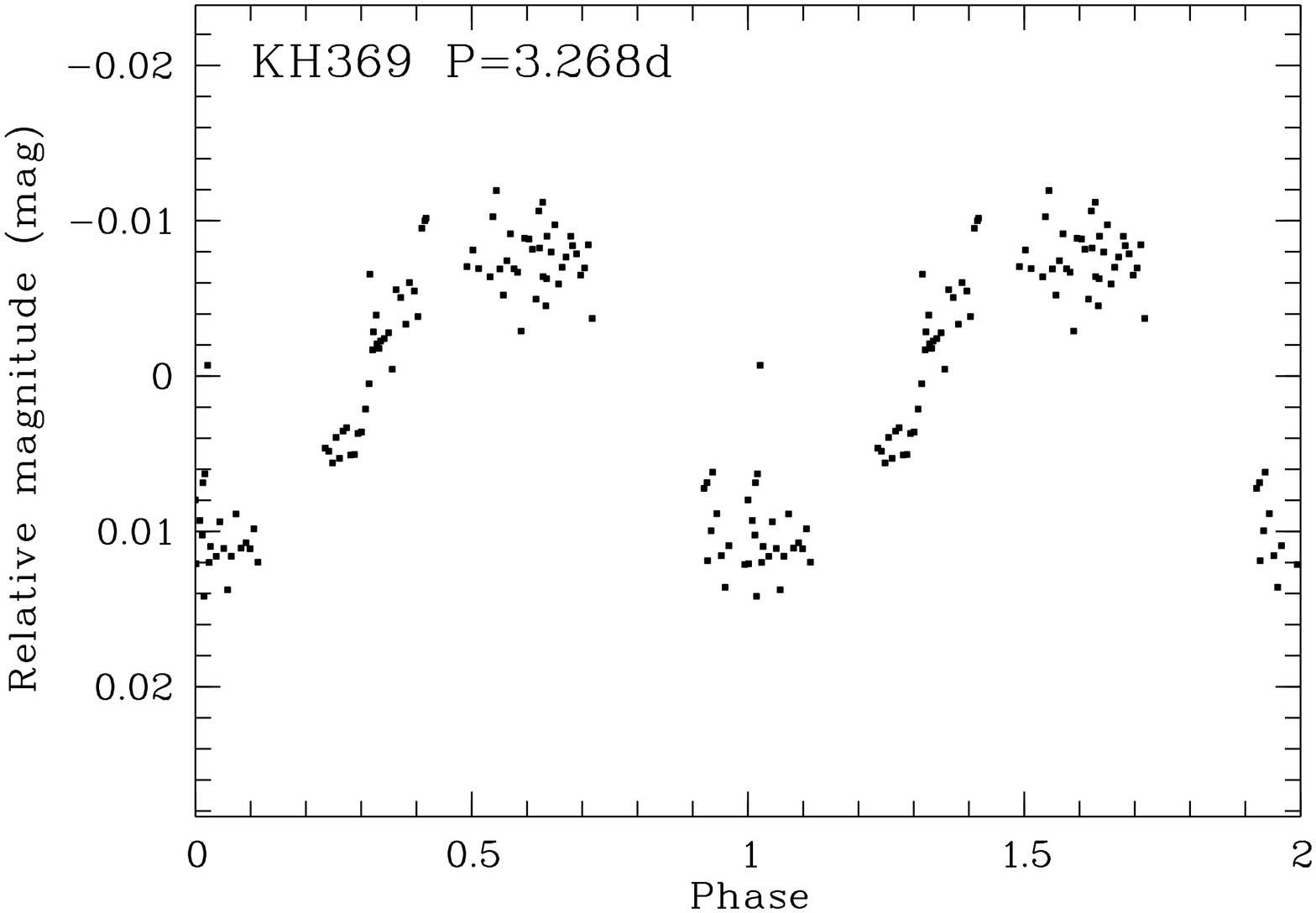} \\
\caption{Phased lightcurves for the 49 objects with periods in the order as listed in Table \ref{periods}, part 2. 
Ids from \citet{2007AJ....134.2340K} and adopted periods are indicated. The most robust periods (flag $\ge 4$)
are plotted with bold symbols. \label{f7}}
\end{figure*}

In Table \ref{periods} we list our final sample of 49 periods. For a period to appear in this table,
it had to be registered at least in two different period search routines. In the final column
of the table we summarise the results of the period search using a four digit flag. Each
digit gives the number of consistent period measurements for this object for each of the four
algorithms in the order as described above, i.e. Monitor, SE2004, CLN, SL. Adding these four digits 
indicates how often a given period has been measured consistently. The phased lightcurves
for these 49 periods are shown in Fig. \ref{f6} and \ref{f7} in the same order as in the table.

In addition, the period search yields 24 periods which are only detected once (sum of the flags in 
Table \ref{periods} would be 1). These mostly have dodgy phaseplots and are considered to be unreliable. 
Moreover, these periods do not follow the trends reported in the mass-period diagram (Sect. 
\ref{rotmass}) and include a substantial sample of long periods $>6$\,d which we do not consider to 
be trustworthy.

The comparison of the four independent period samples shows some noteworthy trends. First, there is a
group of objects for which the same period is reliably recovered by several algorithms (flag $\ge 4$).
While re-assuring, this sample amounts only to one half of the total number of periods we
consider to be trustworthy. This is mostly due to the fact that the string length method is not as
sensitive as the periodogram techniques and misses a substantial number of periods which are measured
with high confidence by multiple periodogram-based algorithms. In addition, we have 5 'string length
outliers', i.e. objects for which consistent periods are measured by multiple periodogram techniques but
the string length methods provides a different result (objects KH791, 569, 603, 957, 894). For all five 
cases, the SL period is close to a harmonic of the periodogram period; the ratios between periodogram period
and SL period are approximately 5/2, 3, 1/2, 3, 4/3. In these cases, we adopt the result from the 
periodogram technique.

When the same object has a period from different periodogram techniques, the agreement is excellent.
We have only three cases with deviating results, two of them appear in the final sample in Table 
\ref{periods} because one period has been measured more then once (KH737, KH912). Excluding these outliers, 
the average deviations are 0.5\%. For comparison, the average deviation between periods from periodograms
and periods from the string length method is, for objects with consistent periods, 1.2\%. In Table
\ref{periods} we adopt the periods determined from periodograms, i.e. the string length result is
used as a consistency check. The substantial samples of objects with only one detection are mainly due to 
the subjective nature of the eyeballing. 

In Table \ref{periods} we additionally give an estimate of the peak-to-peak amplitude for the
periodic lightcurves. This has been determined by a) fitting the lightcurve with a sine function
and b) calculating the peak-to-peak amplitude of this function. By using the sine function instead
of the actual lightcurve, we correct for the effect of the noise on the amplitude. These amplitudes
range from 0.01 to 0.04\,mag, with a few outliers with higher values. They do not show any correlation 
with brightness or period.

\begin{table}
\caption{Periods in Praesepe: object ids from \citet[][HSHJ]{1995A&AS..109...29H}
and \citet[][KH]{2007AJ....134.2340K}, J-band magnitude (from 2MASS, converted to CIT), mass
estimate (based on J-band magnitudes), adopted period, amplitude, flag (for explanation see Sect. 
\ref{comb}). With one exception at 85\% (HSHJ412), all objects have membership probabilities 
$\ge 95\%$ according to KH. The periods are ordered by increasing mass (or decreasing J-band 
magnitude); flags $\ge 4$ indicating the most robust periods are marked in bold.}
\label{periods}
\begin{tabular}{rrccccccccc} 
\hline
HSHJ & KH & J-mag & M ($M_{\odot}$) & $P$ (d) & A (mag) & Flag \\
\hline
  -  &  1117  & 15.95 &  0.12 & 0.683 & 0.024 &      1100  \\ 
412  &  1036  & 15.31 &  0.16 & 1.754 & 0.040 & {\bf 1111} \\ 
  -  &  1013  & 15.20 &  0.16 & 0.693 & 0.041 & {\bf 2222} \\ 
233  &   912  & 15.17 &  0.17 & 0.442 & 0.022 &      1010  \\ 
  -  &  1053  & 15.10 &  0.17 & 0.498 & 0.026 &      0110  \\ 
  -  &  1108  & 15.09 &  0.17 & 0.216 & 0.014 &      1110  \\ 
396  &   940  & 15.01 &  0.18 & 0.593 & 0.009 & {\bf 1111} \\ 
258  &  1029  & 14.99 &  0.18 & 0.550 & 0.013 &      1100  \\ 
423  &   880  & 14.99 &  0.18 & 1.179 & 0.013 & {\bf 0211} \\ 
256  &   975  & 14.99 &  0.18 & 1.754 & 0.010 &      1110  \\ 
  -  &   994  & 14.85 &  0.19 & 0.411 & 0.015 & {\bf 1111} \\ 
  -  &   898  & 14.85 &  0.19 & 1.220 & 0.035 & {\bf 2221} \\ 
  -  &   933  & 14.84 &  0.19 & 0.467 & 0.021 & {\bf 1212} \\ 
  -  &   842  & 14.71 &  0.21 & 0.580 & 0.015 & {\bf 2222} \\ 
  -  &   966  & 14.69 &  0.21 & 0.785 & 0.009 &      1110  \\ 
370  &   894  & 14.64 &  0.22 & 1.689 & 0.010 &      1110  \\ 
376  &   731  & 14.56 &  0.23 & 1.276 & 0.017 & {\bf 2210} \\ 
397  &   793  & 14.54 &  0.23 & 1.232 & 0.025 & {\bf 2222} \\ 
  -  &   791  & 14.50 &  0.23 & 1.511 & 0.011 &      1110  \\ 
  -  &   674  & 14.49 &  0.23 & 2.232 & 0.012 &      1110  \\ 
195  &   702  & 14.37 &  0.25 & 1.155 & 0.011 & {\bf 1210} \\ 
421  &   886  & 14.33 &  0.25 & 0.284 & 0.006 &      1100  \\ 
430  &   789  & 14.33 &  0.25 & 2.242 & 0.021 & {\bf 2221} \\ 
428  &   698  & 14.25 &  0.26 & 1.706 & 0.085 & {\bf 2222} \\ 
291  &   770  & 14.22 &  0.27 & 1.276 & 0.027 & {\bf 2221} \\ 
419  &   647  & 14.20 &  0.27 & 3.996 & 0.010 &      1100  \\ 
229  &   957  & 13.89 &  0.32 & 2.294 & 0.008 & {\bf 1210} \\ 
368  &   612  & 13.88 &  0.32 & 4.464 & 0.025 & {\bf 2220} \\ 
  -  &   824  & 13.87 &  0.32 & 1.348 & 0.017 & {\bf 2212} \\ 
295  &   658  & 13.82 &  0.33 & 2.646 & 0.072 & {\bf 2220} \\ 
289  &   741  & 13.76 &  0.34 & 1.672 & 0.024 & {\bf 2222} \\ 
436  &   714  & 13.74 &  0.34 & 4.869 & 0.017 &      2100  \\ 
305  &   822  & 13.55 &  0.37 & 1.207 & 0.010 & {\bf 2201} \\ 
270  &   557  & 13.55 &  0.37 & 4.291 & 0.015 & {\bf 2210} \\ 
  -  &   737  & 13.54 &  0.37 & 0.509 & 0.029 &      0120  \\ 
267  &   802  & 13.53 &  0.38 & 1.106 & 0.015 & {\bf 1211} \\ 
212  &   624  & 13.30 &  0.42 & 0.526 & 0.018 & {\bf 2020} \\ 
272  &   637  & 13.25 &  0.42 & 3.982 & 0.012 & {\bf 2200} \\ 
  -  &   569  & 13.17 &  0.44 & 1.894 & 0.016 &      1110  \\ 
285  &   676  & 13.12 &  0.45 & 5.755 & 0.029 &      1200  \\ 
303  &   603  & 13.10 &  0.45 & 0.842 & 0.016 &      1200  \\ 
435  &   524  & 12.99 &  0.47 & 3.968 & 0.035 &      0210  \\ 
253  &   735  & 12.86 &  0.49 & 0.812 & 0.007 &      1110  \\ 
425  &   532  & 12.92 &  0.48 & 1.153 & 0.013 &      0101  \\ 
366  &   523  & 12.51 &  0.54 & 1.374 & 0.024 &      1110  \\ 
278  &   361  & 12.16 &  0.60 & 4.274 & 0.050 &      1010  \\ 
249  &   486  & 12.16 &  0.60 & 4.854 & 0.016 &      1110  \\ 
  -  &   234  & 11.28 &  0.75 & 5.682 & 0.007 &      0110  \\ 
357  &   369  & 11.19 &  0.76 & 3.268 & 0.020 &      1110  \\ 
\hline 	    
\end{tabular}	    
\end{table}	      

\section{Rotation vs. mass}
\label{rotmass}

For all objects with periods we estimated masses by comparing the J-band magnitudes from 2MASS, converted to
the CIT system, with the BCAH98 model isochrone for 630\,Myr from \citet{1998A&A...337..403B}. As the magnitudes
are constant within 5\% for ages between 500 and 800\,Myr, these mass estimates do not depend on the particular
choice of the age. The uncertainties in the models might introduce systematic errors. Therefore, our masses should
only be compared with values calculated from the same model tracks. The total sample of cluster members 
covered by our observations contains 21, 23, 21, and 146 objects in the mass bins 0.1-0.2, 0.2-0.3, 0.3-0.4,
and $>0.4\,M_{\odot}$. The 'success rate' (i.e. the fraction of objects for which we are able to measure the 
period) is at 48-61\% for the three lowest mass bins, but only 8.9\% for the higher masses. This indicates
that we are missing a substantial numbers of periods for $M>0.4\,M_{\odot}$, either due to the period 
detection limit at $\sim 6$\,d or due to saturation effects.

\begin{figure}
\includegraphics[width=6cm,angle=-90]{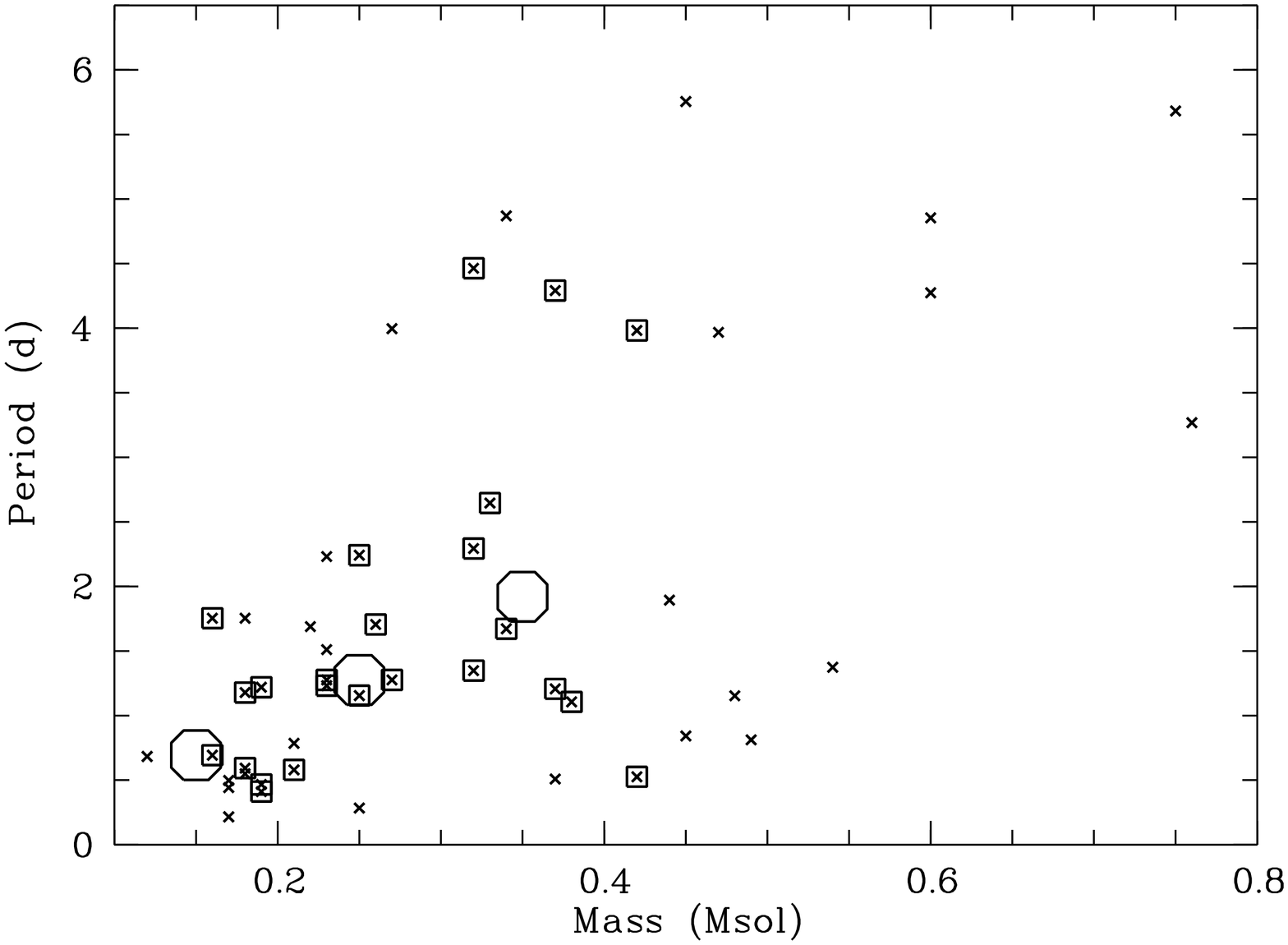} 
\includegraphics[width=6cm,angle=-90]{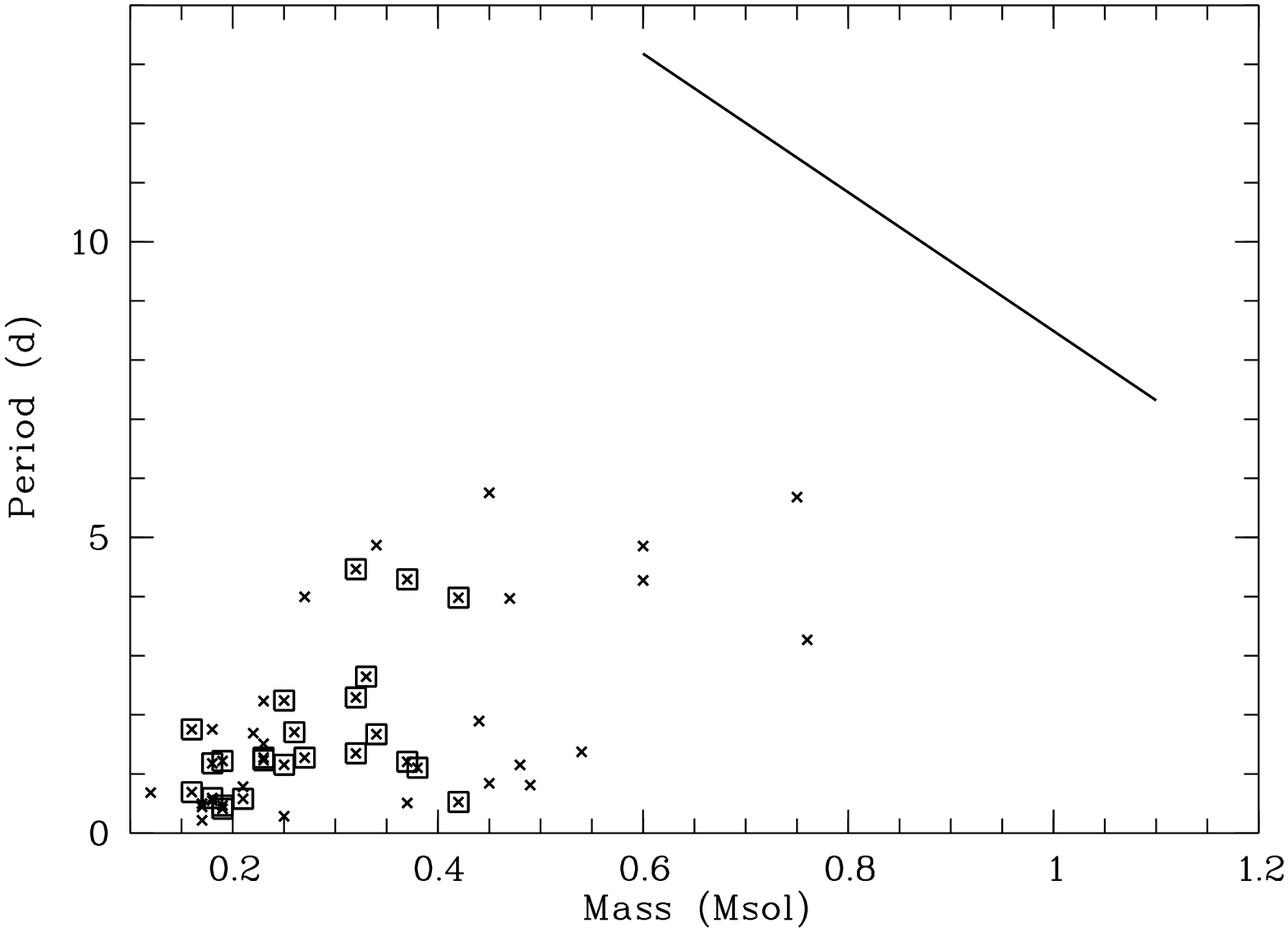} 
\caption{{\bf Upper panel:} Rotation periods in Praesepe from our study plotted vs. mass. Periods with flag $\ge 4$ 
(see Table \ref{periods}) are marked with squares. The median periods in the mass bins 0.1-0.2, 0.2-0.3, and 
0.3-0.4$\,M_{\odot}$ are plotted as octagons.
{\bf Lower panel:} Period-mass relation over the full low-mass range. The solid line shows the period-mass 
relation for low-mass stars in Praesepe, derived from the colour-mass relation given by \citet{2011arXiv1101.1222D}.
 \label{f2}}
\end{figure}

In Fig. \ref{f2} we plot the period sample as a function of object mass; the 24 most robust periods with flag $\ge 4$
are marked with squares. All quantitative results in this Section are derived from these periods. The less robust 
periods with flag of 2-3 generally show the same distribution in this plot. In addition, we overplot 
the median period in the mass bins 0.1-0.2, 0.2-0.3, and 0.3-0.4$\,M_{\odot}$ with large octagons. This clearly 
shows a trend of faster rotation towards lower masses, in line with previous findings
\citep{2005A&A...429.1007S,2007MNRAS.381.1638S}. Below 0.3$\,M_{\odot}$ where the majority of our datapoints
is found there is only one object with $P>2.5$\,d, significantly below our period detection limit (Sect. 
\ref{se2004}). This suggests that very low mass objects in Praesepe are generally fast rotators. The plot also 
reveals that the scatter in the periods in a given mass bin decreases towards lower masses. The total 
spread is 1.34\,d for $M<0.2\,M_{\odot}$, 1.66\,d for $0.2<M<0.3\,M_{\odot}$, and 3.4\,d for 
$0.3<M<0.4\,M_{\odot}$. 

Upon closer inspection Fig. \ref{f2} might indicate a more complex substructure. In particular, we note a dearth
of datapoints for periods between 2.6 and 4.0\,d, which cannot be explained by a lack of sensitivity in the
period search. Furthermore there may be a lack of periods around $M=0.3\,M_{\odot}$ at $P<1$\,d. A larger sample
of periods is needed to verify if these features in the period-mass distribution are real or spurious.

In contrast to the VLM objects discussed here, the more massive stars in Praesepe show a tight correlation
of period vs. mass which is plotted as a solid line in the lower panel of Fig. \ref{f2}. To derive this 
relation, we converted the period-colour relation from \citet{2011arXiv1101.1222D} to period-mass using the 2MASS 
to CIT transformation from \citet{2001AJ....121.2851C} and the linearly fitted colour-mass relation from the 
same BCAH98 isochrone used for our sample. The standard deviation around this relation is 0.46\,d, according
to \citet{2011arXiv1101.1222D}, i.e. the scatter at a given mass is much smaller than for VLM objects. 

The transition from the period-mass relation for 0.6-1.2$\,M_{\odot}$ stars and the VLM regime 
is not fully covered yet by the existing surveys. This regime is difficult to observe because it requires 
deep monitoring over timescales of 2-15\,d. It is clear, however, that between 0.35 and 0.6$\,M_{\odot}$ 
the periods increase strongly with mass, as already stated by \citet{2007MNRAS.381.1638S}. Some of our
objects at $M>0.3\,M_{\odot}$ with relatively long periods might be objects in transition between
the two domains.

Two other properties might have an additional effect on the distribution of datapoints in Fig. \ref{f2},
the configuration of the spots and binarity. Some objects could have specific spot configurations resulting
in an erroneous period measurement. The simplest case for such a configuration is two equally sized spots with 
a longitude difference of 180\,deg. In this scenario these objects would have true rotation periods twice the
measured periods. This could explain some of the scatter in Fig. \ref{f2}.

The effects of binarity are twofold. On one side, we would overestimate their masses. For example, an 
equal-mass binary with 0.35$\,M_{\odot}$ components would be falsely estimated to have a mass of 
$\sim 0.47\,M_{\odot}$. On the other hand, close binaries might be tidally locked and thus faster 
rotating than coeval single stars.

\begin{figure}
\includegraphics[width=6cm,angle=-90]{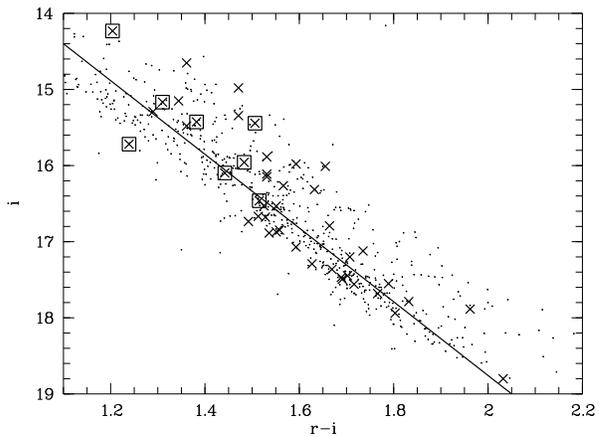} 
\caption{Colour-magnitude diagram for Praesepe members from the KH survey, based on photometry from the
Sloan Digital Sky Survey \citep{2009ApJS..182..543A}. Objects with measured period from Table \ref{periods}
are shown with crosses; slow rotators ($P>3$\,d) are marked with squares. Two objects with periods (KH369 and KH361), 
both slow rotators, are at $i<14$ and $r-i<1.0$ and are not shown here. For two more (KH1013 and KH234) no
valid i-band photometry is available. The solid line marks the assumed limit of the single star sequence.
 \label{f5}}
\end{figure}

As binaries are expected to be brighter than single stars of the same colour, a colour-magnitude diagram as
shown in Fig. \ref{f5} can be used to test for the presence of binaries in our period sample. The distribution
of Praesepe members from KH in colour-magnitude space shows an obvious cumulation at the blue side, interpreted 
as the single star sequence. We approximate this sequence with a straight line and shift it slightly to the red to
estimate the fraction of binaries. Based on this boundary line, the total sample of Praesepe members in Fig.
\ref{f5} has a binary fraction of 351/673 (52\%), almost identical to the value derived by \citet{1999MNRAS.310...87H}
using a similar approach. This ratio is slightly magnitude dependent; for $i<16.5$ we find a higher binary 
fraction of 61\% compared with 45\% for the fainter objects. 

From Fig. \ref{f5} it seems clear that the objects with periods which are brighter than $i\sim 16.5$ exhibit
a significantly higher binary fraction than the fainter ones. We find a binary fraction of 8/24 
($33\pm 12$\%) for $i>16.5$, which is lower, but still consistent with the value for the total sample of members. 
For the objects with $i<16.5$ the fraction is 18/21 ($86\pm^7_{12}\%$), significantly higher than in the
total sample. While the reported binary fractions depend somewhat on the choice of the boundary between
single and binary sequence and should be taken with caution, the {\it increased} binary fraction for the periodic 
objects at $i<16.5$ is robust against variations in this boundary.

This finding might help to explain why the scatter in the periods increases with mass (Fig. \ref{f2}). The 
i-band limit of 16.5\,mag corresponds to a mass of $\sim 0.3\,M_{\odot}$. Above this limit, where we expect
most objects to be binaries, the scatter is roughly twice as large as below. Some 
of them could be spun up tidally locked binaries and thus appear at shorter periods than single stars. It is 
conceivable that most of the single stars at these masses are rotating with periods longer than our detection 
limit, in line with the period-mass correlation seen at 0.6-1.2$\,M_{\odot}$ (solid line in Fig. \ref{f2}). 

Apart from binarity, an alternative explanation for the distribution of the periodic objects in Fig. 
\ref{f5} is the 'blue dwarf phenomenon' described by \citet{2003AJ....126..833S} for the Pleiades and by 
\citet{2009ApJ...691..342H} for M37, a cluster similar in age to Praesepe. The paper by \citet{2009ApJ...691..342H} 
shows that 'at fixed luminosity rapidly rotating late K and early M dwarfs tend to be (...) redder in ($V-I_C$) 
than slowly rotating dwarfs'. This is attributed to increased stellar activity in the fast rotators. Assuming 
that we are missing the slow rotators in the $i<16.5$ range due to our period detection limit at 6\,d, the effect could 
explain why most of the objects with periods in this magnitude range are redder than the single-object isochrone 
in $r-i$. A detailed assessment of these issues requires a more complete period coverage, multi-band photometry 
and/or spectroscopy and is beyond the scope of this paper. 

\section{Rotation vs. age}
\label{rotage}

For the analysis of the rotational evolution, we compare our
period sample in Praesepe with the periods in NGC2516 (age 150\,Myr), as published by 
\citet{2007MNRAS.377..741I}. We select only objects with masses $<0.3\,M_{\odot}$. The NGC2516 sample has 
two outliers with periods $>4$\,d which are excluded. In total, we are working with 96 periods in NGC2516 
and 25 in Praesepe. The lower mass limit in the two period samples is $0.1\,M_{\odot}$. While 
the masses may be systematically off due to model uncertainties, they have been calculated by comparing
photometry from 2MASS with the BCAH tracks in a consistent manner and are thus comparable. 

\begin{figure*}
\includegraphics[width=6.1cm,angle=-90]{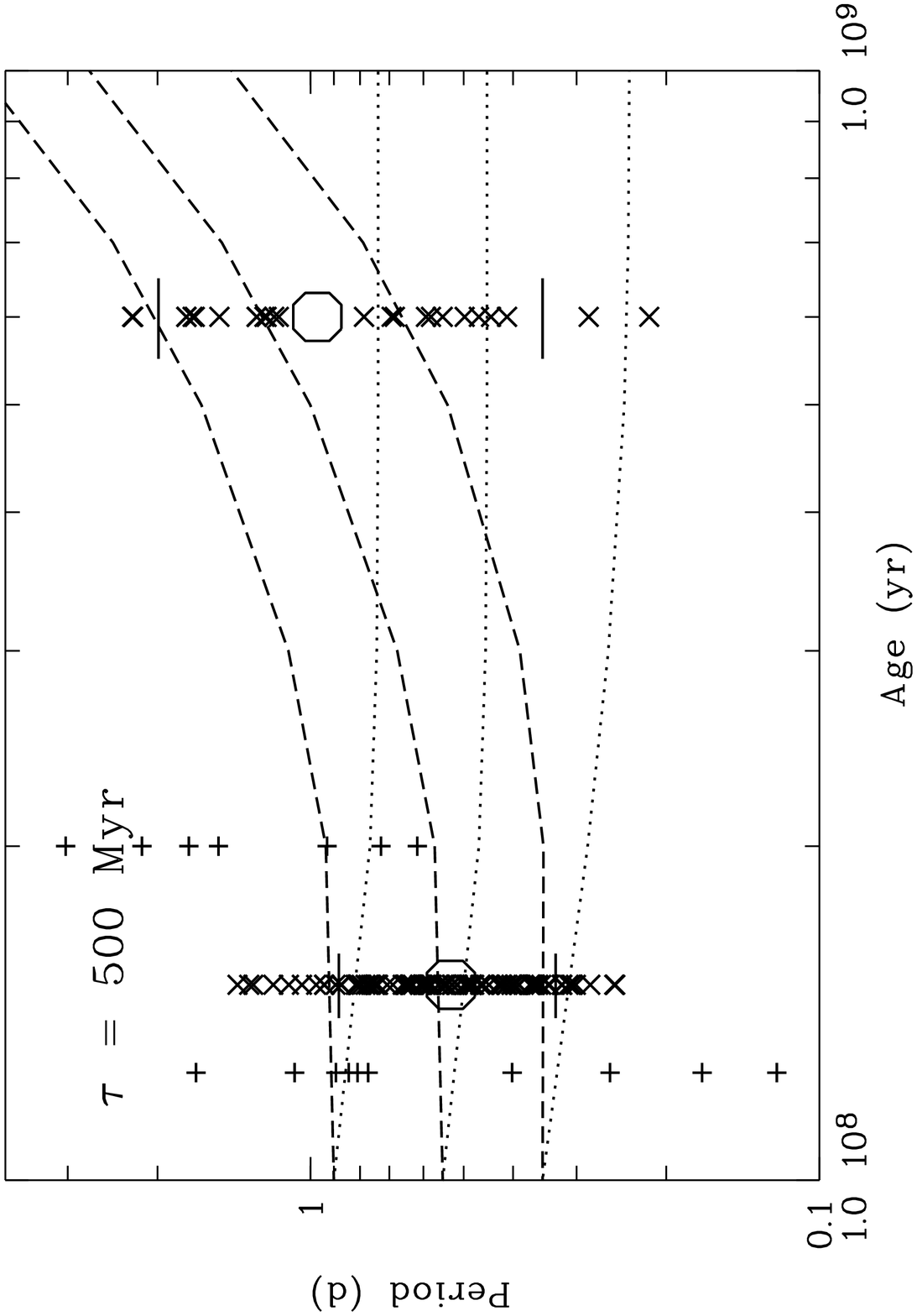} \hfill
\includegraphics[width=6.1cm,angle=-90]{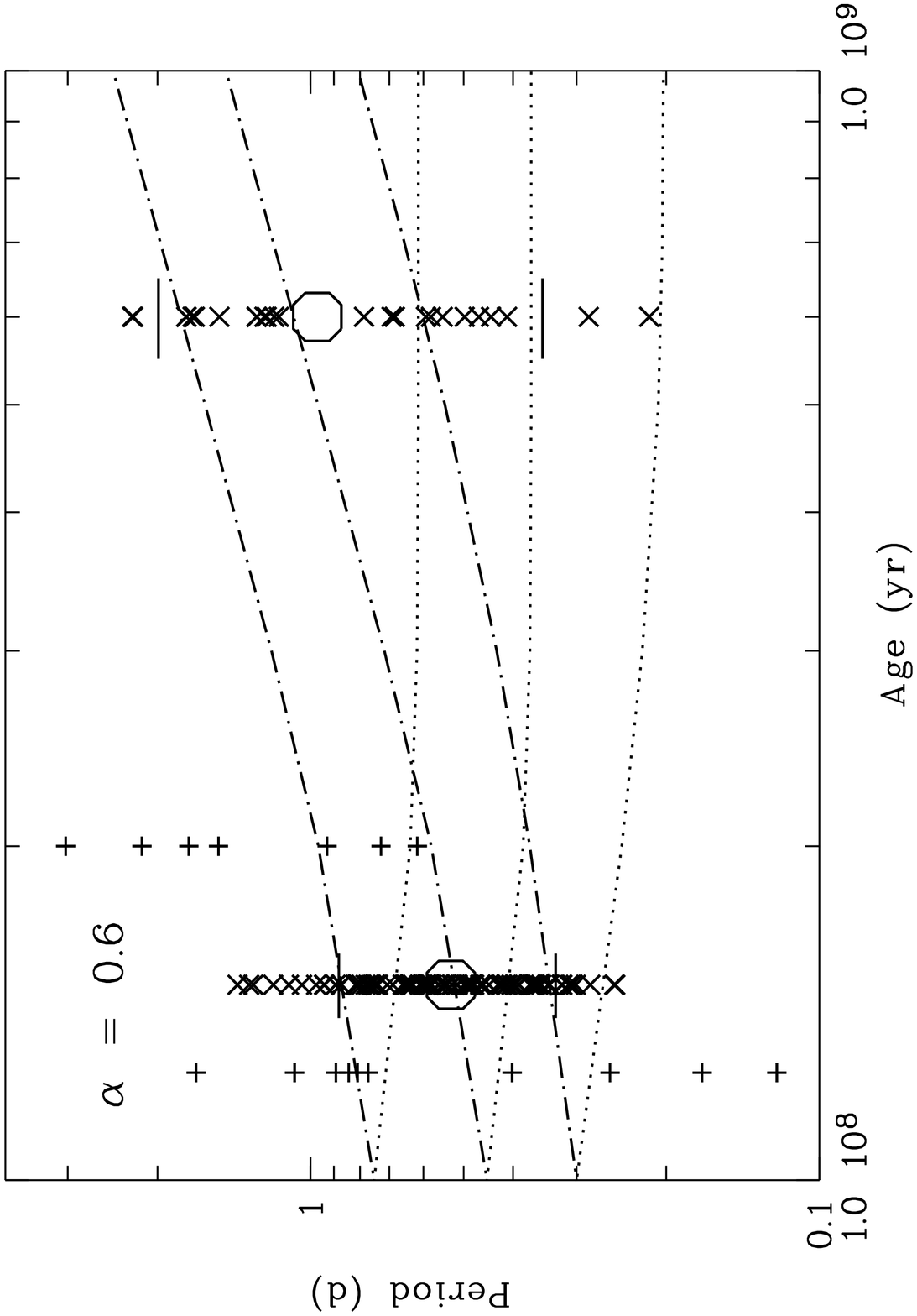} \\
\includegraphics[width=6.1cm,angle=-90]{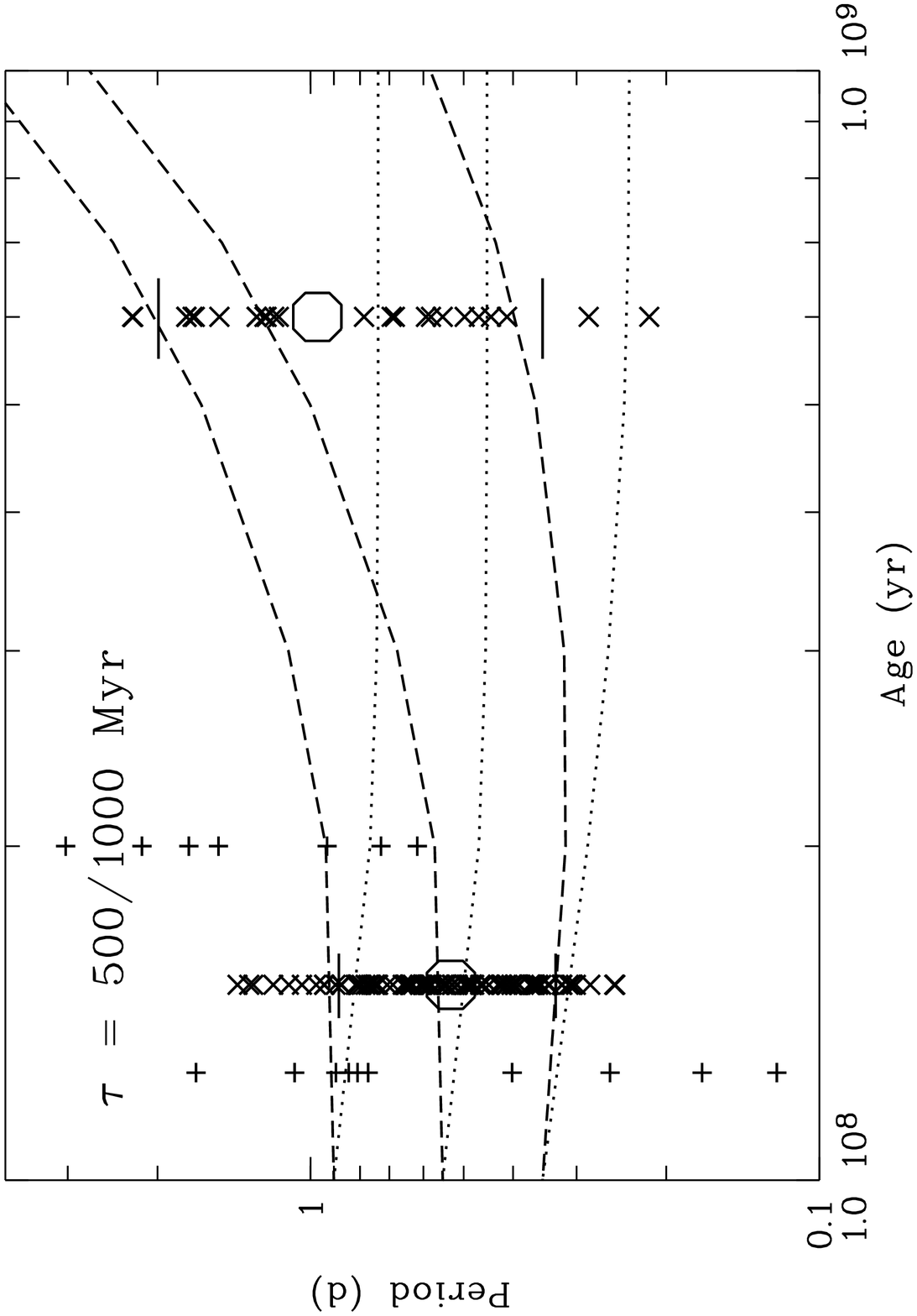} \hfill
\includegraphics[width=6.1cm,angle=-90]{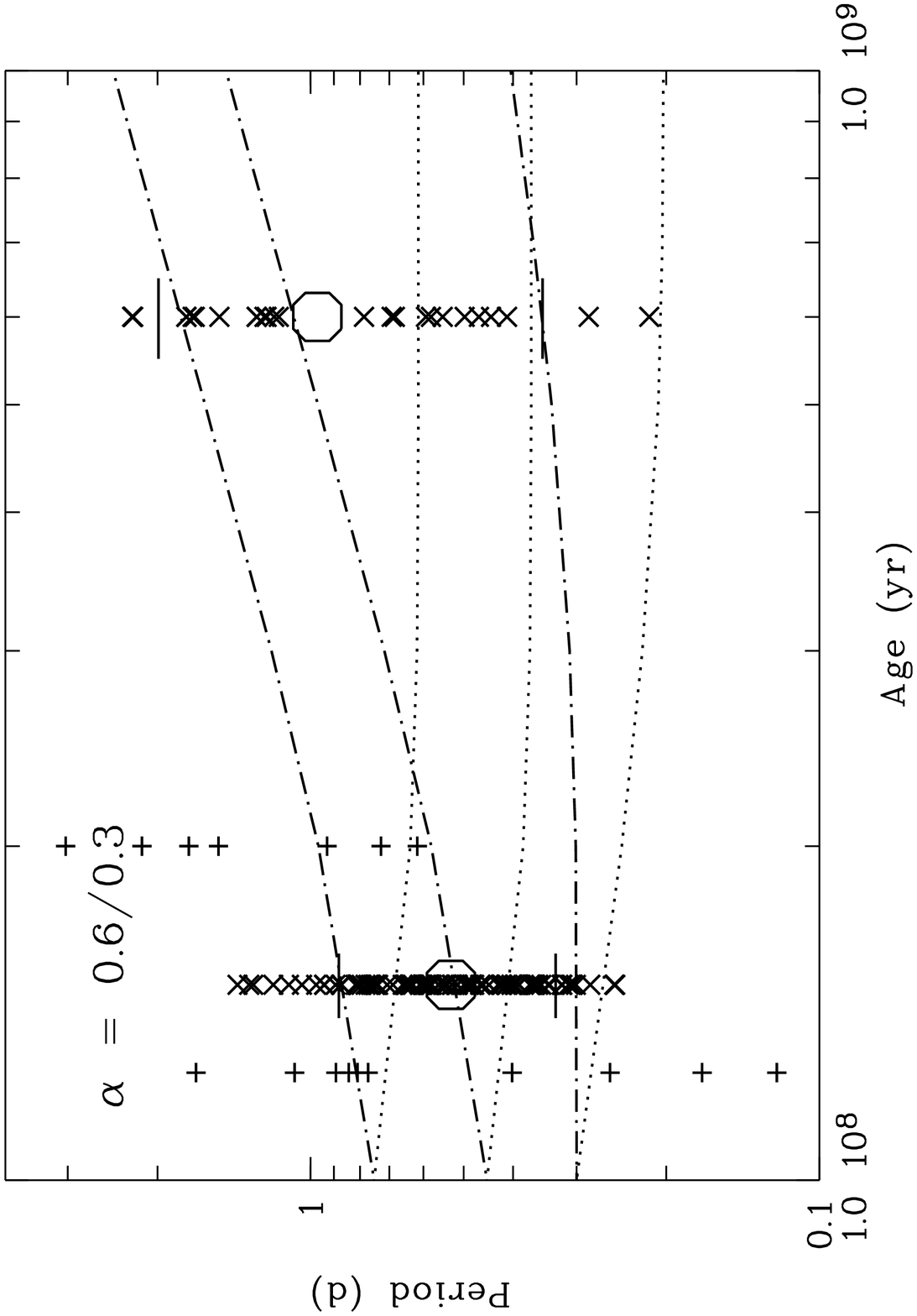} 
\caption{Rotation periods in NGC2516 and Praesepe (crosses) in comparison with spindown tracks. Dashed lines:
exponential spindown; dash-dotted lines: power-law spindown; dotted lines: no spindown. Octagons show
the median of the period distribution, horizontal lines the 10\% and 90\% percentiles of the periods. We 
also overplot periods in the Pleiades and M34 with plusses. \label{f4}}
\end{figure*}

We choose NGC2516 as comparison sample because it is the largest and cleanest VLM period sample
at ages of 100-200\,Myr. Our goal is to isolate the effect of wind braking, therefore we refrain from using
younger clusters where disk braking or contraction might still play a role. The VLM period samples in M34
\citep{2006MNRAS.370..954I} and Pleiades \citep{2004A&A...421..259S} are too small for a meaningful quantitative
analysis, but are consistent with the periods in NGC2516. In M50 the period sample might be severely 
affected by contamination \citep{2009MNRAS.392.1456I}.

The samples in NGC2516 and Praesepe are plotted in Fig. \ref{f4} as a function of age. Median (octagon), 10\%
and 90\% percentiles of the periods (horizontal lines) are overplotted. The uncertainties for the percentiles
are in the range of 0.05/0.1\,d for the 10/90\% limit in Praesepe and 0.02\,d for NGC2516. In both cases, the 
analysis is sensitive to periods from 0.1 to at least 5\,d. Thus, both the upper and lower period limits are 
likely to be reliable. In Praesepe there is a clear gap around 1\,d due to the strongly clumped sampling of 
the time series, which could indicate that the median is in fact slightly higher.

The main caveat, however, is the fact that it is more difficult to find periods at the low-mass end. Combined 
with the period-mass trend, this implies a possible underrepresentation of fast rotators in these samples. Thus, 
the median and the lower limit could be overestimated. This problem likely applies to both clusters in a similar
way; in NGC2516 the completeness drops from 100\% to 20\% around $M=0.15\,M\,{\odot}$ \citep{2007MNRAS.377..741I}.

We aim to reproduce the evolution of the median and the 10\% and 90\% percentiles by simple 
evolutionary tracks. We consider an exponential spindown law $P \propto \exp{(t/\tau)}$ (left panels 
in Fig. \ref{f4}) and a power law $P \propto t^{\alpha}$ (right panels). In the upper row of Fig. \ref{f4} we
attempt to fit the median and upper/lower limit with the same spindown law (i.e. with constant $\tau$ 
or $\alpha$), whereas we allow for varying parameters in the lower row. In addition, we vary the initial 
period $P_{\mathrm{ini}}$ slightly to match the observed periods in NGC2516. Fig. \ref{f4} additionally 
shows the period evolution without any angular momentum loss. 

For all spindown tracks, we take into account the contraction by multiplying with $R/R_{\mathrm{ini}}$,
where we use radii from the BCAH98 isochrones. To mimick the period-mass trend, we use the radii for 
an 0.3$\,M_{\odot}$ object for the period upper limit, 0.2$\,M_{\odot}$ for the median, and $0.1\,M_{\odot}$
for the lower limit. The tracks do not significantly depend on this choice; they are mostly parallel for 
all three masses. As seen in the tracks without spindown, the contraction causes the rotation periods 
to drop by 20-30\%, depending on mass.

This analysis yields two main results: First, the period evolution cannot be explained with zero 
rotational braking (see dotted lines in the left panel of Fig. \ref{f4}). Tracks without spindown predict
the median period in Praesepe to be at $\sim 0.5$\,d, whereas the observed median is at 1.0\,d. Similarly, these
tracks predict an upper and lower period limit significantly lower than observed. Thus, VLM
stars undergo angular momentum losses on the main-sequence.

Second, the spindown tracks with constant parameters cannot simultaneously explain the evolution of upper limit,
median, and lower limit. This is seen for exponential and for power law spindown laws (dashed and dash-dotted
lines in the upper panels of Fig. \ref{f4}). The upper limit and the median can be reasonably well reproduced 
with $\tau = 500$\,Myr or $\alpha = 0.6$, which is consistent with the result from \citet{2007MNRAS.381.1638S}
based on much smaller period samples. However, the lower limit requires different parameters. Simply speaking, the
fast rotators in Praesepe rotate too fast to be explained by the same spindown law as the slower rotators.

There are two ways to explain this second finding. On one side it is possible that we are missing a significant 
population of fast rotators in NGC2516. Based on the similar time sampling and mass coverage in NGC2516 and Praesepe,
this currently seems implausible. However, the small period sample in the Pleiades does contain 2 such objects with
$P<0.2$\,d, out of 9 in total. A larger VLM period sample in a cluster at $\sim 100$\,Myr would be valuable to 
exclude this option.

On the other side we consider the possibility that the spindown is not constant and the lower mass objects (and
faster rotators) are affected by less rotational braking. In the lower panels of Fig. \ref{f4} we show exponential 
and power law tracks with variable $\tau$ and $\alpha$. For the median and the upper limit we take the same parameters 
as before, but for the lower period limit we use $\tau = 1000$\,Myr and $\alpha = 0.3$. The plot shows that this choice of 
parameters matches the data quite well. Thus, it is plausible that the spindown law in the VLM regime is a function 
of stellar mass. 

It should be noted that the choice of $\tau = 1000$\,Myr is really a lower limit; as the exponential spindown is very 
slow at these high values for $\tau$, significantly longer timescales are plausible. To improve the estimate on $\tau$ 
and to distinguish between power law and exponential spindown, it is necessary to test the period evolution on longer 
timescales. This is done by \citet{2010arXiv1011.4909I} using a sample of periods for field stars with 
$0.1<M<0.3\,M_{\odot}$ from the MEarth project\footnote{A few additional periods for this mass regime are available 
from \citet{2007AcA....57..149K}}. Their periods show a wide spread from 0.3 to 150\,d, which rules out the power law
spindown tested in Fig. \ref{f4}. In addition, their analysis yields $\tau = 5-10$\,Gyr for the fastest rotating
VLM objects.

\section{Summary and outlook}

The analysis in Sections \ref{rotmass} and \ref{rotage} can be summarised as follows: 
\begin{enumerate}
\item{At ages of 600\,Myr, VLM objects with masses below $0.3\,M_{\odot}$ are almost exclusively fast rotators with periods 
of $<2.5$\,d. This is in stark contrast to higher mass stars (0.6-1.2$\,M_{\odot}$) which have periods of 7-14\,d and results
in a sharp break in the period-mass relation at 0.3-0.6$\,M_{\odot}$.}
\item{In the VLM regime, the periods as well as the scatter in the periods increases with mass. The scatter is 
significantly larger than for 0.6-1.2$\,M_{\odot}$ stars.}
\item{Between 100 and 600\,Myr VLM objects experience angular momentum losses. However, a single spindown law cannot
explain the evolution of upper and lower period limit simultaneously. Instead, the fast rotators exhibit
less rotational braking than the slow rotators. The exponential spindown timescale increases steeply from 
$\sim $500\,Myr for 0.3$\,M_{\odot}$ to several Gyrs at 0.1$\,M_{\odot}$.}
\end{enumerate}

These results are in line with the currently used models for the rotational evolution. As outlined in 
Sect. \ref{intro}, most recent papers on this subject use a Skumanich-type $P \propto t^{0.5}$ law for the slow 
and an exponential law $P \propto \exp{(t)}$ for the fast rotators. This approach is supported by
our new data. In particular, the periods in Praesepe show that VLM objects at 600\,Myr rotate much faster than 
F-K type stars and have not arrived yet to the Skumanich-type spindown tracks. 

The physical origin of the empirical findings outlined above is a matter of debate. The breakdown of the Skumanich law
around $\sim 0.3-0.6\,M_{\odot}$ can possibly be understood as a consequence of interior structure. Going from 
solar-mass stars down to the VLM regime, the convective zone deepens until the objects become fully convective 
around 0.35$\,M_{\odot}$ \citep{1997A&A...327.1039C}. Assuming that magnetic field generation and properties are 
a function of interior structure, this could qualitatively explain why VLM objects spend longer on the exponential 
track than more massive stars. 

The mass-dependence of the spindown timescale might actually be a dependence on $T_{\mathrm{eff}}$, 
as argued in \citet{2004PhDT.........4S}. With decreasing temperature, the electrical 
conductivity of the photospheric gas drops as well and the coupling between gas and magnetic field becomes less 
efficient. This might affect the mass load of the flux tubes and the efficiency of the stellar wind, and 
explain the strong increase in the spindown timescale at very low masses. It would also provide an explanation 
for the universally fast rotation of the ultracool L dwarfs \citep{2008ApJ...684.1390R}.

The positive period-mass trend at very low masses is already seen in very young clusters 
\citep[e.g.][]{2005A&A...429.1007S}, albeit with much more scatter than in Praesepe. Thus, this feature is likely 
a remnant of the initial conditions. More detailed theoretical work on the magnetic field generation, wind
physics, and their connection to angular momentum loss is required to substantiate this interpretation.

Observationally, the current database still has two major flaws that need to be addressed: a) Too few clusters
have been monitored with good time sampling and depth. As a result, we are lacking observational constraints
in the substellar regime and cannot fully exclude to be affected by bias in the period range and possible 
environmental effects on rotation, as recently reported by \citet{2010MNRAS.403..545L}. b) Follow-up observations are 
required for cluster objects with known rotation periods, to exclude contaminating field stars and obtain 
complementary information about the stellar and magnetic field properties. 

\section*{Acknowledgments}
We thank the referee, Joel Hartman, for a constructive report that helped to improve the paper.
AS would like to thank Philippe Delorme who made his results available prior to publication.
BS is supported by RoPACS, a Marie Curie Initial Training Network funded by the European Commission's 
Seventh Framework Programme. Part of this work was funded by the Science Foundation Ireland
through grant no. 10/RFP/AST2780 to AS. 
 
\newcommand\aj{AJ} 
\newcommand\actaa{AcA} 
\newcommand\araa{ARA\&A} 
\newcommand\apj{ApJ} 
\newcommand\apjl{ApJ} 
\newcommand\apjs{ApJS} 
\newcommand\aap{A\&A} 
\newcommand\aapr{A\&A~Rev.} 
\newcommand\aaps{A\&AS} 
\newcommand\mnras{MNRAS} 
\newcommand\pasa{PASA} 
\newcommand\pasp{PASP} 
\newcommand\pasj{PASJ} 
\newcommand\solphys{Sol.~Phys.} 
\newcommand\nat{Nature} 
\newcommand\bain{Bulletin of the Astronomical Institutes of the Netherlands}

\bibliographystyle{mn2e}
\bibliography{aleksbib}

\label{lastpage}

\end{document}